\RequirePackage{ifpdf}
\ifpdf 
\documentclass[pdftex]{sigma}
\else
\documentclass{sigma}
\fi

\begin{document}
\allowdisplaybreaks

\renewcommand{\PaperNumber}{006}

\def\C{\mathbb C}
\def\R{\mathbb R}
\def\Z{\mathbb Z}
\def\N{\mathbb N}
\def\O{\Omega}
\def\p{\partial}
\def\r{\rangle}
\def\l{\langle}
\def\p{\partial}

\def\tr{\operatorname{tr}}
\def\diag{\operatorname{diag}}

\FirstPageHeading

\ShortArticleName{Orbit Functions}

\ArticleName{Orbit Functions}

\Author{Anatoliy KLIMYK~$^\dag$ and Jiri PATERA~$^\ddag$}
\AuthorNameForHeading{A.~Klimyk and J.~Patera}

\Address{$^\dag$~Bogolyubov Institute for Theoretical Physics,
       14-b  Metrologichna Str., Kyiv, 03143 Ukraine}

\EmailD{\href{mailto:aklimyk@bitp.kiev.ua}{aklimyk@bitp.kiev.ua}}

\Address{$^\ddag$~Centre de Recherches Math\'ematiques,
         Universit\'e de Montr\'eal,\\
$\phantom{^\ddag}$~C.P.6128-Centre ville,
         Montr\'eal, H3C\,3J7, Qu\'ebec, Canada}
\EmailD{\href{mailto:patera@crm.umontreal.ca}{patera@crm.umontreal.ca}}

\ArticleDates{Received January 04, 2006; Published online January
19, 2006}

\Abstract{In the paper, properties of orbit functions are reviewed
and further developed. Orbit functions on the Euclidean space
$E_n$ are symmetrized exponential functions. The symmetrization is
fulfilled by a Weyl group corresponding to a Coxeter--Dynkin
diagram. Properties of such functions will be described. An orbit
function is the contribution to an irreducible character of a
compact semisimple Lie group $G$ of rank $n$ from one of its Weyl
group orbits. It is shown that values of orbit functions are
repeated on copies of the fundamental domain $F$ of the affine
Weyl group (determined by the initial Weyl group) in the entire
Euclidean space $E_n$. Orbit functions are solutions of the
corresponding Laplace equation in $E_n$, satisfying the Neumann
condition on the boundary of $F$. Orbit functions determine a
symmetrized Fourier transform and a transform on a finite set of
points.}


\Classification{33-02; 33E99; 42C15; 58C40}

\section{Introduction}
In this paper we consider orbit functions --- multivariable
functions having many beautiful and useful properties and
applicable both in mathematics and engineering. For this reason,
they certainly can be treated as special functions \cite{P04},
although there is no generally accepted definition of special
functions \cite{KV, M, Vi}.

Orbit functions are closely related to finite groups $W$ of
geometric symmetries generated by reflection transformations $r_i$
(that is, such that $r_i^2=1$), $i=1,2,\ldots ,n$, of an
$n$-dimensional Euclidean space $E_n$ with respect to
$(n-1)$-dimensional subspaces containing the origin. Namely, we
take a point $\lambda \in E_n$ (in a certain coordinate system)
and act upon $\lambda$ using all elements of the group $W$. If
$O(\lambda)$ is the orbit of the point $\lambda$, that is the set
of all different points of the form $w\lambda$, $w\in W$, then the
orbit function, determined by the point $\lambda$, coincides with
\[
\phi_\lambda(x)=\sum_{\mu\in O(\lambda)} e^{2\pi{\rm i}\langle
\mu,x\rangle},
\]
where $\langle\mu,x\rangle$ is the scalar product on $E_n$.
Clearly, these functions are invariant with respect to the group
$W$: $\phi_\lambda(wx)=\phi_\lambda(x)$, $w\in W$. This is the
main property of orbit functions, a property which contributes in
great part to the usefulness of orbit functions in applications.
An orbit function $\phi_\lambda(x)$ can be understood as a
symmetrized (by means of the group $W$) exponential function.

In this paper, we consider only orbit functions related to the
symmetry groups $W$, which are Weyl groups of semisimple Lie
groups (semisimple Lie algebras). Such orbit functions are closely
related to irreducible representations of the corresponding
semisimple compact Lie groups $G$. Namely, for $\lambda$ with
integral coordinates, they are constituents of traces (characters)
of finite dimensional representations of $G$. Although characters
contain all (or almost all) information about the corresponding
irreducible representations, they are seldom used as special
functions due to the fact that a construction of characters is
rather complicated, whereas orbit functions have a much simpler
structure.

Orbit functions can be considered as a certain modification of
monomial symmetric (Laurent) polynomials $m_\lambda(y)$ which are
determined by the formula $m_\lambda(y)=\sum\limits_{\mu\in
O(\lambda)} y^\mu$, where $y=(y_1,y_2$, $\ldots ,y_n)$, $\mu
=(\mu_1,\mu_2,\ldots ,\mu_n)$ and $y^\mu=y_1^{\mu_1}\cdots
y_n^{\mu_n}$. Namely, replacing $e^{2\pi{\rm i}x_i}$ by $y_i$,
$i=1,2,\ldots ,n$, in the definition of an orbit function gives a
corresponding monomial symmetric polynomial. It is generally
accepted that monomial symmetric polynomials play a very important
role in the theory of symmetric polynomials \cite{Mac1, Mac2,
Mac3, KVl} (Schur polynomials, Macdonald symmetric polynomials,
Jacobi polynomials of many variables, etc.).

If we were to expand on the practical applicability of orbit
functions, it would be necessary to reference works relevant to
our present topic, and to consider some `abstract' and rather
challenging problems \cite{MP87, MP84, MMP85, MMP86, PS, GP}.

The motivation for this article is the recent recognition of the
fact that orbit functions are indispensable to the vast
generalization of the discrete cosine transform \cite{AP}, which
has proven to be very useful in recent years \cite{R}. In
addition, it appears that it is difficult to overestimate the
future role of orbit functions in some applications requiring data
compression \cite{AP}.

A generalization of the discrete cosine transform based on orbit
functions (discrete orbit function transform) can be useful in
models of quantum field theory on discrete lattices (especially
when such models admit a discrete symmetry).

Our present goal is to bring together, in full generality, diverse
facts about orbit functions, many of which are not found in the
literature, although they often are straightforward consequences
of known facts.

In general, for a given transformation group $W$ of the Euclidean
space $E_n$, most of the properties of orbit functions described
in this paper are implications of properties either of orbits of
the group $W$ (for description of reflection groups see, for
example, \cite{Kane} and \cite{Hem-1}), or of the irreducible
characters of the corresponding compact semisimple Lie group $G$
of rank $n$ (for an exposition of the theory of finite dimensional
representations of semisimple Lie groups and their characters see,
for example, \cite{Hem-2}).

The title of the article refers to the problem which was solved in
\cite{Pi} for an equilateral triangle, which is the fundamental
domain of the affine Weyl group $W^{\rm aff}$, corresponding to
the simple Lie algebra $A_2$ and to the compact group $SU(3)$.
Through the use of an entirely different method~\cite{pat}, orbit
functions provide a solution to this problem for any compact
simple Lie group. It was shown that orbit functions are
eigenfunctions of the $n$-dimensional Laplace operator on the
simplexes, which are fundamental domains of compact simple Lie
groups, with the Neumann boundary value requirement (normal
derivatives are zero). The eigenvalues are given explicitly.

For each transformation group $W$, the orbit functions
$\phi_\lambda$ with integral $\lambda$ form a complete orthogonal
basis in the space of symmetric (with respect to $W$) polynomials
in $e^{2\pi{\rm i}x_j}$ or in the Hilbert space obtained by
closing this space with respect to an appropriate scalar product.

In the case where the group $W$ is a direct product of its
subgroups, say $W=W_1\times W_2$, the fundamental domain is the
Cartesian product of fundamental domains for $W_1$ and $W_2$.
Similarly, the orbit functions of $W$ are products of orbit
functions of $W_1$ and $W_2$. Hence it suffices to carry out our
considerations for groups $W$ which cannot be represented as a
product of its subgroups (that is, for such $W$ for which a
Coxeter--Dynkin diagram is connected).

It is generally accepted that the $n$-dimensional exponential
function $e^{2\pi {\rm i}(m_1x_1+\cdots +m_nx_n)}$ can be
considered for integral values of $m_1,m_2,\ldots,m_n$. Then it is
a function on a torus, and determines a decomposition into Fourier
series. When the function is considered for any real values of
$m_1,m_2,\ldots,m_n$, then it is a function on the whole space
$\mathbb R$ and determines a decomposition into Fourier integrals
of exponential functions. Orbit functions have similar properties.
For integral~$\lambda$ (that is, when coordinates of $\lambda$ in
a certain coordinate system are integers), orbit functions are
functions on a torus admitting a symmetry with respect to the Weyl
group $W$. They determine a decomposition into a sum of orbit
functions (symmetrized Fourier series). If we admit any real
values for coordinates of $\lambda$, then they are functions on
$\mathbb R$ (admitting a symmetry with respect to $W$). We then
have to replace decompositions into series by decompositions into
integrals of orbit functions. The first case is more interesting
for applications. Nevertheless, we consider in our exposition a
general case of orbit functions, with emphasis on an integral
$\lambda$.

The symmetrized Fourier transform (the orbit function transform)
is of importance in dif\-ferent branches of pure and applied
mathematics in the context of theories which admit discrete
symmetries. In particular, such symmetries are often encountered
in theoretical and mathe\-ma\-ti\-cal physics.

In this paper we deal only with orbit functions symmetric with
respect to Weyl groups. There exist antisymmetric orbit functions
\cite{P-05}. They will be reviewed in a separate paper.

In this article, many examples of dimensions 2 and 3 are shown
because they are likely to be used more often. Hopefully, they can
be understood without the general arguments presented in this
paper by directly verifying their properties through explicit
calculation.

However, since our conclusions are to be general as to type of the
transformation group $W$, it is imperative to profit from the
uniformity of the pertinent parts of the theory of geometric
symmetry groups: many general facts of the theory are recalled
without explanation and with only a few references to the
literature.

\renewcommand{\theequation}{\arabic{section}.\arabic{equation}}

\section{Weyl group orbits}
\label{weyl}

\subsection{Coxeter--Dynkin diagrams and Cartan matrices}\label{DD}

As indicated in the Introduction, the sets of orbit functions on
the $n$-dimensional Euclidean space $E_n$ are determined by finite
transformation groups $W$, generated by reflections $r_i$,
$i=1,2,\ldots ,n$ (a characteristic property of reflections is the
equality $r^2_i=1$). We are interested in groups $W$ which are
Weyl groups of semisimple Lie groups (semisimple Lie algebras).
Such Weyl groups, together with the corresponding systems of
reflections $r_i$, $i=1,2,\ldots ,n$, are determined by
Coxeter--Dynkin diagrams (for the theory of such groups, see
\cite{Kane} and \cite{Hem-1}). There are 4 series and 5 separate
simple Lie algebras, each with a uniquely determined Weyl groups
$W$. They are denoted as
\[
A_n\ (n\geq1),\quad B_n\ (n\geq3),\quad C_n\ (n\geq2),\quad D_n\
(n\geq4), \quad E_6,\quad E_7,\quad E_8,\quad F_4,\quad G_2.
\]
These Lie algebras correspond to connected Coxeter--Dynkin
diagrams. The index below denotes a number of simple roots. This
number is called a rank of the corresponding simple Lie algebra.

Semisimple Lie algebras (direct sums of simple Lie subalgebras)
correspond to Coxeter--Dynkin diagrams, which in turn consist of
connected parts corresponding to simple Lie subalgebras; these
parts are not connected with each other (for a description of the
correspondence between simple Lie algebras and Coxeter--Dynkin
diagrams, see, for example, \cite{Hem-2}). Thus, it is sufficient
to describe the Coxeter--Dynkin diagrams that correspond to simple
Lie algebras. They have the form

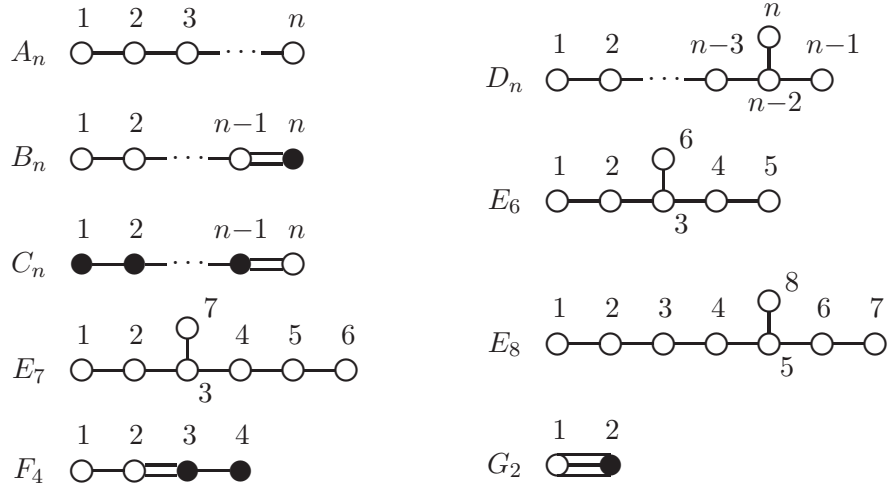
\begin{center}

\parbox{.6\linewidth}{\setlength{\unitlength}{2pt}
\def\kr{\circle{4}}
\def\cr{\circle*{4}}
\thicklines
\hspace*{-10pt}\begin{picture}(180,90)

\put(0,80){\makebox(0,0){${A_n}$}} \put(10,80){\kr} \put(9,85){1}
\put(20,80){\kr} \put(19,85){2} \put(30,80){\kr} \put(29,85){3}
\put(50,80){\kr} \put(49,85){$n$} \put(12,80){\line(1,0){6}}
\put(22,80){\line(1,0){6}} \put(32,80){\line(1,0){4}}
\put(37,80){$\ldots$} \put(44,80){\line(1,0){4}}

\put(0,60){\makebox(0,0){${B_n}$}}   
\put(10,60){\kr} \put(9,65){1}   \put(20,60){\kr} \put(19,65){2}
\put(40,60){\kr}  \put(35,65){$n{-}1$}  \put(50,60){\cr}
\put(49,65){$n$} \put(12,60){\line(1,0){6}}
\put(22,60){\line(1,0){4}} \put(27,60){$\ldots$}
\put(34,60){\line(1,0){4}} \put(42,61){\line(1,0){6}}
\put(42,59){\line(1,0){6}}

\put(0,40){\makebox(0,0){${C_n}$}}   
\put(10,40){\cr}  \put(9,45){1}   \put(20,40){\cr}
\put(19,45){2} \put(40,40){\cr}  \put(35,45){$n{-}1$}
\put(50,40){\kr}  \put(49,45){$n$} \put(12,40){\line(1,0){6}}
\put(22,40){\line(1,0){4}}  \put(27,40){$\ldots$}
\put(34,40){\line(1,0){4}} \put(42,41){\line(1,0){6}}
\put(42,39){\line(1,0){6}}

\put(90,75){\makebox(0,0){${D_n}$}}    
\put(100,75){\kr}  \put(99,80){1}  \put(110,75){\kr}
\put(109,80){2}
  \put(130,75){\kr} \put(125,80){$n{-}3$}
\put(140,75){\kr}  \put(136,69){$n{-}2$}  \put(150,75){\kr}
\put(147,80){$n{-}1$} \put(140,83){\kr}  \put(139,87){$n$}
\put(102,75){\line(1,0){6}} \put(112,75){\line(1,0){4}}
\put(117,75){$\ldots$} \put(124,75){\line(1,0){4}}
\put(132,75){\line(1,0){6}} \put(142,75){\line(1,0){6}}
\put(140,77){\line(0,1){4}}

\put(90,52){\makebox(0,0){${E_6}$}}     
\put(100,52){\kr} \put(99,57){1}  \put(110,52){\kr}
\put(109,57){2} \put(120,52){\kr} \put(122,46){3}
\put(130,52){\kr}  \put(129,57){4} \put(140,52){\kr}
\put(139,57){5}  \put(120,60){\kr} \put(123,62){6}
\put(102,52){\line(1,0){6}} \put(112,52){\line(1,0){6}}
\put(122,52){\line(1,0){6}} \put(132,52){\line(1,0){6}}
\put(120,54){\line(0,1){4}}

\put(0,20){\makebox(0,0){${E_7}$}}   
\put(10,20){\kr} \put(9,25){1}  \put(20,20){\kr}  \put(19,25){2}
\put(30,20){\kr}  \put(32,14){3}  \put(40,20){\kr}  \put(39,25){4}
\put(50,20){\kr}  \put(49,25){5}  \put(60,20){\kr}  \put(59,25){6}
\put(30,28){\kr}   \put(33,30){7} \put(12,20){\line(1,0){6}}
\put(22,20){\line(1,0){6}} \put(32,20){\line(1,0){6}}
\put(42,20){\line(1,0){6}} \put(52,20){\line(1,0){6}}
\put(30,22){\line(0,1){4}}

\put(90,25){\makebox(0,0){${E_8}$}} \put(100,25){\kr}
\put(99,30){1}  \put(110,25){\kr}  \put(109,30){2}
\put(120,25){\kr}  \put(119,30){3}  \put(130,25){\kr}
\put(129,30){4} \put(140,25){\kr}  \put(142,19){5}
\put(150,25){\kr}  \put(149,30){6} \put(160,25){\kr}
\put(159,30){7}  \put(140,33){\kr}  \put(143,35){8}
\put(102,25){\line(1,0){6}} \put(112,25){\line(1,0){6}}
\put(122,25){\line(1,0){6}} \put(132,25){\line(1,0){6}}
\put(142,25){\line(1,0){6}} \put(152,25){\line(1,0){6}}
\put(140,27){\line(0,1){4}}

\put(0,1){\makebox(0,0){${F_4}$}}   
\put(10,1){\kr}   \put(9,6){1}   \put(20,1){\kr}  \put(19,6){2}
\put(30,1){\cr}  \put(29,6){3}   \put(40,1){\cr} \put(39,6){4}
\put(12,1){\line(1,0){6}} \put(22,2){\line(1,0){6}}
\put(22,0){\line(1,0){6}} \put(32,1){\line(1,0){6}}
\put(90,2){\makebox(0,0){${G_2}$}} \put(100,2){\kr} \put(99,7){1}
\put(110,2){\cr}  \put(109,7){2} \put(102,2){\line(1,0){6}}
\put(100,4){\line(1,0){10}} \put(100,0){\line(1,0){10}}
\end{picture}}
\end{center}

\bigskip

A diagram gives a certain non-orthogonal basis
$\{\alpha_1,\dots,\alpha_n\}$ in the Euclidean space $E_n$. Each
node is associated with a basis vector $\alpha_k$, called a {\it
simple root}. A direct link between two nodes indicates that the
corresponding basis vectors are not orthogonal. Conversely, the
absence of a direct link between nodes implies orthogonality of
the corresponding vectors. Single, double, and triple links
indicate that relative angles between the corresponding two simple
roots are $2\pi/3$, $3\pi/4$, $5\pi/6$, respectively. There may be
only two cases: all simple roots are of the same length or there
are only two different lengths of simple roots. In the first case
all simple roots are denoted by white nodes. In the case of two
lengths, shorter roots are denoted by black nodes and longer ones
by white nodes. Lengths of roots are determined uniquely up to a
common constant. For the cases $B_n, C_n,$ and $F_4$, the squared
longer root length is double the squared shorter root length. For
$G_2$, the squared longer root length is triple the squared
shorter root length.

It is necessary to fix an order of simple roots. For this reason,
roots on Coxeter--Dynkin diagrams are numbered.

To each Coxeter--Dynkin diagram there corresponds a Cartan matrix
$M$, consisting of the entries
\begin{gather}\label{Mmatrix}
M_{jk}=\frac{2\l\alpha_j ,\alpha_k\r}
            {\l\alpha_k ,\alpha_k\r},\qquad
            j,k\in\{1,2,\ldots,n\},
\end{gather}
where $\l x, y\r$ denotes a scalar product of $x,y\in E_n$. All
numbers $M_{ij}$ are integers. Cartan matrices of simple Lie
algebras are given in many books (see, for example, \cite{BMP}).
We give them for the cases of ranks $2$ and $3$ because of their
frequent usage later on:
\[
 A_2 :\ \left(
 \begin{array}{cc}
 2&-1\\ -1&2\
 \end{array} \right) ,\qquad
C_2 :\  \left(
 \begin{array}{cc}
 2&-1\\ -2&2\
 \end{array} \right) ,\qquad
 G_2 :\
 \left(
 \begin{array}{cc}
 2&-3\\ -1&2\
 \end{array} \right) ,
 \]  \[
 A_3 :\
\left(
 \begin{array}{ccc}
 2&-1&0\\ -1&2&-1\\ 0&-1&2
 \end{array} \right) ,\qquad
 B_3 :\
\left(
 \begin{array}{ccc}
 2&-1&0\\ -1&2&-2\\ 0&-1&2
 \end{array} \right) ,\qquad
 C_3 :\
\left(
 \begin{array}{ccc}
 2&-1&0\\ -1&2&-1\\ 0&-2&2
 \end{array} \right) .
 \]

The lengths of the basis vectors $\alpha_i$ are fixed by the
corresponding Coxeter--Dynkin diagram up to a constant. We adopt
the standard choice of Lie theory, namely
\[
\l\alpha ,\alpha\r=2
\]
for the longer (whenever there are two lengths) of the basis
vectors. The Coxeter--Dynkin diagram differentiates long and short
simple roots by node color.

\subsection{Weyl group}\label{reflect}

The Coxeter--Dynkin diagram uniquely determines the corresponding
transformation group of~$E_n$, generated by the reflections $r_i$,
$i=1,2,\ldots ,n$. Namely, the transformation $r_i$, corresponding
to a simple root $\alpha_i$, is a reflection with respect to
$(n-1)$-dimensional linear subspace (hyperplane) of $E_n$
(containing the origin) orthogonal to $\alpha_i$. It is well-known
that such reflections are given by the formula
\begin{gather}\label{reflection}
r_ix=x-\frac{2\l x, \alpha_i\r}{\l\alpha_i, \alpha_i\r}\alpha_i\,,
\qquad i= 1,2,\ldots,n,\qquad x\in E_n .
\end{gather}
Each reflection $r_i$ can be thought as attached to the $i$-th
node of the corresponding Coxeter--Dynkin diagram.

The finite group $W$, generated by the reflections $r_i$,
$i=1,2,\ldots ,n$, is called a {\it Weyl group} corresponding to a
given Coxeter--Dynkin diagram. If a Weyl group $W$ corresponds to
a Coxeter--Dynkin diagram of a simple Lie algebra $L$, then this
Weyl group is denoted by $W(L)$. The properties of Weyl groups are
well known. The orders (numbers of elements) of Weyl groups are
given by the formulas
 \begin{gather*}
|W(A_n)|=(n+1)!,\qquad |W(B_n)|=|W(C_n)|=2^nn!,\qquad
                              |W(D_n)|=2^{n-1}n!,\\
|W(E_6)|=51\,840,\qquad |W(E_7)|=2\ 903\,040,\qquad
                                    |W(E_8)|=696\,729\ 600,
\\
|W(F_4)|=1\,152,\qquad |W(G_2)|=12.
\end{gather*}
In particular,
\begin{gather}
|W(A_2)|=6,\qquad |W(C_2)|=8,\qquad |W(A_3)|=24,\qquad
|W(C_3)|=48.
\end{gather}

\subsection{Roots and weights}\label{roots}
A Coxeter--Dynkin diagram determines a system of simple roots in
the Euclidean space $E_n$. An~action of elements of the Weyl group
$W$ upon simple roots leads to a finite system of vectors, which
is invariant with respect to $W$. A set of all these vectors is
called a {\it system of roots} associated with a given
Coxeter--Dynkin diagram. It is denoted by $R$. As we see, a system
of roots~$R$ is calculated from the simple roots by a
straightforward algorithm. Root systems~$R$ which correspond to
Coxeter--Dynkin diagrams in 2-dimensional Euclidean spaces are
shown below:
 \bigskip

\centerline{Root system of $A_2$:}
 \bigskip

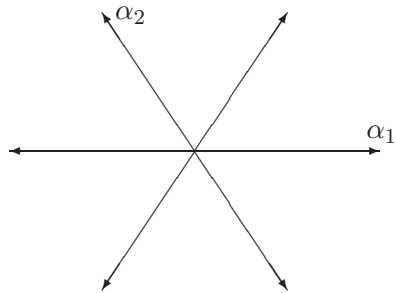
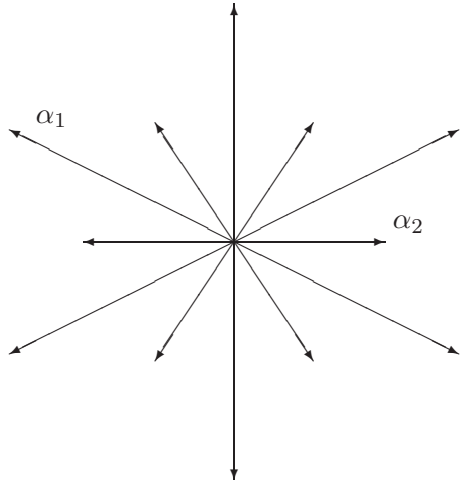
\centerline{
\begin{picture}(140,140)
\put(75,75){\vector(-1,0){70}} \put(75,75){\vector(-2,3){35}}
 \put(45,125){$\alpha_2$}
\put(75,75){\vector(2,3){35}} \put(75,75){\vector(1,0){70}}
 \put(140,80){$\alpha_1$}
\put(75,75){\vector(-2,-3){35}} \put(75,75){\vector(2,-3){35}}
\end{picture}  }

\newpage

\centerline{Root system of $C_2$:}
 \bigskip

 \centerline{
\begin{picture}(150,150)
\put(75,75){\vector(-1,0){70}} \put(75,75){\vector(-1,1){70}}
 \put(15,145){$\alpha_2$}
\put(75,75){\vector(0,1){70}}
 \put(145,80){$\alpha_1$}
 \put(75,75){\vector(1,1){70}}
\put(75,75){\vector(1,0){70}} \put(75,75){\vector(1,-1){70}}
\put(75,75){\vector(0,-1){70}} \put(75,75){\vector(-1,-1){70}}
\end{picture}  }

\bigskip

\centerline{Root system of $G_2$:}
 \medskip

 \centerline{  \qquad \quad
\begin{picture}(180,180)
\put(75,75){\vector(-1,0){57}}
 \put(0,120){$\alpha_1$}
\put(75,75){\vector(-2,3){30}} \put(75,75){\vector(2,3){30}}
\put(75,75){\vector(1,0){57}} \put(75,75){\vector(-2,-3){30}}
\put(75,75){\vector(2,-3){30}} \put(75,75){\vector(2,1){85}}
\put(75,75){\vector(2,-1){85}} \put(75,75){\vector(-2,1){85}}
\put(75,75){\vector(-2,-1){85}} \put(75,75){\vector(0,1){90}}
 \put(135,80){$\alpha_2$}
\put(75,75){\vector(0,-1){90}}
\end{picture}  }

\bigskip
\bigskip

It is proven (see, for example, \cite{Hem-2}) that roots of $R$
are linear combinations of simple roots with integral
coefficients. Moreover, there exist no roots which are linear
combinations of simple roots $\alpha_i$, $i=1,2,\ldots ,n$, both
with positive and negative coefficients. Therefore, the set of
roots $R$ can be represented as a union $R=R_+\cup R_-$, where
$R_+$ (respectively $R_-$) is the set of roots which are linear
combinations of simple roots with positive (negative)
coefficients. The set $R_+$ (the set $R_-$) is called a {\it set
of positive (negative) roots}. The root systems of simple Lie
algebras of rank $\leq12$ are shown, for example, in~\cite{BMP}.

The set of all linear combinations
\[
Q=\left\{ \sum_{i=1}^n a_i\alpha_i \ | \ a_i\in {\mathbb
Z}\right\}\equiv \bigoplus_i {\mathbb Z}\alpha_i
\]
is called a {\it root lattice} corresponding to a given
Coxeter--Dynkin diagram. Its subset
\[
Q_+=\left\{ \sum_{i=1}^n a_i\alpha_i \ | \
a_i=0,1,2,\ldots\right\}
\]
is called a {\it positive root lattice}.

 To each root $\alpha\in R$
there corresponds a coroot $\alpha^\vee$ defined by the formula
\[
\alpha^\vee =\frac{2\alpha}{\l \alpha,\alpha\r} .
\]
It is easy to see that $\alpha^{\vee\vee} =\alpha$. The set of all
linear combinations
\[
Q^\vee=\left\{ \sum_{i=1}^n a_i\alpha^\vee_i \ | \ a_i\in {\mathbb
Z}\right\}\equiv \bigoplus_i {\mathbb Z}\alpha^\vee_i
\]
is called a {\it coroot lattice} corresponding to a given
Coxeter--Dynkin diagram. The subset
\[
Q^\vee_+=\left\{ \sum_{i=1}^n a_i\alpha^\vee_i \ | \
a_i=0,1,2,\ldots\right\}
\]
is called a {\it positive coroot lattice}.

As noted above, the set of simple roots $\alpha_i$, $i=1,2,\ldots
,n$, is a basis of the space $E_n$. In addition to the
$\alpha$-basis, it is convenient to introduce the $\omega$-basis,
$\omega_1,\omega_2,\ldots ,\omega_n$ (also called the {\it basis
of fundamental weights}). The two bases are dual to each other in
the following sense:
\begin{gather}\label{kronecker}
\frac{2\l\alpha_j ,\omega_k\r} {\l\alpha_j,\alpha_j\r}\equiv
\l\alpha^\vee_j ,\omega_k\r =\delta_{jk},\qquad
j,k\in\{1,\ldots,n\}.
\end{gather}
The $\omega$-basis (as the $\alpha$-basis) is not orthogonal.

Note that the factor $2/\l\alpha_j,\alpha_j\r$ can take only three
values. Indeed, with the standard normalization of root lengths,
we have
\begin{gather*}
\frac2{\l\alpha_k,\alpha_k\r}
 =1 \qquad\text{for roots of}\quad
                A_n,\ D_n,\ E_6,\ E_7,\ E_8,
\\
\frac2{\l\alpha_k,\alpha_k\r}=1  \qquad
 \text{for long roots of}\quad B_n,\ C_n,\ F_4,\ G_2,
\\
\frac2{\l\alpha_k,\alpha_k\r}=2
 \qquad\text{for short roots of}\quad B_n,\ C_n,\ F_4,
\\
\frac2{\l\alpha_k,\alpha_k\r}= 3 \qquad\text{for short root
of}\quad G_2 .
\end{gather*}

The $\alpha$- and $\omega$-bases are related by the Cartan matrix
\eqref{Mmatrix} and by its inverse:
\begin{gather}\label{bases}
\alpha_j=\sum_{k=1}^nM_{jk}\omega_k,\qquad
\omega_j=\sum_{k=1}^n\big(M^{-1}\big)_{jk}\,\alpha_k .
\end{gather}
For ranks 2 and 3 the inverse Cartan matrices are of the form
\begin{gather*}
 A_2 :\ \frac 13 \left(
 \begin{array}{cc}
 2&1\\ 1&2\
 \end{array} \right) ,\qquad
C_2 :\  \left(
 \begin{array}{cc}
 1&1/2\\ 1&1\
 \end{array} \right) ,\qquad
 G_2 :\
 \left(
 \begin{array}{cc}
 2&3\\ 1&2\
 \end{array} \right) ,
 \\
 A_3 :\ \frac 14
\left(
 \begin{array}{ccc}
 3&2&1\\ 2&4&2\\ 1&2&3
 \end{array} \right) ,\qquad
 B_3 :\  \frac 12
\left(
 \begin{array}{ccc}
 2&2&2\\ 2&4&4\\ 1&2&3
 \end{array} \right) ,\qquad
 C_3 :\ \frac 12
\left(
 \begin{array}{ccc}
 2&2&1\\ 2&4&2\\ 2&4&3
 \end{array} \right) .
 \end{gather*}

Later on we will need to calculate the scalar product $\l x, y\r$
where $x$ and $y$ are given by coordinates $x_i$ and $y_i$
relative to the $\omega$-basis. It is given by the formula
\begin{gather}\label{matr}
\l x, y\r
     =\frac12\sum_{j,k=1}^n
            x_jy_k(M^{-1})_{jk}\l\alpha_k\mid\alpha_k\r
 = xM^{-1}Dy^T  \equiv  xS\,y^T,
\end{gather}
where $D$ is the diagonal matrix ${\rm diag}\, (\frac 12 \l\alpha
_1 | \alpha_1\r ,\ldots ,\frac 12 \l \alpha_n | \alpha_n \r )$.
Matrices $S$, called `quadratic form matrices', are shown in
\cite{BMP} for all connected Coxeter--Dynkin diagrams.

The sets $P$ and $P_+$, defined as
 \[
P=\Z\omega_1+\cdots+\Z\omega_n \ \supset\
P_+=\Z^{\geq0}\omega_1+\cdots+\Z^{\geq0}\omega_n,
 \]
are respectively called the {\it weight lattice} and the {\it cone
of dominant weights}. The set $P$ can be characterized as a set of
all $\lambda\in E_n$ such that $2\langle \lambda,\alpha_j\rangle /
\langle \alpha_j,  \alpha_j \rangle \equiv \langle \lambda,
\alpha^\vee_j\rangle \in \Z$ for all simple roots~$\alpha_j$.
Clearly, $Q\subset P$. Moreover, the additive group $P/Q$ is
finite.

The smallest non-vanishing dominant weights $\omega_k$,
$k=1,2,\ldots,n$, defined by \eqref{kronecker}, are called {\it
fundamental weights}. Throughout this article we will often use
the following notation for weights in the $\omega$-basis:
\begin{gather}\label{weight}
\lambda=\sum_{j=1}^n a_j\omega_j=(a_1\ a_2\ \cdots\ a_n).
\end{gather}
If $x=\sum\limits_{j=1}^n b_j\alpha^\vee_j$, then
 \[
\l \lambda,x\r =\sum_{j=1}^n a_jb_j.
\]

There is a unique highest (long) root $\xi$ and a unique highest
short root $\xi_s$. A highest root can be written in the form
\begin{gather}\label{highestroot}
\xi=\sum_{i=1}^nm_i\alpha_i=\sum_{i=1}^n m_i\frac{\langle
\alpha_i,\alpha_i\rangle}{2} \alpha_i^\vee\equiv \sum_{i=1}^n
q_i\alpha_i^\vee .
\end{gather}
Coefficients $m_i$ and $q_i$ can be considered as attached to the
$i$-th node of the Coxeter--Dynkin diagram. They are called {\it
marks\/} and {\it comarks}. They are often listed in the
literature (see, for example, \cite{BMP} and \cite{MPR90}). In
root systems with two lengths of roots, namely in $B_n$, $C_n$,
$F_4$ and~$G_2$, the highest (long) root $\xi$ is the following:
\begin{alignat}{3}
B_n\ &:&\quad \xi &=& (0\,1\,0\,\cdots\,0)
               &=\alpha_1+2\alpha_2+2\alpha_3+\cdots+2\alpha_n,\\
C_n\ &:&\ \xi     &=&\ (2\,0\,\cdots\,0)
               &=2\alpha_1+2\alpha_2+\cdots+2\alpha_{n-1}+\alpha_n,\\
F_4\ &:&\ \xi     &=& (1\,0\,0\,0)
               &=2\alpha_1+3\alpha_2+4\alpha_3+2\alpha_4,\\
G_2\ &:&\ \xi     &=& (1\,0)   &= 2\alpha_1+3\alpha_2.
\end{alignat}
In case of $A_n$, $D_n$ and $E_n$, all roots are of the same
length and we have
\begin{alignat}{3}
A_n\ &:&\quad \xi &=& (1\,0\,\cdots\,0\,1)
               &=\alpha_1+\cdots+\alpha_n,\label{eq13}\\
D_n\ &:&\ \xi     &=& (0\,1\,0\,\cdots\,0)
               &=\alpha_1+2\alpha_2+\cdots+2\alpha_{n-2}+
               \alpha_{n-1}+\alpha_n,\\
E_6\ &:&\ \xi     &=&(0\, 0\, \cdots\, 0\,
1)&=\alpha_1+2\alpha_2+3\alpha_3+2\alpha_4+\alpha_5+2\alpha_6,
               \\
E_7\ &:&\ \xi     &=&(1\, 0\,0\,\cdots\,
0)&=2\alpha_1+3\alpha_2+4\alpha_3+3\alpha_4+2\alpha_5+\alpha_6+2\alpha_7,
      \\
E_8\ &:&\ \xi     &=&(1\,0\, 0\,\cdots\,
0)&=2\alpha_1+3\alpha_2+4\alpha_3+5\alpha_4+6\alpha_5+4\alpha_6+2\alpha_7+
3\alpha_8.\label{eq17}
\end{alignat}
Note that for highest root $\xi$ we have
\begin{gather}\label{highest}
\xi^\vee =\xi .
\end{gather}
Moreover, if all simple roots are of the same length, then
 \[
\alpha_i^\vee =\alpha_i.
 \]
For this reason,
 \[
(q_1,q_2,\ldots,q_n)=(m_1,m_2,\ldots,m_n).
 \]
for $A_n, D_n$ and $E_n$. Formulas \eqref{eq13}--\eqref{highest}
determine these numbers. For short roots $\alpha_i$ of $B_n$, $C_n$
and $F_4$ we have $\alpha_i^\vee=2\alpha_i$.  For short root
$\alpha_2$ of $G_2$ we have $\alpha_2^\vee=3\alpha_2$. For this
reason,
 \begin{gather*}
(q_1,q_2,\ldots,q_n)=(1,2,\ldots, 2,1)\qquad {\rm for}\quad  B_n,
\\
(q_1,q_2,\ldots,q_n)=(1,1,\ldots, 1,1)\qquad {\rm for}\quad  C_n,
\\
(q_1,q_2,q_3,q_4)=(2,3, 2,1)\qquad {\rm for}\quad  F_4,
\\
(q_1,q_2)=(2,1)\qquad {\rm for}\quad  G_2.
\end{gather*}

With the help of the highest root $\xi$, it is possible to
construct an {\it extended root system} (which is also called an
{\it affine root system}). Namely, if $\alpha_1,\alpha_2,\ldots,
\alpha_n$ is a set of all simple roots, then the roots
\[
\alpha_0:=-\xi,\alpha_1,\alpha_2,\ldots, \alpha_n
\]
constitute a set of simple roots of the corresponding extended
root system. Taking into account the orthogonality
(non-orthogonality) of the root $\alpha_0$ to other simple roots,
the diagram of an~extended root system can be constructed (which
is an extension of the corresponding Coxeter--Dynkin diagram; see,
for example, \cite{MPR90}). Note that for all simple Lie algebras
(except for $A_n$) only one simple root is orthogonal to the root
$\alpha_0$. In the case of $A_n$, the two simple roots $\alpha_1$
and $\alpha_n$ are not orthogonal to $\alpha_0$.

\subsection{Weyl group orbits}\label{orbi}

The $(n-1)$-dimensional linear subspaces of $E_n$, orthogonal to
positive roots and containing the origin, divide the space $E_n$
into connected parts, which are called {\it Weyl chambers}. The
number of such chambers is equal to the order of the corresponding
Weyl group $W$. Elements of the Weyl group $W$ permute these
chambers. There exists a single chamber $D_+$ such that
 \[
\l\alpha_i, x \r \ge 0,\qquad x\in D_+,\qquad i=1,2,\ldots ,n.
 \]
It is called the {\it dominant Weyl chamber}.

Clearly, the cone of dominant weights $P_+$ belongs to the
dominant Weyl chamber $D_+$. (Note that it is not the case for the
set $Q_+$.) Moreover, $P\cap D_+=P_+$.

Let $y$ be an arbitrary element of the Euclidean space $E_n$. We
act upon $y$ by all elements of the Weyl group $W$, thus creating
the set of elements $wy$, $w\in W$. There may exist coinciding
elements in this set. We extract from this set all distinct points
(we denote them by $y_1, y_2,\ldots, y_m$). The set of these
points is called an orbit of $y$ with respect to the Weyl group
$W$ (or a Weyl group orbit that contains $y$). Since $Wy_i$ is the
same as $Wy_j$ for all $i,j=1,2,\ldots,m$, the orbit of $y$ is
thus the same as the orbit of any other point $y_i$.

The orbit of a point $y\in E_n$ is denoted by $O(y)$ or $O_W(y)$.
The number of elements contained in an orbit $O(y)$ is referred to
as its size, and is denoted $|O(y)|$. One has two extremal values
of the orbit size
 \[
|O(0)|=1,\qquad |O(y)|=|W|,
 \]
where $0$ is the origin of $E_n$ and $y$ is an intrinsic point of
any Weyl chamber (that is, it is invariant only by the trivial
transformation from the Weyl group $W$).

Each Weyl chamber contains only one point from any given orbit
$Q(y)$. In particular, in an orbit there exists only one point in
the dominant Weyl chamber $D_+$. An orbit is usually denoted by
this particular point in $D_+$, that is, the notation $O(y)$ means
that $y\in D_+$.

Let $W_y$ be a subgroup of $W$, consisting of all elements $w\in
W$, such that $wy=y$. It is called the {\it stabilizer} of the
point $y$. We have
 \[
|O(y)|=\frac{|W|}{|W_y|}.
 \]
If $y\in D_+$, then $W_y$ is the subgroup generated by all
reflections $r_i$, which correspond to the simple roots
$\alpha_i$, such that $r_iy=y$.

\subsection[Orbits of $A_1$, $A_1\times A_1$, $A_2$, $B_2$, $G_2$]{Orbits of
$\boldsymbol{A_1}$, $\boldsymbol{A_1\times A_1}$,
$\boldsymbol{A_2}$, $\boldsymbol{B_2}$,
$\boldsymbol{G_2}$}\label{orbits}

There is always an orbit of size 1, consisting of the single
weight zero. We consider all the other orbits. Assuming $(a,b)\ne
(0,0)$, $a\ge 0, b\ge 0$, here is a list of the remaining orbits
and their contents in the $\omega$-basis:
\begin{align}
A_1:\quad
  &O(a)\ni(a),\ (-a)\\
A_1\times A_1\,:\quad
  &O(a\ 0)\ni(a\ 0),\ (-a\ 0)\,,\qquad O(0\ b)\ni(0\ b),\ (0\ {-}b),\\
  &O(a\ b)\ni(a\ b),\ ({-}a\ b),\ (a\ {-}b),\ ({-}a\ {-}b)\\
A_2\,:\quad
           &O(a\ 0)\ni (a\ 0),\ ({-}a\ a),\ (0\ {-}a),\\
           &O(0\ b)\ni (0\ b),\ (b\ {-}b),\ ({-}b\ 0),\\
           &O(a\ b)\ni(a\ b),\ ({-}a\ a{+}b),\ (a{+}b\ {-}b),\notag\\
           &\qquad\qquad (-b\ -a),\ ( {-}a{-}b\ a),\ (b\ {-}a{-}b).
\end{align}
In the cases of $C_2$ and $G_2$ (where the second simple root is
the longer one for $C_2$ and the shorter one for $G_2$) we have
\begin{align}
C_2\,:\quad
  &O(a\ 0)\ni\pm(a\ 0),\ \pm({-}a\ a),\qquad
      O(0\ b)\ni\pm(0\ b),\ \pm(2b\ {-}b) \notag\\
  &O(a\ b)\ni\pm(a\ b),\ \pm({-}a\ a{+}b),\ \pm(a{+}2b\ {-}b),\
           \pm(a{+}2b\ {-}a{-}b),\\
G_2\,:\quad
  &O(a\ 0)\ni\pm(a\ 0),\ \pm({-}a\ 3a),\ \pm(2a\ {-}3a),\\
  &O(0\ b)\ni\pm(0\ b),\ \pm(b\ {-}b),\ \pm({-}b\ 2b),\\
  &O(a\ b)\ni\pm(a\ b),\ \pm({-}a\ 3a{+}b),\ \pm(a{+}b\ {-}b),\notag\\
  &\phantom{O(a\ b)\ni}{}\pm(2a{+}b\ {-}3a{-}b),\pm({-}a{-}b\ 3a{+}2b),\
  \pm({-}2a{-}b\   3a{+}2b).
\end{align}

\subsection{Geometric interpretation of orbits}\label{polytopes}

The elements of $O(y)$ can be interpreted as vertices of a
polytope generated by reflections, starting from a single point,
in $n$ mirrors intersecting at the origin. For a general method of
describing such polytopes and their faces of all dimensions in
$\R^n$ see \cite{CKPS}. For example, orbits in $2$ dimensions can
be interpreted as planar polygons.

In particular, the non-zero roots of Lie algebra $A_2$ form the
orbit $O(1\,1)$, which corresponds to the vertices of the regular
hexagon. Roots of all other rank 2 Lie algebras belong to the
union of two different orbits. Namely, $O(2\,0)\cup O(0\,2)$ in
the case of $A_1\times A_1$; for $C_2$ it is the orbit of the long
roots $O(2\,0)$ and the orbit of the short roots $O(0\,1)$. The
long roots of $G_2$ form the hexagon $O(1\,0)$, while the short
roots form the hexagon $O(0\,1)$.

The $W$-orbits in $3$ dimensions include many common polytopes
(\cite{CKPS} and \cite{MP92}) with the notable exceptions of the
icosahedral polytopes: icosahedron, dodecahedron, buckyball
(`soccerball'), and others.

\setcounter{equation}{0}

\section[Lattices, Weyl groups and orbits for $A_n$, $B_n$, $C_n$, $D_n$]{Lattices, Weyl
groups and orbits for $\boldsymbol{A_n}$, $\boldsymbol{B_n}$,
$\boldsymbol{C_n}$, $\boldsymbol{D_n}$} \label{weylABCD}

In the case of Coxeter--Dynkin diagrams $A_{n-1}$, $B_n$, $C_n$,
$D_n$, root and weight lattices, Weyl groups and orbits are simply
described using an orthogonal coordinate system in $E_n$. Let us
present these coordinate systems.

\subsection[The case of $A_n$]{The case of $\boldsymbol{A_n}$}\label{An}

For this case it is convenient to describe root and weight
lattices, Weyl group and orbits in the subspace of the Euclidean
space $E_{n+1}$, given by the equation
 \[
x_1+x_2+\cdots +x_{n+1}=0,
 \]
where $x_1,x_2,\ldots ,x_{n+1}$ are the orthogonal coordinates of
a point $x\in E_{n+1}$. The unit vectors in directions of these
coordinates are denoted by ${\boldsymbol e}_j$, respectively.
Clearly, ${\boldsymbol e}_i\bot {\boldsymbol e}_j$, $i\ne j$. The
set of roots is given by the vectors
 \[
\alpha_{ij}={\boldsymbol e}_i-{\boldsymbol e}_j, \qquad i\ne j.
 \]
The roots
 \[
\alpha_{ij}={\boldsymbol e}_i-{\boldsymbol e}_j, \qquad i< j,
 \]
are positive and the roots
 \[
\alpha_i\equiv \alpha_{i,i+1}={\boldsymbol e}_i-{\boldsymbol
e}_{i+1},\qquad i=1,2,\ldots ,n,
 \]
constitute the system of simple roots.

If $x=\sum\limits_{i=1}^{n+1} x_i{\boldsymbol e}_i$,
$x_1+x_2+\cdots +x_{n+1}=0$, is a point of $E_{n+1}$, then it
belongs to the dominant Weyl chamber $D_+$ if and only if
 \[
x_1\ge x_2\ge \cdots \ge x_{n+1}.
 \]
Indeed, if this condition is fulfilled, then $\langle
x,\alpha_i\rangle =x_i-x_{i+1}\geq 0$, $i=1,2,\ldots, n$, and $x$
is dominant. Conversely, if $x$ is dominant, then $\langle
x,\alpha_i\rangle \geq 0$, thus fulfilling the prescribed
condition.

If $\lambda =\sum\limits_{i=1}^n \lambda_i \omega_i$, then the
coordinates $\lambda_i$ in the $\omega$-basis are connected with
the orthogonal coordinates $m_j$ of
$\lambda=\sum\limits_{i=1}^{n+1} m_i{\bf e}_i$ by the formulas
 \begin{alignat*}{12}
 m_1& =&\frac{n}{n+1} \lambda_1
 &+&\frac{n-1}{n+1}\lambda_2&+&\frac{n-2}{n+1}\lambda_3
 +&\cdots &+&\frac{2}{n+1}\lambda_{n-1}&+&\frac{1}{n+1}\lambda_n,\\
 m_2& =&-\frac{1}{n+1} \lambda_1
 &+&\frac{n-1}{n+1}\lambda_2&+&\frac{n-2}{n+1}\lambda_3
 +&\cdots &+&\frac{2}{n+1}\lambda_{n-1}&+&\frac{1}{n+1}\lambda_n,\\
 m_3& =&-\frac{1}{n+1} \lambda_1
 &-&\frac{2}{n+1}\lambda_2&+&\frac{n-2}{n+1}\lambda_3
 +&\cdots &+&\frac{2}{n+1}\lambda_{n-1}&+&\frac{1}{n+1}\lambda_n,\\
  \cdots && \cdots && \cdots   && \cdots && \cdots && \cdots \\
 m_{n}& =&-\frac{1}{n+1} \lambda_1
 &-&\frac{2}{n+1}\lambda_2&-&\frac{3}{n+1}\lambda_3
 -&\cdots &-&\frac{n-1}{n+1}\lambda_{n-1}&+&\frac{1}{n+1}\lambda_n,\\
 m_{n+1}& =&-\frac{1}{n+1} \lambda_1
 &-&\frac{2}{n+1}\lambda_2&-&\frac{3}{n+1}\lambda_3
 -&\cdots
 &-&\frac{n-1}{n+1}\lambda_{n-1}&-&\frac{n}{n+1}\lambda_n.
 \end{alignat*}
 The inverse formulas are
 \begin{gather}\label{dif-A}
 \lambda_i=m_i-m_{i+1},\qquad i=1,2,\ldots ,n.
  \end{gather}

By means of the formula
 \begin{gather}\label{refl}
 r_\alpha \lambda=\lambda -\frac{2\l \lambda,\alpha\r}{\l \alpha,
 \alpha \r}\alpha ,
  \end{gather}
for the reflection respective to the hyperplane, orthogonal to a
root $\alpha$, we can find that the reflection $r_{\alpha_{ij}}$
acts upon the vector $\lambda=\sum\limits_{i=1}^{n+1}
m_i{\boldsymbol e}_i$, given by orthogonal coordinates, by
permuting the coordinates $m_i$ and $m_j$. Indeed, when $i>j$ we
have
 \begin{gather*}
\lambda -\frac{2\langle \lambda,\alpha_{ij} \rangle} {\langle
\alpha_{ij},\alpha_{ij} \rangle}\alpha_{ij} =\sum_{s=1}^{n+1}
m_s{\boldsymbol e}_s -(m_i-m_j){\boldsymbol e}_i+(m_i-m_j){\boldsymbol e}_j\\
 \phantom{\lambda -\frac{2\langle \lambda,\alpha_{ij} \rangle} {\langle
\alpha_{ij},\alpha_{ij} \rangle}\alpha_{ij}}{} =m_1{\boldsymbol
e}_1+\cdots +m_j{\boldsymbol e}_i+\cdots +m_i{\boldsymbol
e}_j+\cdots +m_{n+1}{\boldsymbol e}_{n+1}.
 \end{gather*}
Thus, $W(A_n)$ consists of all permutations of the orthogonal
coordinates $m_1,m_2,\ldots ,m_{n+1}$ of a~point $\lambda$, that
is, $W(A_n)$ coincides with the symmetric group $S_{n+1}$.

Sometimes (for example, if we wish the coordinates to be integers
or non-negative integers), it is convenient to introduce the
orthogonal coordinates $x_1,x_2,\ldots,$ $x_{n+1}$ for $A_n$ in
such a way that
 \[
x_1+x_2+\cdots +x_{n+1}=m,
 \]
where $m$ is some fixed real number. They are obtained from the
previous orthogonal coordinates by adding the same number
$m/(n+1)$ to each coordinate. Then, as one can see from
\eqref{dif-A}, $\omega$-coordinates $\lambda_i=x_i-x_{i+1}$ and
the Weyl group $W$ remain unchanged.

The orbit $O(\lambda)$, $\lambda=(m_1,m_2,\ldots,m_{n+1})$,
$m_1\ge m_2\ge \cdots \ge m_{n+1}$, consists of all {\it
different} points
 \[
(m_{i_1},m_{i_2},\ldots,m_{i_{n+1}})
 \]
obtained from $(m_1,m_2,\ldots,m_{n+1})$ by permutations. The
subgroup $W_\lambda$ of $W$ is a product of subgroups, consisting
of permutations of collections of coinciding $m$'s. For example,
the orbit $O(m,0,\ldots,0)$ (with $n+1$ orthogonal coordinates)
consists of $n+1$ points $(0,\ldots,0, m,0,\ldots,0)$, where $m$
is in $j$-th place, $j=1,2,\ldots,n+1$. The subgroup $W_\lambda$
in this case is isomorphic to the symmetric group $S_n$.

The orbit $O(m_1,m_2,0,\ldots,0)$, $m_1\ne m_2$, consists of
$n(n+1)$ points of the form
 \[
(0,\ldots,0,m',0,\ldots,0,m'',0,\ldots,0),
 \]
where $(m',m'')=(m_1,m_2)$ or $(m',m'')=(m_2,m_1)$. The subgroup
$W_\lambda$ is isomorphic to the symmetric group $S_{n-1}$. If
$m_1=m_2$, then this orbit consists of $n(n+1)/2$ elements and the
subgroup $W_\lambda$ is isomorphic to $S_{n-1}\times S_2$.

If we have an orbit with points given in orthogonal coordinates,
it is easy, by using formula~\eqref{dif-A}, to express those
points in $\omega$-coordinates.

\subsection[The case of $B_n$]{The case of $\boldsymbol{B_n}$}\label{Bn}

Orthogonal coordinates of a point $x\in E_n$ are denoted by
$x_1,x_2,\ldots ,x_n$. We denote by ${\boldsymbol e}_i$ the
corresponding unit vectors in directions of these coordinates.
Then the set of roots of $B_n$ is given by the vectors
 \[
\alpha_{\pm i,\pm j}=\pm {\boldsymbol e}_i\pm {\boldsymbol e}_j,
\qquad i\ne j, \qquad  \alpha_{\pm i}=\pm {\boldsymbol e}_i,\qquad
i=1,2,\ldots ,n
 \]
(all combinations of signs must be taken). The roots
 \[
\alpha_{i,\pm j}={\boldsymbol e}_i\pm {\boldsymbol e}_j, \qquad i<
j,\qquad \alpha_{i}={\boldsymbol e}_i,\qquad i=1,2,\ldots ,n,
 \]
are positive and the $n$ roots
 \[
\alpha_i:={\boldsymbol e}_i-{\boldsymbol e}_{i+1},\qquad
i=1,2,\ldots ,n-1, \qquad \alpha_n={\boldsymbol e}_n
 \]
constitute the system of simple roots.

It is easy to see that if $\lambda=\sum\limits_{i=1}^{n}
m_i{\boldsymbol e}_i$ is a point of $E_{n}$, then this point
belongs to the dominant Weyl chamber $D_+$ if and only if
 \[
m_1\ge m_2\ge \cdots \ge m_{n}\ge 0.
 \]

If $\lambda =\sum\limits_{i=1}^n \lambda_i \omega_i$, then the
coordinates $\lambda_i$ in the $\omega$-basis are connected with
the orthogonal coordinates $m_j$ of $\lambda=\sum\limits_{i=1}^{n}
m_i{\boldsymbol e}_i$ by the formulas
 \begin{alignat*}{11}
 m_1 &=&\lambda_1&+&\lambda_2&+&\cdots &+&\lambda_{n-1}&+&\tfrac 12
 \lambda_n, \\
 m_2&=&         && \lambda_2&+&\cdots &+&\lambda_{n-1}&+&\tfrac 12
 \lambda_n, \\
 \cdots &&\cdots && \cdots && \cdots   && \cdots  && \cdots \\
 m_{n}&=&         &&    &&   &&  &&\tfrac 12  \lambda_n,
 \end{alignat*}
 The inverse formulas are
\begin{gather}\label{dif-B}
  \lambda_i=m_i-m_{i+1},\qquad i=1,2,\ldots ,n-1,\qquad
 \lambda_n=2m_n.
 \end{gather}

By means of the formula \eqref{refl} we find that the reflection
$r_{\alpha}$ acts upon orthogonal coordinates of the vector
$\lambda=\sum\limits_{i=1}^{n} m_i{\boldsymbol e}_i$ by permuting
the $i$-th and $j$-th coordinates if $\alpha=\pm ({\boldsymbol
e}_i - {\boldsymbol e}_j)$, as the permutation of the $i$-th and
$j$-th coordinates and a change of their signs if $\alpha=\pm
({\boldsymbol e}_i + {\boldsymbol e}_j)$, and as a change of a
sign of $i$-th coordinate if $\alpha=\pm {\boldsymbol e}_i$. Thus,
the Weyl group $W(B_n)$ consists of all permutations of the
orthogonal coordinates $m_1,m_2,\ldots ,m_{n}$ of a point
$\lambda$ with possible sign alternations for some of them.

The orbit $O(\lambda)$, $\lambda=(m_1,m_2,\ldots,m_{n})$, $m_1\ge
m_2\ge \cdots \ge m_{n}\ge 0$, consists of all {\it different}
points
 \[
(\pm m_{i_1}, \pm m_{i_2},\ldots,\pm m_{i_{n+1}})
 \]
(each combination of signs is possible) obtained from $(m_1,m_2,
\ldots,m_{n})$ by permutations and alternations of signs. The
subgroup $W_\lambda$ of $W$ is a product of the Weyl group
$W(B_r)$, where $r$ is the number of 0's in the weight
$(m_1,m_2,\ldots,m_{n})$, and of the subgroups, consisting of
permutations of collections of coinciding non-vanishing $m$'s. For
example, the orbit $O(m,0,\ldots,0)$, $m\ne 0$, consists of $2n$
points $(0,\ldots,0 ,\pm m,0,\ldots$, $0)$, where $\pm m$ is on
$j$-th place, $j=1,2,\ldots,n$. The subgroup $W_\lambda$ in this
case is isomorphic to the Weyl group $W(B_{n-1})$.

The orbit $O(m_1,m_2,0,\ldots,0)$, $m_1\ne m_2$, consists of
$4n(n-1)$ points of the form
 \[
(0,\ldots,0,\pm m',0,\ldots,0,\pm m'',0,\ldots,0),
 \]
where $(m',m'')=(m_1,m_2)$ or $(m',m'')=(m_2,m_1)$. The subgroup
$W_\lambda$ is isomorphic to the Weyl group $W(B_{n-2})$. If
$m_1=m_2$, then this orbit consists of $2n(n-1)$ elements.

If we have an orbit with points given in orthogonal coordinates,
it is easy, by using formula~\eqref{dif-B}, to transform it to
$\omega$-coordinates.

\subsection[The case of $C_n$]{The case of $\boldsymbol{C_n}$}\label{Cn}

In the orthogonal coordinate system of the Euclidean space $E_{n}$
the set of roots of $C_n$ is given by the vectors
 \[
\alpha_{\pm i,\pm j}=\pm {\boldsymbol e}_i\pm {\boldsymbol e}_j,
\qquad i\ne j, \qquad  \alpha_{\pm i}=\pm 2{\boldsymbol
e}_i,\qquad i=1,2,\ldots ,n,
 \]
where ${\boldsymbol e}_i$ is the unit vector in the direction of
the $i$-th coordinate $x_i$ (all combinations of signs must be
taken into account). The roots
 \[
\alpha_{i,\pm j}={\boldsymbol e}_i\pm {\boldsymbol e}_j, \qquad i<
j,\qquad \alpha_{i}=2{\boldsymbol e}_i,\qquad i=1,2,\ldots ,n,
 \]
are positive and the $n$ roots
 \[
\alpha_i:={\boldsymbol e}_i-{\boldsymbol e}_{i+1},\qquad
i=1,2,\ldots ,n-1, \qquad \alpha_n=2{\boldsymbol e}_n
 \]
constitute the system of simple roots.

It is easy to see that if $\lambda=\sum\limits_{i=1}^{n}
m_i{\boldsymbol e}_i$ is a point of $E_{n}$, then it belongs to
the dominant Weyl chamber $D_+$ if and only if
 \[
m_1\ge m_2\ge \cdots \ge m_{n}\ge 0.
 \]

If $\lambda =\sum\limits_{i=1}^n \lambda_i \omega_i$, then the
coordinates $\lambda_i$ in the $\omega$-basis are connected with
the coordinates $m_j$ of $\lambda=\sum\limits_{i=1}^{n}
m_i{\boldsymbol e}_i$ by the formulas
 \begin{alignat*}{11}
 m_1 &=&\lambda_1&+&\lambda_2&+&\cdots &+&\lambda_{n-1}&+&
 \lambda_n,\\
 m_2&=&  &&\lambda_2&+&\cdots &+&\lambda_{n-1}&+& \lambda_n,\\
 \cdots &&\cdots && \cdots && \cdots   && \cdots  && \cdots \\
 m_{n}&=&        &&      &&        &&     && \lambda_n .
 \end{alignat*}
 The inverse formulas are
 \begin{gather}\label{dif-C}
 \lambda_i=m_i-m_{i+1},\qquad i=1,2,\ldots ,n-1,\qquad
 \lambda_n=m_n.
 \end{gather}

By means of the formula \eqref{refl} we find that the reflection
$r_{\alpha}$ acts upon orthogonal coordinates of the vector
$\lambda=\sum\limits_{i=1}^{n} m_i{\boldsymbol e}_i$ by permuting
the $i$-th and $j$-th coordinates if $\alpha=\pm ({\boldsymbol
e}_i - {\boldsymbol e}_j)$, as the permutation of $i$-th and
$j$-th coordinates and a change of their signs if $\alpha=\pm
({\boldsymbol e}_i + {\boldsymbol e}_j)$, and as a~change of a
sign of the $i$-th coordinate if $\alpha=\pm 2{\boldsymbol e}_i$.
Thus, the Weyl group $W(C_n)$ consists of all permutations of the
orthogonal coordinates $m_1,m_2,\ldots ,m_{n}$ of a point
$\lambda$ with sign alternations for some of them, that is, this
Weyl group is isomorphic to the Weyl group $W(B_n)$.

The orbit $O(\lambda)$, $\lambda=(m_1,m_2,\ldots,m_{n})$, $m_1\ge
m_2\ge \cdots \ge m_{n}\ge 0$, consists of all {\it different}
points
 \[
(\pm m_{i_1}, \pm m_{i_2},\ldots,\pm m_{i_{n+1}})
 \]
(each combination of signs is possible) obtained from $(m_1,m_2,
\ldots,m_{n})$ by permutations and alternations of signs. The
subgroup $W_\lambda$ of $W$ is a product of the Weyl group
$W(C_r)$, where $r$ is the number of 0's in the weight
$(m_1,m_2,\ldots,m_{n})$, and of the subgroups consisting of
permutations of collections of coinciding non-vanishing $m$'s. For
example, the orbit $O(m,0,\ldots,0)$ consists of $2n$ points
$(0,\ldots,0,$ $\pm m,0,\ldots,0)$, where $\pm m$ is in the $j$-th
place, $j=1,2,\ldots,n$. The subgroup $W_\lambda$ is isomorphic to
the Weyl group $W(C_{n-1})$.

The orbit $O(m_1,m_2,0,\ldots,0)$, $m_1\ne m_2$, consists of
$4n(n-1)$ points of the same form as in the case of $B_n$. The
subgroup $W_\lambda$ is isomorphic to the Weyl group $W(C_{n-2})$.
If $m_1=m_2$, then this orbit consists of $2n(n-1)$ elements.

If we have an orbit with points given in orthogonal coordinates,
it is easy, by using formula~\eqref{dif-C}, to transform it to
$\omega$-coordinates.

\subsection[The case of $D_n$]{The case of $\boldsymbol{D_n}$}\label{Dn}

In the orthogonal coordinate system of the Euclidean space $E_{n}$
the set of roots of $D_n$ is given by the vectors
 \[
\alpha_{\pm i,\pm j}=\pm {\boldsymbol e}_i\pm {\boldsymbol e}_j,
\qquad i\ne j,
 \]
where ${\boldsymbol e}_i$ is the unit vector in the direction of
the $i$-th coordinate (all combinations of signs must be taken
into account). The roots
 \[
\alpha_{i,\pm j}={\boldsymbol e}_i\pm {\boldsymbol e}_j, \qquad i<
j,
 \]
are positive and the $n$ roots
 \[
\alpha_i:={\boldsymbol e}_i-{\boldsymbol e}_{i+1},\qquad
i=1,2,\ldots ,n-1, \qquad \alpha_n={\boldsymbol e}_{n-1}+
{\boldsymbol e}_n
 \]
constitute the system of simple roots.

If $\lambda=\sum\limits_{i=1}^{n} m_i{\boldsymbol e}_i$ is a point
of $E_{n}$, then it belongs to the dominant Weyl chamber $D_+$ if
and only if
 \[
m_1\ge m_2\ge \cdots \ge m_{n-1}\ge |m_n|.
 \]

If $\lambda =\sum\limits_{i=1}^n \lambda_i \omega_i$, then the
coordinates $\lambda_i$ in the $\omega$-basis are connected with
the coordinates $m_j$ of $\lambda=\sum\limits_{i=1}^{n}
m_i{\boldsymbol e}_i$ by the formulas
  \begin{alignat*}{13}
 m_1&=&\lambda_1&+&\lambda_2&+&\cdots &+&\lambda_{n-2}&+&\tfrac 12
 (\lambda_{n-1} &+&\lambda_n),\\
 m_2&=&   &&  \lambda_2&+&\cdots &+&\lambda_{n-2}&+&\tfrac 12
 (\lambda_{n-1}&+&  \lambda_n),\\
 \cdots &&\cdots && \cdots && \cdots && \cdots  && \cdots &&\cdots\\
 m_{n-1}&=&     &&         &&        &&        &&
 \tfrac 12 (\lambda_{n-1}&+& \lambda_n),\\
 m_{n}&=&     &&         &&        &&        &&
 \tfrac 12 (\lambda_{n-1}&-& \lambda_n),
  \end{alignat*}
 The inverse formulas are
 \begin{gather}\label{dif-D}
 \lambda_i=m_i-m_{i+1},\quad i=1,2,\ldots ,n-2,\qquad
 \lambda_{n-1}=m_{n-1}+m_n,\qquad \lambda_{n}=m_{n-1}-m_n .\!\!\!
  \end{gather}

By means of the formula \eqref{refl} for the reflection $r_\alpha$
we find that $r_{\alpha}$ acts upon orthogonal coordinates of the
vector $\lambda=\sum\limits_{i=1}^{n} m_i{\boldsymbol e}_i$ by
permuting the $i$-th and $j$-th coordinates if $\alpha=\pm
({\boldsymbol e}_i - {\boldsymbol e}_j)$, as the permutation of
$i$-th and $j$-th coordinates and the change of their signs if
$\alpha=\pm ({\boldsymbol e}_i + {\boldsymbol e}_j)$. Thus, the
Weyl group $W(D_n)$ consists of all permutations of the orthogonal
coordinates $m_1,m_2,\ldots ,m_{n}$ of $\lambda$ with sign
alternations for an even number of coordinates.

The orbit $O(\lambda)$, $\lambda=(m_1,m_2,\ldots,m_{n})$, $m_1\ge
m_2\ge \cdots \ge m_{n-1}\ge |m_{n}|$, consists of all {\it
different} points
 \[
(\pm m_{i_1}, \pm m_{i_2},\ldots,\pm m_{i_{n+1}})
 \]
obtained from $(m_1,m_2,\ldots,m_{n})$ by permutations and
alternations of even number of signs. The subgroup $W_\lambda$ of
$W$ is a product of the Weyl group $W(D_r)$, where $r$ is the
number of 0's in the weight $(m_1,m_2,\ldots,m_{n})$, and of the
subgroups consisting of permutations of collections of coinciding
non-vanishing $m$'s. For example, the orbit $O(m,0,\ldots,0)$
consists of $2n$ points $(0,\ldots,0 ,\pm m,0,\ldots,0)$. The
subgroup $W_\lambda$ is isomorphic to the Weyl group $W(D_{n-1})$.

The orbit $O(m_1,m_2,0,\ldots,0)$, $m_1\ne m_2$, consists of
$4n(n-1)$ points of the same form as in the case of $B_n$. The
subgroup $W_\lambda$ is isomorphic to the Weyl group $W(D_{n-2})$.
If $m_1=m_2$, then this orbit consists of $2n(n-1)$ elements.

If we have an orbit with points given in orthogonal coordinates,
it is easy, by using formula~\eqref{dif-D}, to transform it to
$\omega$-coordinates.

\subsection[Orbits of $A_3$]{Orbits of $\boldsymbol{A_3}$}\label{Orb2}

Orbits for $A_3$, $B_3$ and $C_3$ can be calculated by means of
the description of Weyl groups $W(A_3)$, $W(B_3)$ and $W(C_3)$ in
the orthogonal coordinate systems. Below we give results of such
calculations. The points $\lambda$ of the orbits are given in the
$\omega$-coordinates $(a\,b\,c)$, that is,
$\lambda=a\omega_1+b\omega_2+c\omega_3$.

The orbit $O(a\ b\ c)$, $a>0$, $b>0$, $c>0$, of $A_3$ contains the
points
 \begin{gather*}
O(a\ b\ c)\ni (a\ b\ c), (a{+}b\ {-}b\ b{+}c), (a{+}b\ c\
{-}b{-}c),
   (a\ b{+}c\ {-}c), (a{+}b{+}c\ {-}c\ {-}b),
\\
\phantom{O(a\ b\ c)\ni{}}{} (a{+}b{+}c\ {-}b{-}c\ b), ({-}a\
a{+}b\ c), ({-}a\ a{+}b{+}c\ {-}c), (b\ {-}a{-}b\ a{+}b{+}c),
\\
\phantom{O(a\ b\ c)\ni{}}{} (b{+}c\ {-}a{-}b{-}c\ a{+}b),
({-}a{-}b\ a\ b{+}c), ({-}b\ {-}a\ a{+}b{+}c)
\end{gather*}
and the points, contragredient to these points, where the
contragredient of the point $(a'\ b'\ c')$ is $({-}c'\ {-}b'\
{-}a')$. We also have
 \begin{gather*}
O(a\ b\ 0)\ni (a\ b\ 0), (a{+}b\ {-}b\ b), (a{+}b\ 0\ {-}b),
   ({-}a\ a{+}b\ 0), ({-}a{-}b\ a\ b),
\\
\phantom{O(a\ b\ 0)\ni{}}{} (b\ {-}a{-}b\ a{+}b)\ {\rm and\
contragredient\  points};
\\
O(a\ 0\ c)\ni (a\ 0\ c), (a\ c\ {-}c), (a{+}c\ {-}c\ 0),
   ({-}a\ a\ c), (0\ {-}a\ a{+}c),
\\
\phantom{O(a\ 0\ c)\ni{}}{} ({-}a\ a{+}c\ {-}c)\ {\rm and\
contragredient\  points};
\\
O(0\ b\ c)\ni (0\ b\ c), (b\ {-}b\ b{+}c), (0\ b{+}c\ {-}c),
   (b{+}c\ {-}b{-}c\ b), ({-}b\ 0\ b{+}c),
\\
\phantom{O(0\ b\ c)\ni{}}{} (b\ c\ {-}b{-}c)\ {\rm and\
contragredient\  points};
\\
O(a\ 0\ 0)\ni (a\ 0\ 0), ({-}a\ a\ 0), (0\ 0\ {-}a),
   (0\ {-}a\ a);
\\
O(0\ b\ 0)\ni (0\ b\ 0), (b\ {-}b\ b), (b\ 0\ {-}b), ({-}b\ b\
{-}b),
    (0\ {-}b\ 0), ({-}b\ 0\ b);
\\
O(0\ 0\ c)\ni (0\ 0\ c), (0\ c\ {-}c), (c\ {-}c\ 0),
   ({-}c\ 0\ 0).
\end{gather*}

\subsection[Orbits of $B_3$]{Orbits of $\boldsymbol{B_3}$} \label{Orb3}
As in the previous case, the points $\lambda$ of the orbits are
given by the coordinates $(a\ b\ c)$, where
$\lambda=a\omega_1+b\omega_2+c\omega_3$. The orbit $O(a\ b\ c)$,
$a>0$, $b>0$, $c>0$, of $B_3$ contains the points
 \begin{gather*}
O(a\ b\ c)\ni  (a\ b\ c),  (a{+}b\ {-}b\ 2b{+}c),  ({-}a\ a{+}b\
c), (b\ {-}a{-}b\ 2a{+}2b{+}c),
 \\
 \phantom{O(a\ b\ c)\ni{} }{}
({-}a{-}b\ a\ 2b{+}c), ({-}b\ {-}a\ 2a{+}2b{+}c), (a\ b{+}c\
{-}c),
 (a{+}b{+}c\ {-}b{-}c\ 2b{+}c),
 \\
  \phantom{O(a\ b\ c)\ni{} }{}
 ({-}a\ a{+}b{+}c\ {-}c),  (b{+}c\ {-}a{-}b{-}c\ 2a{+}2b{+}c),
 ({-}a{-}b{-}c\ a\ 2b{+}c),
 \\
  \phantom{O(a\ b\ c)\ni {}}{}
({-}b{-}c\ {-}a\ 2a{+}2b{+}c), ({-}a{-}2b{-}c\ b\
c),({-}a{-}b{-}c\ {-}b\ 2b+c),
\\
  \phantom{O(a\ b\ c)\ni{} }{}
(a{+}2b{+}c\ {-}a{-}b{-}c\ c), (b\ a{+}b{+}c\ {-}2a{-}2b{-}c),
(a{+}b{+}c\ {-}a{-}2b{-}c\ 2b{+}c),
 \\
  \phantom{O(a\ b\ c)\ni{} }{}
({-}b\ a{+}2b{+}c\ {-}2a{-}2b{-}c),({-}a{-}2b{-}c\ b{+}c\ {-}c),
({-}a{-}b\ {-}b{-}c\ 2b{+}c),
 \\
  \phantom{O(a\ b\ c)\ni{} }{}
(a{+}2b{+}c\ {-}a{-}b\ {-}c),(b{+}c\ a{+}b\ {-}2a{-}2b{-}c),
(a{+}b\ {-}a{-}2b{-}c\ 2b{+}c),
 \\
 \phantom{O(a\ b\ c)\ni{} }{}
({-}b{-}c\ a{+}2b{+}c\ {-}2a{-}2b{-}c)
\end{gather*}
and also all these points with the minus sign. For the orbits
$O(a\ b\ 0)$, $O(a\ 0\ c)$ and $O(0\ b\ c)$ we have
 \begin{gather*}
O(a\ b\ 0)\ni \pm (a\ b\ 0), \pm (a{+}b\ {-}b\ 2b), \pm ({-}a\
a{+}b\ 0), \pm (b\ {-}a{-}b\ 2a{+}2b),
 \\
\phantom{O(a\ b\ 0)\ni}{} \pm ({-}a{-}b\ a\ 2b), \pm ({-}b\ {-}a\
2a{+}2b),\pm ({-}a{-}2b\ b\ 0), \pm ({-}a{-}b\ {-}b\ 2b),
 \\
\phantom{O(a\ b\ 0)\ni}{} \pm (a{+}2b\ {-}a{-}b\ 0), \pm (b\
a{+}b\ {-}2a{-}2b), \pm (a{+}b\ {-}a{-}2b\ 2b), \pm ({-}b\ a{+}2b\
{-}2a{-}2b);\!
\\
O(a\ 0\ c)\ni \pm (a\ 0\ c), \pm ({-}a\ a\ c), \pm (0\ {-}a\
2a{+}c), \pm (a\ c\ {-}c),
 \\
\phantom{O(a\ 0\ c)\ni}{} \pm (a{+}c\ {-}c\ c), \pm ({-}a\ a{+}c\
{-}c),\pm (c\ {-}a{-}c\ 2a{+}c), \pm ({-}a{-}c\ a\ c),
 \\
\phantom{O(a\ 0\ c)\ni}{} \pm ({-}c\ {-}a\ 2a{+}c), \pm ({-}a{-}c\
0\ c), \pm (a{+}c\ {-}a{-}c\ c), \pm (0\ a{+}c\ {-}2a{-}c);\!
\\
 O(0\ b\ c)\ni \pm (0\ b\ c), \pm (b\ {-}b\ 2b{+}c), \pm ({-}b\ 0\
2b{+}c), \pm (0\ b{+}c\ {-}c),
 \\
\phantom{O(0\ b\ c)\ni}{} \pm (b{+}c\ {-}b{-}c\ 2b{+}c), \pm
({-}b{-}c\ 0\ 2b{+}c),\pm ({-}2b{-}c\ b\ c), \pm ({-}b{-}c\ {-}b\
2b{+}c),
 \\
\phantom{O(0\ b\ c)\ni}{} \pm (2b{+}c\ {-}b{-}c\ c), \pm (b\
b{+}c\ {-}2b{-}c), \pm (b{+}c\ {-}2b{-}c\ 2b{+}c), \pm ({-}b\
2b{+}c\ {-}2b{-}c).
\end{gather*}
The orbits $O(a\ 0\ 0)$, $O(0\ b\ 0)$ and $O(0\ 0\ c)$ consist of
the points
\begin{gather*}
O(a\ 0\ 0)\ni \pm (a\ 0\ 0), \pm (a\ {-}a\ 0), \pm (0\ a\ {-}2a);
\\
O(0\ b\ 0)\ni \pm (0\ b\ 0), \pm (b\ {-}b\ 2b), \pm ({-}b\ 0\ 2b),
\pm ({-}2b\ b\ 0),  \pm ({-}b\ {-}b\ 2b), \pm (b\ {-}2b\ 2b);
\\
O(0\ 0\ c)\ni \pm (0\ 0\ c), \pm (c\ {-}c\ c), \pm (0\ c\ {-}c),
\pm ({-}c\ 0\ c) .
\end{gather*}

\subsection[Orbits of $C_3$]{Orbits of $\boldsymbol{C_3}$}\label{Orb4}

Again, the points $\lambda$ of the orbits are given by the
coordinates $(a\ b\ c)$, where
$\lambda=a\omega_1+b\omega_2+c\omega_3$. The orbit $O(a\ b\ c)$,
$a>0$, $b>0$, $c>0$, of $C_3$ contains the points
\begin{gather*}
O(a\ b\ c)\ni  (a\ b\ c),  (a{+}b\ {-}b\ b{+}c),  ({-}a\ a{+}b\
c), (b\ {-}a{-}b\ a{+}b{+}c),
\\
\phantom{O(a\ b\ c)\ni{}}{} ({-}a{-}b\ a\ b{+}c), ({-}b\ {-}a\
a{+}b{+}c), (a\ b{+}2c\ {-}c),
 (a{+}b{+}2c\ {-}b{-}2c\ b{+}c),
\\
\phantom{O(a\ b\ c)\ni{}}{}
 ({-}a\ a{+}b{+}2c\ {-}c),  (b{+}2c\ {-}a{-}b{-}2c\
 a{+}b{+}c), ({-}a{-}b{-}2c\ a\ b{+}c),
\\
\phantom{O(a\ b\ c)\ni{}}{} ({-}b{-}2c\ {-}a\ a{+}b{+}c),
({-}a{-}2b{-}2c\ b\ c), ({-}a{-}b{-}2c\ {-}b\ b{+}c),
\\
\phantom{O(a\ b\ c)\ni{}}{} (a{+}2b{+}2c\ {-}a{-}b{-}2c\ c), (b\
a{+}b{+}2c\ {-}a{-}b{-}c),(a{+}b{+}2c\ {-}a{-}2b{-}2c\ b{+}c),
\\
\phantom{O(a\ b\ c)\ni{}}{} ({-}b\ a{+}2b{+}2c\
{-}a{-}b{-}c),({-}a{-}2b{-}2c\ b{+}2c\ {-}c),
 ({-}a{-}b\ {-}b{-}2c\ b{+}c),
\\
\phantom{O(a\ b\ c)\ni{}}{} (a{+}2b{+}2c\ {-}a{-}b\ {-}c),(b{+}2c\
a{+}b\ {-}a{-}b{-}c), (a{+}b\ {-}a{-}2b{-}2c\ b{+}c),
\\
\phantom{O(a\ b\ c)\ni{}}{} ({-}b{-}2c\ a{+}2b{+}2c\ {-}a{-}b{-}c)
\end{gather*}
and also all these points with the minus sign. For the orbits
$O(a\ b\ 0)$, $O(a\ 0\ c)$ and $O(0\ b\ c)$ we have
 \begin{gather*}
O(a\ b\ 0)\ni \pm (a\ b\ 0), \pm (a{+}b\ {-}b\ b), \pm ({-}a\
a{+}b\ 0), \pm (b\ {-}a{-}b\ a{+}b),
\\
\phantom{O(a\ b\ 0)\ni}{} \pm ({-}a{-}b\ a\ b), \pm ({-}b\ {-}a\
a{+}b),\pm ({-}a{-}2b\ b\ 0), \pm ({-}a{-}b\ {-}b\ b),
 \\
\phantom{O(a\ b\ 0)\ni}{} \pm (a{+}2b\ {-}a{-}b\ 0), \pm (b\
a{+}b\ {-}a{-}b), \pm (a{+}b\ {-}a{-}2b\ b), \pm ({-}b\ a{+}2b\
{-}a{-}b);
\\
O(a\ 0\ c)\ni \pm (a\ 0\ c), \pm ({-}a\ a\ c), \pm (0\ {-}a\
a{+}c), \pm (a\ 2c\ {-}c),
\\
\phantom{O(a\ 0\ c)\ni}{} \pm (a{+}2c\ {-}2c\ c), \pm (a{+}2c\
{-}a{-}2c\ c), \pm (0\ a{+}2c\ {-}a{-}c), \pm ({-}a\ a{+}2c\
{-}c),
\\
\phantom{O(a\ 0\ c)\ni}{} \pm (2c\ {-}a{-}2c\ a{+}c), \pm
({-}a{-}2c\ a\ c), \pm ({-}2c\ {-}a\ a{+}c), \pm  ({-}a{-}2c\ 0\
c);
\\
 O(0\ b\ c)\ni \pm (0\ b\ c), \pm (b\ {-}b\ b{+}c), \pm ({-}b\ 0\
b{+}c), \pm (0\ b{+}2c\ {-}c),
\\
\phantom{O(0\ b\ c)\ni}{} \pm (b{+}2c\ {-}b{-}2c\ b{+}c), \pm
({-}b{-}2c\ 0\ b{+}c), \pm ({-}2b{-}2c\ b\ c), \pm ({-}b{-}2c\ -b\
b{+}c),
\\
\phantom{O(0\ b\ c)\ni}{} \pm (2b{+}2c\ {-}b{-}2c\ c), \pm (b\
b{+}2c\ {-}b{-}c), \pm (b{+}2c\ {-}2b{-}2c\ b{+}c),
\\
\phantom{O(0\ b\ c)\ni}{}
 \pm ({-}b\ 2b{+}2c\ {-}b{-}c).
\end{gather*}
The orbits $O(a\ 0\ 0)$, $O(0\ b\ 0)$ and $O(0\ 0\ c)$ consist of
the points
\begin{gather*}
O(a\ 0\ 0)\ni \pm (a\ 0\ 0), \pm (a\ {-}a\ 0), \pm (0\ a\ {-}a);
\\
O(0\ b\ 0)\ni \pm (0\ b\ 0), \pm (b\ {-}b\ b), \pm (b\ 0\ {-}b),
\pm (2b\ {-}b\ 0),  \pm ({-}b\ {-}b\ b), \pm (b\ {-}2b\ b);
\\
O(0\ 0\ c)\ni \pm (0\ 0\ c), \pm (0\ 2c\ {-}c), \pm (2c\ {-}2c\
c), \pm (2c\ 0\ {-}c) .
\end{gather*}

\setcounter{equation}{0}

\section{Operations on orbits}
\label{operat}

In this section we consider some operations with orbits:
decomposition of products of orbits into orbits and decomposition
of $W$-orbits into $W'$-orbits. These operations are a base for
corresponding operations with orbit functions.

\subsection{Decomposition of products of orbits}\label{Decomp1}

The product $O(\lambda)\otimes O(\lambda')$ of two orbits
$O(\lambda)$ and $O(\lambda')$ is the set of all points of the
form $\lambda_1+\lambda_2$, where $\lambda_1\in O(\lambda)$ and
$\lambda_2\in O(\lambda')$. Since a set of points
$\lambda_1+\lambda_2$, $\lambda_1\in O(\lambda)$, $\lambda_2\in
O(\lambda')$, is invariant with respect to action of the
corresponding Weyl group, each product of orbits is decomposable
into a sum of orbits. If $\lambda=0$, then $O(\lambda)\otimes
O(\lambda')=O(\lambda')$. If $\lambda'=0$, then $O(\lambda)\otimes
O(\lambda')=O(\lambda)$. In what follows we assume that
$\lambda\ne 0$ and $\lambda'\ne 0$. Let us consider decompositions
of products of orbits.

\medskip

\noindent {\bf Example.} {\it Orbits of $A_1$.} If $a\in E_1$,
then the orbit of this point $O(a)$ consists of two points~$a$
and~$-a$. It is easy to see that for the product $O(a)\otimes
O(b)$, $a\ne b$, we have
\begin{gather*}
O(a)\otimes O(b)
\equiv\{a,-a\}\otimes\{b,-b\} =\{a+b,-a-b\}\cup\{|a-b|,-|a-b|\}\\
\phantom{O(a)\otimes O(b)}{}=O(a+b)\cup O(|a-b|).
\end{gather*}

However, decomposition of products of orbits in higher dimension
of Euclidean spaces is not a simple task. We first consider some
general results on decomposition.

Let $O(\lambda)=\{ w\lambda | w\in W/W_\lambda\}$ and $O(\mu)=\{
w\mu | w\in W/W_\mu\}$ be two orbits. Then
 \begin{gather}
O(\lambda)\otimes O(\mu)= \{ w\lambda +w'\mu | w\in
W/W_\lambda,w'\in W/W_\mu\}\nonumber
 \\ \label{decom}  \qquad
{}=\{ w\lambda{+}w_1\mu | w{\in} W/W_\lambda\}\cup \{
w\lambda{+}w_2\mu | w{\in} W/W_\lambda\}\cup \cdots \cup \{
w\lambda{+}w_s\mu | w{\in} W/W_\lambda\},
 \end{gather}
where $w_1,w_2,\ldots ,w_s$ is the set of elements of $W/W_\mu$.
Since a product of orbits is invariant with respect to $W$, for
decomposition of the product $O(\lambda)\otimes O(\mu)$ into
separate orbits it is necessary to take dominant elements from
each term of the right hand side of \eqref{decom}. That is,
$O(\lambda)\otimes O(\mu)$ is a union of the orbits, determined by
points from
 \begin{gather}
 D(\{ w\lambda+w_1\mu \,|\, w\in W/W_\lambda\}),\quad D(\{
w\lambda+w_2\mu \,|\, w\in W/W_\lambda\}),\quad \ldots ,\nonumber\\
\qquad\qquad D(\{ w\lambda+w_s\mu \,|\, w\in
W/W_\lambda\}),\label{decom1}
 \end{gather}
where $D(\{ w\lambda+w_i\mu \,|\, w\in W/W_\lambda\})$ means the
set of dominant elements in the collection $\{ w\lambda+w_i\mu
\,|\, w\in W/W_\lambda\}$.

\begin{proposition}\label{proposition1} The product $O(\lambda)\otimes O(\mu)$
consists only of orbits of the form $O(|w\lambda+\mu|)$, $w\in
W/W_\lambda$, where $|w\lambda+\mu|$ is a dominant weight of the
orbit containing $w\lambda+\mu$. Moreover, each such orbit
$O(|w\lambda+\mu|)$, $w\in W/W_\lambda$, belongs to the product
$O(\lambda)\otimes O(\mu)$.
\end{proposition}

\begin{proof} For each dominant element $w\lambda+w_i\mu$ from
\eqref{decom1} there exists an element $w''\in W$ such that $w''(
w\lambda+w_i\mu)=w'\lambda+\mu$. It means that $w\lambda+w_i\mu$
is of the form $|w'\lambda+\mu|$, $w'\in W/W_\lambda$. Conversely,
take any element $w\lambda+\mu$, $w\in W/W_\lambda$. It belongs to
the product $O(\lambda)\otimes O(\mu)$. This means that
$|w\lambda+\mu|$ also belongs to this product. Therefore, the
orbit $O(|w\lambda+\mu|)$ is contained in $O(\lambda)\otimes
O(\mu)$. The proposition is proved.
\end{proof}

It follows from Proposition \ref{proposition1} that for
decomposition of the product $O(\lambda)\otimes O(\mu)$ into
separate orbits we have to take all elements $w\lambda+\mu$, $w\in
W/W_\lambda$, and to find the corresponding dominant elements
$|w\lambda+\mu |$, $w\in W/W_\lambda$. It may seem that
 \begin{gather} \label{decom2}
O(\lambda)\otimes O(\mu)=\bigcup_{w\in W/W_\lambda}
O(|w\lambda+\mu |).
 \end{gather}
However, this is not true. The simplest counterexample is when
$\mu=0$. Then according to this formula $O(\lambda)\otimes
O(\mu)=O(\lambda)\cup O(\lambda) \cup \cdots \cup O(\lambda)$
($|W/W_\lambda|$ times). However, as we know, $O(\lambda)\otimes
O(\mu)=O(\lambda)$.

Proposition  \ref{proposition1} states that instead of
\eqref{decom2} we have
 \begin{gather} \label{decom-0}
O(\lambda)\otimes O(\mu) \subseteq \bigcup_{w\in W/W_\lambda}
O(|w\lambda+\mu |).
 \end{gather}
Note that some orbits on the right hand side can coincide. Some
orbits can be contained in the decomposition of $O(\lambda)\otimes
O(\mu)$ with a multiplicity. The most difficult problem under
consideration of products of orbits is to find these
multiplicities.

\begin{proposition}\label{proposition2}  Let $O(\lambda)$ and $O(\mu)$ be orbits
such that $\lambda\ne 0$ and $\mu\ne 0$, and let all elements
$w\lambda+\mu$, $w\in W/W_\lambda$, be strictly dominant (that is,
they are dominant and do not belong to any wall of Weyl dominant
chamber). Then
 \begin{gather} \label{decom3}
O(\lambda)\otimes O(\mu)=\bigcup_{w\in W/W_\lambda}
O(w\lambda+\mu).
 \end{gather}
 \end{proposition}

\begin{proof} Under the conditions of the proposition the set of
elements $w\lambda+\mu$, $w\in W/W_\lambda$, are contained in the
first set of \eqref{decom1} if $w_1$ coincides with the identical
transformation. Moreover, $W_{w\lambda+\mu}=\{ 1\}$ for all
elements $w\in W/W_\lambda$. Then $\mu$ is a strictly dominant
element, that is, $W_\mu=\{ 1\}$. Let us show that the collection
\eqref{decom1} contains only one non-empty set $ D(\{ w\lambda+\mu
| w\in W/W_\lambda\})$. Indeed, let $w\lambda+w_i\mu$, $w_i\ne 1$,
be a dominant element. Then
 \[
w_i^{-1}(w\lambda+w_i\mu)=w_i^{-1}w\lambda+\mu
 \]
is not a dominant element since $W_{w_i^{-1}w\lambda+\mu}=\{ 1\}$.
This contradicts the conditions of the proposition and, therefore,
the collection \eqref{decom1} contains only one non-empty set. The
set of orbits, corresponding to the points of $D(\{
w\lambda+\mu|w\in W/W_\lambda \} )$, coincides with the right hand
side of \eqref{decom3}. The proposition is proved.
\end{proof}

Under the conditions of Proposition \ref{proposition2}, elements
$w\lambda+w'\mu$, $w\in W/W_\lambda$, $w'\in W$, are pairwise
different, and all dominant elements of $O(\lambda)\otimes O(\mu)$
are contained in one set of \eqref{decom1}. In general, some
elements $|w\lambda+\mu |$ with fixed $w$ can be contained in
several sets of \eqref{decom1}, that is, some orbits from
Proposition 1 can be contained in $O(\lambda)\otimes O(\mu)$ with
a multiplicity $m_{|w\lambda+\mu |}>1$. Then there exist dominant
elements $w'\lambda+w_{i}\mu$ and $w''\lambda+w_{j}\mu$ from
different sets of \eqref{decom1} such that
 \[
w'\lambda+w_{i}\mu=w''\lambda+w_{j}\mu=| w\lambda+\mu|
 \]
for some $w\in W/W_\lambda$ (one of these elements can coincide
with $w\lambda+\mu$ if $w\lambda+\mu$ is dominant).

\begin{proposition}\label{proposition3} Let $O(\lambda)$ and $O(\mu)$ be orbits
such that $\lambda\ne 0$ and $\mu\ne 0$, and let all elements
$w\lambda+\mu$, $w\in W/W_\lambda$, be dominant. Then
 \[
O(\lambda)\otimes O(\mu)=\bigcup_{w\in W/W_\lambda}
n_{w\lambda+\mu} O(w\lambda+\mu),
 \]
where $n_{w\lambda+\mu}=|W_{w\lambda+\mu}|$.
\end{proposition}

\begin{proof} Since $\lambda\ne 0$, all elements $w\lambda+\mu$,
$w\in W/W_\lambda$, can be dominant if and only if $W_\mu=\{ 1\}$.
Then on the right hand side of \eqref{decom} there are $|W|$
terms. If the element $w\lambda+\mu$ is strictly dominant, that
is, $W_{w\lambda+\mu}=\{ 1\}$, then this element is met only in
one term. This means that $n_{w\lambda+\mu}=1$.  If $w\lambda+\mu$
is placed on some Weyl wall, then it is met in
$n_{w\lambda+\mu}=|W_{w\lambda+\mu}|$ terms. Therefore, there are
$n_{w\lambda+\mu}$ orbits $O(w\lambda+\mu)$ in the decomposition
of $O(\lambda)\otimes O(\mu)$. The proposition is proved.
\end{proof}

The highest and the lowest components of the decomposition of
$O(\lambda)\otimes O(\mu)$ can be easily found. The highest one is
$O(\lambda+\mu)$. The lowest one is $O(|\lambda+\widetilde{\mu} |
)=O(|\widetilde{\lambda}+\mu |)$, where the tilde denotes the
lowest weight of the orbit and, as before,
$|\lambda+\widetilde{\mu} |$ denotes a dominant element of the
orbit containing $\lambda+\widetilde{\mu}$ (it often happens that
$\lambda+\widetilde{\mu}$ is not dominant). The highest component
$O(\lambda+\mu)$ is always with multiplicity 1 in the
decomposition. The lowest one may not be unique.

By Proposition \ref{proposition1}, all other orbits in the
decomposition of $O(\lambda)\otimes O(\mu)$ are placed between the
orbits $O(\lambda+\mu)$ and $O(|\lambda+\widetilde{\mu} | )$. This
means that dominant elements $|w\lambda+\mu|$ of other orbits
$O(|w\lambda+\mu|)$, $w\in W/W_\lambda$, are in the intersection
of the sets $\{ \lambda+\mu-\sum\limits_i m_i\alpha_i\; |\; m_i\in
{\mathbb N}\}$ and $\{ |\widetilde{\lambda}+\mu|+\sum\limits_i
m_i\alpha_i\; | \; m_i\in {\mathbb N}\}$, where $\{
\alpha_1,\alpha_2,\ldots ,\alpha_n\}$ is a set of simple roots.

\begin{proposition}\label{proposition4}  If $W_\mu=\{ 1\}$ and none of the points
$w\lambda+\mu$, $w\in W$, lie on some Weyl chamber, then
 \[
O(\lambda)\otimes O(\mu)=\bigcup_{w\in W/W_\lambda}
O(|w\lambda+\mu|).
 \]
\end{proposition}

\begin{proof} For the product $O(\lambda)\times O(\mu)$ the inclusion
\eqref{decom-0} takes place. Each orbit $O(|w\lambda+\mu|)$, $w\in
W$, has $|W|$ elements and is contained in $O(\lambda)\times
O(\mu)$. Therefore, numbers of elements in both sides of
\eqref{decom-0} coincide. This means that the inclusion
\eqref{decom-0} is in fact an equality. This proves the
proposition.
\end{proof}

We formulate 2 conjectures concerning decomposition of products of
orbits.

 \begin{conjecture}\label{conjecture1}  Let $O(\lambda)$ and $O(\mu)$,
$\lambda\ne 0$, $\mu\ne 0$, be orbits, and let $\mu$ be strictly
dominant. If for some $w\in W$ the element $w\lambda+\mu$ is
strictly dominant, then the multiplicity of $O(w\lambda+\mu )$ in
$O(\lambda)\otimes O(\mu)$ is $1$.
\end{conjecture}

 \begin{conjecture}\label{conjecture2}
 Let $O(\lambda)$ and $O(\mu)$,
$\lambda\ne 0$, $\mu\ne 0$, be orbits, and let a stabilizer
subgroup of $\mu$ be generated by the reflections $r_{i_1},\ldots,
r_{i_s}$, corresponding to simple roots $\alpha_{i_1},\ldots,
\alpha_{i_s}$. If for each $w\in W/W_\lambda$ we have $\langle
w\lambda+\mu,\alpha_j \rangle >0$, $j\not\in \{ i_1,i_2,\ldots
,i_s\}$, then
 \[
O(\lambda)\otimes O(\mu)=\bigcup_{w\in W/W_\lambda} O(w\lambda+\mu
).
 \]
\end{conjecture}

At the end of this subsection we formulate the following method
for decomposition of products $O(\lambda)\otimes O(\mu)$, which
follows from the statement of Proposition~\ref{proposition1}. On
the first step we shift all points of the orbit $O(\lambda)$ by
$\mu$. As a result, we obtain the set of points $w\lambda+\mu$,
$w\in W/W_\lambda$. On the second step, we map non-dominant
elements of this set by elements of the Weyl group~$W$ to the
dominant chamber. On this step we obtain the set $|w\lambda+\mu|$,
$w\in W/W_\lambda$. Then according to
Proposition~\ref{proposition1}, $O(\lambda)\otimes O(\mu)$
consists of the orbits $O(|w\lambda+\mu|)$. On the third step, we
determine multiplicities of these orbits, taking into account the
above propositions or making direct calculations.

\subsection{Decomposition of products for rank 2}\label{Decomp2}

Here we give examples of decompositions of products of orbits for
the cases $A_2$, $B_2$ and $G_2$. Orbits for these cases are
placed on the plane. Therefore, decompositions can be done by
geometrical calculations on this plane. The cases of $A_2$ and
$B_2$ can be easily considered by using for points of orbits
orthogonal coordinates from Section~\ref{weylABCD}. For these
coordinates the corresponding Weyl groups have a simple
description and this give the possibility to make calculations in
a simple manner.

For the case of $A_2$ we have
\begin{alignat*}{3}
&A_2\,:\, &O(a\ 0)\otimes O(b\ 0) &= O(a{+}b\ 0)\cup O({-}a{+}b\
a)
                                                 &&    \quad (a<b)\\
            &&O(a\ 0)\otimes O(a\ 0) &= O(2a\ 0)\cup 2O(0\ a)&&\\
            &&O(a\ 0)\otimes O(0\ b) &= O(a\ b)\cup O(0\ {-}a{+}b)
                                                 &&    \quad (a<b)\\
            &&O(a\ 0)\otimes O(0\ a) &= O(a\ a)\cup 3O(0\ 0)&&\\
            &&O(a\ 0)\otimes O(b\ b) &= O(a{+}b\ b)\cup O(a{-}2b\ b)
                       \cup O(a{-}b\ 2b) && \quad (a>2b)\\
            &&O(a\ 0)\otimes O(b\ b) &= O(a{+}b\ b)\cup O(2b{-}a\ a{-}b)
                       \cup O(a{-}b\ 2b)&& \quad (2b>a>b)\\
            &&O(a\ 0)\otimes O(b\ b) &= O(a{+}b\ b)\cup O(b{-}a\ a{+}b)
                       \cup O(b\ b{-}a)&& \quad (b>a)\\
            &&O(1\ 1)\otimes O(1\ 1) &= O(2\ 2)\cup 2O(0\ 3)
             \cup 2O(3\ 0)\cup 2O(1\ 1)\\
            &&    & \ \ \ \cup 6 O(0\ 0).
\end{alignat*}

Similar products of $C_2$ orbits are of the form
\begin{alignat*}{3}
&C_2\,:\, &O(a\ 0)\otimes O(b\ 0) &= O(a{+}b\ 0)\cup O(a{-}b\ 0)
                                      \cup O(a{-}b\ b)\quad &&(a>b)\\
            &&O(a\ 0)\otimes O(a\ 0) &= O(2a\ 0)\cup 2O(0\ 2a)
                                      \cup 4O(0\ 0)\quad &&\\
            &&O(0\ a)\otimes O(0\ b) &= O(0\ a{+}b)\cup O(2b\ a{-}b)
                                      \cup O(0\ a{-}b)\quad &&(a>b)\\
            &&O(0\ a)\otimes O(0\ a) &= O(0\ 2a)\cup 2O(2a\ 0)
                                      \cup 4O(0\ 0)\quad &&\\
            &&O(a\ 0)\otimes O(0\ b) &= O(a\ b)\cup O(a{-}2b\ b)
                                          \quad &&(a>2b)\\
            &&O(a\ 0)\otimes O(0\ b) &= O(a\ b)\cup O(2b{-}a\ a{-}b)
                                          \quad &&(2b>a>b)\\
            &&O(a\ 0)\otimes O(0\ b) &= O(a\ b)\cup O(a\ b{-}a)
                                          \quad &&(b>a)\\
            &&O(a\ 0)\otimes O(0\ a) &= O(a\ a)\cup 2O(a\ 0)&&{}\\
            &&O(a\ 0)\otimes O(0\ 2a) &= O(a\ 2a)\cup O(a\ a)&&{}\\
            &&O(2a\ 0)\otimes O(0\ a) &= O(2a\ a)\cup 2O(0\ a).&&
\end{alignat*}

Finally a few products of $G_2$ orbits:
\begin{alignat*}{2}
&G_2\,:\, &O(a\ 0)\otimes O(b\ 0) &= O(a{+}b\ 0)\cup O(b{-}a\ 3a)
                 \cup O(2a{-}b\ 3b{-}3a)\\
            &  & & \qquad  \cup O(b{-}a\ 0)\qquad \qquad   (a<b<2a)\\
            &&O(a\ 0)\otimes O(b\ 0) &= O(a{+}b\ 0)\cup O(b{-}a\ 3a)
                    \cup O(b{-}2a\ 3a)\\
             &  &  & \qquad  \cup O(b{-}a\ 0)\quad \qquad (b>2a)\\
            &&O(a\ 0)\otimes O(a\ 0) &= O(2a\ 0)\cup 2O(0\ 3a)
                    \cup 2O(a\ 0)\cup 6O(0\ 0)\\
            &&O(a\ 0)\otimes O(2a\ 0) &= O(3a\ 0)\cup O(a\ 0)
                     \cup O(a\ 3a) \cup 2O(0\ 3a)\\
            &&O(0\ a)\otimes O(0\ b) &= O(0\ a{+}b)\cup O(a\ b{-}a)
                 \cup O(b{-}a\ 2a{-}b)\\
            &  &  & \qquad  \cup O(0\ b{-}a)  \qquad \qquad (a<b<2a)\\
            &&O(0\ a)\otimes O(0\ b) &= O(0\ a{+}b)\cup O(0\ b{-}a)
                   \cup O(a\ b{-}a)\\
            &  &  & \qquad  \cup O(a\ b{-}2a)\quad \qquad  (b>2a)\\
            &&O(0\ a)\otimes O(0\ a) &= O(0\ 2a)\cup 2O(a\ 0)
                    \cup 2O(0\ a)\cup 6O(0\ 0)\\
            &&O(0\ a)\otimes O(0\ 2a) &= O(0\ 3a)\cup 2O(a\ 0)
                     \cup O(a\ a)  \cup O(0\ a) .
\end{alignat*}

Let us note that the most complicated case for decomposition of
products of orbits is for $E_8$. Products of all 36 pairs of
orbits of fundamental weights of $E_8$ are fully decomposed
in~\cite{GP}, as a sideline to the decomposition of products of
fundamental representations of $E_8$.

\subsection[Decomposition of $W$-orbits into $W'$-orbits]{Decomposition of $\boldsymbol{W}$-orbits into
$\boldsymbol{W'}$-orbits}\label{Decomp3}

Let $R$ be a root system with the Weyl group $W$. Let $R'$ be
another root system such that $R'$ is a subset of $R$. Then the
Weyl group $W'$ of $R'$ can be considered as a subgroup of $W$.

Let $O_W(\lambda)$ be a $W$-orbit. The set of elements of
$O_W(\lambda)$ is invariant with respect to $W'$, that is,
$w'O_W(\lambda)=O_W(\lambda)$, $w'\in W'$. This means that
$O_W(\lambda)$ consists of $W'$-orbits. It is an important problem
to represent $O_W(\lambda)$ as a union of $W'$-orbits. Properties
of such representations depend on the root systems $R$ and $R'$
(or on the Weyl groups $W$ and $W'$). We distinguish two cases:
\medskip

{\bf Case 1:} {\it The root systems $R$ and $R'$ span vector
spaces of the same dimension.} In this case the Weyl chambers for
$W$ are smaller than the Weyl chambers for $W'$. Moreover, each
Weyl chamber of $W'$ consists of $|W/W'|$ chambers of $W$. Let
$D_+$ be a dominant Weyl chamber of the root system $R$. Then a
dominant Weyl chamber of $W'$ consists of $W$-chambers  $w_iD_+$,
$i=1,2,\ldots ,k$, $k=|W/W'|$, where $w_i$, $i=1,2,\ldots ,k$, are
representatives of all cosets in $W/W'$. If $\lambda$ do not lie
on any wall of the dominant Weyl chamber $D_+$, then
 \[
O_W(\lambda)=\bigcup_{i=1}^k O_{W'}(w_i\lambda),
 \]
where $O_{W'}$ are $W'$-orbits. When representing $\lambda$ by
coordinates in the $\omega$-basis it is necessary to take into
account that the coordinates of the same point in $\omega$-bases
related to the root systems~$R$ and $R'$ are different. There
exist matrices mapping coordinates in these different
$\omega$-bases (see Subsection~\ref{Proj-mat} below).
\medskip

{\bf Case 2:} {\it The root systems $R$ and $R'$ span vector
spaces of different dimensions.} This case is more complicated. In
order to represent $O_W(\lambda)$ as a union of $W'$-orbits, it is
necessary to take the projection of the points $\mu$ of
$O_W(\lambda)$ to the vector subspace $E_{n'}$ spanned by $R'$ and
to select in that set of projected points dominant points with
respect to the root system $R'$. Note that, under projection,
different points of $O_W(\lambda)$ can give the same point in
$E_{n'}$. This leads to coinciding $W'$-orbits in the
representation of $O_W(\lambda)$ as a union of $W'$-orbits.

\subsection[Decomposition of $W(A_n)$-orbits into
$W(A_{n-1})$-orbits]{Decomposition of $\boldsymbol{W(A_n)}$-orbits
into $\boldsymbol{W(A_{n-1})}$-orbits}\label{W-A_n}

For such decomposition it is convenient to represent orbit
elements in an orthogonal coordinate system (see Subsection
\ref{An}). Let $O(\lambda)\equiv O(m_1,\ldots,m_{n+1})$ be a
$W(A_n)$-orbit with dominant element $\lambda=(m_1,m_2,\ldots,
m_{n+1})$. The orthogonal coordinates $m_1,m_2,\ldots,m_{n+1}$
satisfy the condition $m_1+m_2+\cdots +m_{n+1}=0$. However, as we
noted in Subsection~\ref{An}, we may add to all coordinates $m_i$
the same real number, since under this the $\omega$-coordinates
$\lambda_i=m_i-m_{i+1}$, $i=1,2,\ldots, n$ do not change. This
note is important under decomposition of $W(A_n)$-orbits into
$W(A_{n-1})$-orbits since, under this the condition $x_1+x_2+
\cdots +x_n=0$ can be violated. This does not change the result
for $\omega$-coordinates.

Let us suppose that the dominant element $\lambda=(m_1,m_2,\ldots,
m_{n+1})$ does not lie on a wall of the dominant Weyl chamber,
that is,
 \[
m_1>m_2>\cdots >m_n>m_{n+1}.
 \]
The orbit $O(\lambda)$ consists of all points
  \begin{gather} \label{A_n-1}
w(m_1,m_2,\ldots
,m_{n+1})=(m_{i_1},m_{i_2},\ldots,m_{i_{n+1}}),\qquad w\in W(A_n),
 \end{gather}
where $(i_1,i_2,\ldots ,i_{n+1})$ is a permutation of the numbers
$1,2,\ldots,n+1$. Points of $O(\lambda)$ belong to the vector
space $E_{n+1}$. We restrict these points to the vector subspace
$E_n$, spanned by the simple roots $\alpha_1,\alpha_2,\ldots,
\alpha_{n-1}$ of $A_n$, which form a set of simple roots of
$A_{n-1}$. This restriction then reduces to removing the last
coordinate $m_{i_{n+1}}$ in points
$(m_{i_1},m_{i_2},\ldots,m_{i_{n+1}})$ of the orbit~$O(\lambda)$
(see \eqref{A_n-1}). As a result, we obtain the set of points
  \begin{gather} \label{A_n-2}
(m_{i_1},m_{i_2},\ldots,m_{i_{n}})
  \end{gather}
from the points \eqref{A_n-1}. The point \eqref{A_n-2} is dominant
if and only if
 \[
m_{i_1}\ge m_{i_2}\ge \cdots \ge m_{i_{n}}
 \]
(in fact, for the points \eqref{A_n-1} equalities here are
excluded). It is easy to see that (after restriction to $E_{n}$,
that is, if we remove the last coordinate) the set \eqref{A_n-1}
contains the following set of dominant elements:
 \[
(m_1,\ldots,m_{i-1},\hat{m_i},m_{i+1},\ldots, m_{n+1}),\qquad
i=1,2,\ldots,n+1,
 \]
where a hat over $m_i$ means that the coordinate $m_i$ must be
omitted. Thus, {\it the $W(A_n)$-orbit $O(m_1,m_2,\ldots,m_{n+1})$
with $m_1>m_2>\cdots >m_{n+1}$ consists of the following
$W(A_{n-1})$-orbits:
 \begin{gather} \label{or}
O(m_1,\ldots,m_{i-1},\hat{m_i},m_{i+1},\ldots, m_{n+1}),\qquad
i=1,2,\ldots,n+1.
 \end{gather}
All these orbits are contained in $O(m_1,m_2,\ldots,m_{n+1})$ with
multiplicity~$1$.} By using formula~\eqref{dif-A} it is easy to
represent the dominant elements of~\eqref{or} in
$\omega$-coordinates.

\subsection[Decomposition of $W(A_{n-1})$-orbits into
$W(A_{p-1}){\times} W(A_{q-1})$-orbits, $p{+}q=n$]{Decomposition
of $\boldsymbol{W(A_{n-1})}$-orbits\\ into
$\boldsymbol{W(A_{p-1})\times W(A_{q-1})}$-orbits,
$\boldsymbol{p{+}q=n}$}\label{W-A_p}

Again we use orthogonal coordinates to represent orbit elements.
We take in the system of simple roots $\alpha_1,\alpha_2,\ldots,
\alpha_{n-1}$ of $A_{n-1}$ two parts:
$\alpha_1,\alpha_2,\ldots,\alpha_{p-1}$ and
$\alpha_{p+1},\alpha_{p+2},\ldots,\alpha_{p+q-1}\equiv\alpha_{n-1}$.
The first part determines $W(A_{p-1})$ and the second part
generates $W(A_{q-1})$. Again we suppose that the orthogonal
coordinates $m_1,m_2,\ldots,m_n$ of $\lambda$, determining the
$W(A_{n-1})$-orbit $O(\lambda)$, satisfy the condition
 \[
m_1>m_2>\cdots >m_n,
 \]
that is, $\lambda$ does not lie on a wall of the dominant chamber.
The orbit $O(\lambda)$ consists of all points
  \begin{gather} \label{A_n-3}
w(m_1,m_2,\ldots ,m_{n})=(m_{i_1},m_{i_2},\ldots,m_{i_{n}}),\qquad
w\in W(A_{n-1}),
 \end{gather}
where $(i_1,i_2,\ldots ,i_{n})$ is a permutation of the numbers
$1,2,\ldots,n$. We restrict points \eqref{A_n-3} to the vector
subspace $E_{p}\times E_{q}$ spanned by the simple roots
$\alpha_1,\alpha_2,\ldots,\alpha_{p-1}$ and
$\alpha_{p+1},\alpha_{p+2},\ldots,\alpha_{n-1}$, respectively.
Under restriction  to $E_p\times E_q$ the point \eqref{A_n-3}
becomes the point
 \[
(m_{i_1},m_{i_2},\ldots,m_{i_{p}})(m_{i_{p+1}},m_{i_{p+2}},\ldots,m_{i_{n}}).
 \]

In order to determine the set of $W(A_{p-1})\times
W(A_{q-1})$-orbits contained in $O(\lambda)$ we have to choose in
\eqref{A_n-3} all elements for which
 \[
m_{i_1}>m_{i_2}>\cdots > m_{i_{p}}, \qquad
m_{i_{p+1}}>m_{i_{p+2}}>\cdots >m_{i_{n}}.
 \]
To find this set we have to take all possible sets of numbers
$m_{i_1},m_{i_2},\ldots,m_{i_{p}}$ in the set
$m_1,m_2,\ldots,m_n$, such that $m_{i_1}>m_{i_2}>\cdots
>m_{i_{p}}$. Let $\Sigma$ denotes the collection of such sets.
Then $O(\lambda)$ {\it consists of $W(A_{p-1})\times
W(A_{q-1})$-orbits
 \[
O((m_{i_1},m_{i_2},\ldots,m_{i_{p}})
(m_{j_1},m_{j_2},\ldots,m_{j_{q}})),\qquad
(m_{i_1},m_{i_2},\ldots,m_{i_{p}})\in \Sigma,
 \]
where $(m_{j_1},m_{j_2},\ldots,m_{j_{q}})$ is a supplement of
$(m_{i_1},m_{i_2},\ldots,m_{i_{p}})$ in the set
$(m_1,m_2,\ldots,m_n)$, taken in such an order that
$m_{j_1}>m_{j_2}>\cdots >m_{j_{q}}$. Each of these
$W(A_{p-1})\times W(A_{q-1})$-orbits is contained in $O(\lambda)$
only once.}

\subsection[Decomposition of $W(B_{n})$-orbits into
$W(B_{n-1})$-orbits and of $W(C_{n})$-orbits into
$W(C_{n-1})$-orbits]{Decomposition of
$\boldsymbol{W(B_{n})}$-orbits into
$\boldsymbol{W(B_{n-1})}$-orbits\\ and of
$\boldsymbol{W(C_{n})}$-orbits into
$\boldsymbol{W(C_{n-1})}$-orbits}\label{W-C_n}

Decomposition of $W(B_{n})$-orbits and decomposition of
$W(C_{n})$-orbits are achieved in the same way. For this reason,
we give a proof only in the case of $W(C_{n})$-orbits.

We know that $C_n$ is defined by simple roots
$\alpha_1,\alpha_2,\ldots,\alpha_n$. The roots
$\alpha_2,\ldots,\alpha_n$ is the set of simple roots of
$C_{n-1}$. They span the subspace $E_{n-1}$.

In order to determine elements $\lambda$ of $E_n$ we use the
orthogonal coordinates $m_1,m_2,\ldots,m_n$ (see
Subsection~\ref{Cn}). Then $\lambda$ is dominant if $m_1\ge m_2\ge
\cdots \ge m_n\ge 0$. We assume that $\lambda$ satisfies the
condition
 \[
m_1>m_2>\cdots >m_n>0.
 \]
Then the orbit $O(\lambda)$ consists of all points
  \begin{gather} \label{C_n-1}
w(m_1,m_2,\ldots ,m_{n})=(\pm m_{i_1},\pm m_{i_2},\ldots,\pm
m_{i_{n}}),\qquad w\in W(C_n),
 \end{gather}
where $(i_1,i_2,\ldots ,i_{n})$ is a permutation of the numbers
$1,2,\ldots,n$ (all combinations of signs are possible).

Restricting the elements \eqref{C_n-1} to the vector subspace
$E_{n-1}$, defined above, is reduced to removing the first
coordinate $\pm m_{i_1}$ in \eqref{C_n-1}. As a result, we obtain
from the set  \eqref{C_n-1} the collection
 \[
(\pm m_{i_2},\pm m_{i_3},\ldots,\pm m_{i_{n}}),\qquad w\in W(C_n).
 \]
Only the points $(m_{i_2}, m_{i_3},\ldots,m_{i_{n-1}},\pm
m_{i_{n}})$ may be dominant. Moreover, such a point is dominant if
and only if
 \[
 m_{i_2}> m_{i_3}>\cdots > m_{i_{n}}.
 \]
Therefore, under restriction of points \eqref{C_n-1} to $E_{n-1}$,
we obtain the following $W(C_{n-1})$-dominant elements:
  \begin{gather} \label{C_n-2}
(m_1,m_2,\ldots ,m_{i-1},\hat{m_i},m_{i+1},\ldots ,m_{n}),\qquad
i=1,2,\ldots,n,
  \end{gather}
where a hat over $m_i$ means that the coordinate $m_i$ must be
omitted. Moreover, the element \eqref{C_n-2} with fixed $i$ can be
obtained from two elements in \eqref{C_n-1}, namely, from
$(m_1,m_2,\ldots ,m_{i-1},\pm m_i,$ $m_{i+1},\ldots ,m_{n})$.

Thus, {\it the $W(C_n)$-orbits $O(m_1,m_2,\ldots,m_n)$ with
$m_1>m_2>\cdots >m_n>0$ consists of the following
$W(C_{n-1})$-orbits:
 \[
O(m_1,m_2,\ldots ,m_{i-1},\hat{m_i},m_{i+1},\ldots ,m_{n}),\qquad
i=1,2,\ldots,n.
 \]
Each such $W(C_{n-1})$-orbit is contained in
$O(m_1,m_2,\ldots,m_n)$ with multiplicity~$2$.}

For $W(B_n)$-orbits we have a similar assertion: {\it A
$W(B_n)$-orbits $O(m_1,m_2,\ldots, m_n)$ with $m_1>m_2>\cdots
>m_n>0$ consists of $W(B_{n-1})$-orbits
 \[
O(m_1,m_2,\ldots ,m_{i-1},\hat{m_i},m_{i+1},\ldots ,m_{n}),\qquad
i=1,2,\ldots,n,
 \]
and each such orbit is contained in the decomposition with
multiplicity~$2$.}

\subsection[Decomposition of $W(C_{n})$-orbits into
$W(A_{p-1})\times W(C_{q})$-orbits, $p+q=n$]{Decomposition of
$\boldsymbol{W(C_{n})}$-orbits\\ into
$\boldsymbol{W(A_{p-1})\times W(C_{q})}$-orbits,
$\boldsymbol{p+q=n}$}\label{W-C_p}

If $\alpha_1,\alpha_2,\ldots,\alpha_n$ are simple roots for $C_n$,
then $\alpha_1,\alpha_2,\ldots,\alpha_{p-1}$ are simple roots for
$A_{p-1}$ (they can be embedded into the linear subspace $E_{p}$)
and $\alpha_{p+1},\alpha_{p+2},\ldots,\alpha_n$ are simple roots
for $C_q$ (they generate the linear subspace $E_{q}$).

We use orthogonal coordinates for elements of $E_n$ and consider
$W(C_{n})$-orbits $O(\lambda)$ with $\lambda=(m_1,m_2,\ldots,m_n)$
such that $m_1>m_2>\cdots >m_n>0$. The orbit $O(\lambda)$ consists
of all points \eqref{C_n-1}. Restriction of these points to the
vector subspace $E_{p}\times E_q$ reduces to splitting the set of
coordinates \eqref{C_n-1} into two parts:
  \begin{gather} \label{C_n-3}
(\pm m_{i_1},\pm m_{i_2},\ldots ,\pm m_{i_p})(\pm
m_{i_{p+1}},\ldots ,\pm m_{i_n}) .
  \end{gather}
Due to the condition $m_1>m_2>\cdots >m_n>0$, these elements do
not lie on walls of the $W(A_{p-1})\times W(C_{q})$-chambers. We
have to choose dominant elements (with respect to the Weyl group
$W(A_{p-1})\times W(C_{q})$) in the set of points \eqref{C_n-3}.
The conditions for elements of $E_{p}$ and $E_q$ to be dominant
implies that only the elements
  \[
( m_{i_1},\ldots, m_{i_j},-m_{i_{j+1}}, \ldots ,-m_{i_p})(
m_{i_{p+1}},\ldots , m_{i_n}) , \qquad j=0,1,2,\ldots, p,
 \]
satisfying
 \[
 m_{i_1}>m_{i_2}>\cdots > m_{i_j},\qquad m_{i_{j+1}}<m_{i_{j+2}}<
 \cdots <m_{i_p}, \qquad m_{i_{p+1}}> m_{i_{p+2}}>\cdots >m_{i_n},
 \]
are dominant. Moreover, each such point is contained in the
$W(C_n)$-orbit $O(\lambda)$ only once. This assertion completely
determines the list of $W(A_{p-1})\times W(C_q)$-orbits in
$W(C_n)$-orbit $O(\lambda)$. All $W(A_{p-1})\times W(C_q)$-orbits
are contained in $O(\lambda)$ with multiplicity 1.

\subsection[Decomposition of $W(D_{n})$-orbits into
$W(D_{n-1})$-orbits]{Decomposition of
$\boldsymbol{W(D_{n})}$-orbits into
$\boldsymbol{W(D_{n-1})}$-orbits}\label{W-D_n}

If $\alpha_1,\alpha_2,\ldots,\alpha_n$ is the set of simple roots
of $D_n$, then the roots $\alpha_2,\ldots,\alpha_n$ are simple
roots of~$D_{n-1}$. The last roots span the subspace $E_{n-1}$. It
is assumed that $n>4$.

For elements $\lambda$ of $E_n$ we again use orthogonal
coordinates $m_1,m_2,\ldots,m_n$. Then $\lambda$ is dominant if
$m_1\ge m_2\ge \cdots \ge m_{n-1}\ge |m_n|$. We assume that
$\lambda$ satisfies the condition
 \[
m_1>m_2>\cdots >m_n>0,
 \]
that is, $\lambda$ does not lie on a wall of the dominant Weyl
chamber. Then the orbit $O(\lambda)$ consists of all points
  \begin{gather} \label{D_n-1}
w(m_1,m_2,\ldots ,m_{n})=(\pm m_{i_1},\pm m_{i_2},\ldots,\pm
m_{i_{n}}),\qquad w\in W(D_n),
 \end{gather}
where $(i_1,i_2,\ldots ,i_{n})$ is a permutation of the numbers
$1,2,\ldots,n $ and there is an even number of minus signs.
Restricting the elements of \eqref{D_n-1} to the subspace
$E_{n-1}$ is reduced to removing the first coordinate $\pm
m_{i_1}$ in \eqref{D_n-1}. As a result, we obtain from the set of
points \eqref{D_n-1} the collection
 \[
(\pm m_{i_2},\pm m_{i_3},\ldots,\pm m_{i_{n}}),\qquad w\in W(D_n),
 \]
where the number of minus signs may either be even or odd,
depending on a sign of the removed number $\pm m_{i_1}$. Only the
points of the form $(m_{i_2}, m_{i_3},\ldots,m_{i_{n-1}},\pm
m_{i_{n}})$ may be dominant. Moreover, such a point is dominant if
and only if
 \[
 m_{i_2}> m_{i_3}>\cdots > m_{i_{n}}>0.
 \]
Thus, if we restrict the points \eqref{D_n-1} to $E_{n-1}$ we
obtain the following $W(D_{n-1})$-dominant elements:
  \begin{gather} \label{D_n-2}
(m_1,m_2,\ldots ,m_{i-1},\hat{m_i},m_{i+1},\ldots ,\pm
m_{n}),\qquad i=1,2,\ldots,n,
  \end{gather}
where a hat over $m_i$ means that the coordinate $m_i$ must be
omitted (if $i=n$ then, instead of~$\pm m_{n}$, we have to take
$\pm m_{n-1}$ in \eqref{D_n-2}). Moreover, the element
\eqref{D_n-2} with fixed $i$ can be obtained only from one element
in \eqref{D_n-1}, namely, from element $(m_1,m_2,\ldots
,m_{i-1},\pm m_i,m_{i+1},\ldots$, $\pm m_{n})$, where at $m_i$ and
$m_n$ a sign is the same.

Thus, {\it the $W(D_n)$-orbits $O(m_1,m_2,\ldots,m_n)$ with
$m_1>m_2>\cdots >m_{n}>0$ consists of the following
$W(D_{n-1})$-orbits:
 \[
O(m_1,m_2,\ldots ,m_{i-1},\hat{m_i},m_{i+1},\ldots ,\pm
m_{n}),\qquad i=1,2,\ldots,n.
 \]
Each such $W(D_{n-1})$-orbit is contained in
$O(m_1,m_2,\ldots,m_n)$ with multiplicity $1$.}

It is shown similarly that {\it the $W(D_n)$-orbits
 \[
O(m_1,m_2,\ldots,m_{n-1},-m_n), \qquad m_1>m_2>\cdots
>m_{n}>0
 \]
consists of the same $W(D_{n-1})$-orbits as the $W(D_n)$-orbits
$O(m_1,\ldots,m_{n-1},m_n)$ with the same set of numbers
$m_1,m_2,\ldots,m_n$.}

\subsection[Decomposition of $W(D_{n})$-orbits into
$W(A_{p-1})\times W(D_{q})$-orbits, $p+q=n$, $q\ge
4$]{Decomposition of $\boldsymbol{W(D_{n})}$-orbits\\ into
$\boldsymbol{W(A_{p-1})\times W(D_{q})}$-orbits,
$\boldsymbol{p+q=n}$, $\boldsymbol{q\ge 4}$}\label{W-D_p}

If $\alpha_1,\alpha_2,\ldots,\alpha_n$ are simple roots of $D_n$,
then $\alpha_1,\alpha_2,\ldots,\alpha_{p-1}$ are simple roots of
$A_{p-1}$ (they can be embedded into the linear subspace $E_{p}$)
and $\alpha_{p+1},\alpha_{p+2},\ldots,\alpha_n$ are simple roots
of $D_q$ (they generate the linear subspace $E_{q}$).

We use orthogonal coordinates in $E_n$ and consider
$W(D_{n})$-orbits $O(\lambda)$ with an element
$\lambda=(m_1,m_2,\ldots,m_n)$ such that $m_1>m_2>\cdots >m_n>0$.
The orbit $O(\lambda)$ consists of all points~\eqref{D_n-1}.
Restricting these points to the vector subspace $E_{p}\times E_q$
is reduced to splitting the set of coordinates \eqref{D_n-1} into
two parts:
  \begin{gather} \label{D_n-3}
(\pm m_{i_1},\pm m_{i_2},\ldots ,\pm m_{i_p})(\pm
m_{i_{p+1}},\ldots ,\pm m_{i_n}) .
  \end{gather}
Due to the condition $m_1>m_2>\cdots >m_n>0$, these elements do
not lie on walls of the dominant $W(A_{p-1})\times
W(D_{q})$-chamber. We have to choose dominant elements (with
respect to the Weyl group $W(A_{p-1})\times W(D_{q})$) in the set
of points \eqref{D_n-3}. The conditions for elements of $E_{p}$
and $E_q$ to be dominant implies that the elements
 \[
( m_{i_1},\ldots, m_{i_j},-m_{i_{j+1}}, \ldots ,-m_{i_p})(
m_{i_{p+1}},\ldots ,\pm m_{i_n}) , \qquad j=0,1,2,\ldots, p,
 \]
satisfying the conditions
 \[
 m_{i_1}>m_{i_2}>\cdots > m_{i_j},\ \ m_{i_{j+1}}<m_{i_{j+2}}<
 \cdots <m_{i_p}, \ \ m_{i_{p+1}}> m_{i_{p+2}}>\cdots >m_{i_n},
 \]
are dominant if and only if the number of minus signs is even.
Moreover, each such point is contained in the $W(D_n)$-orbit
$O(\lambda)$ only once. These assertions completely determine the
list of $W(A_{p-1})\times W(D_{q})$-orbits in $W(D_n)$-orbit
$O(\lambda)$. All $W(A_{p-1})\times W(D_{q})$-orbits are contained
in $O(\lambda)$ with multiplicity 1.

\subsection{\bf Decomposition by means of a projection matrix}\label{Proj-mat}
In above examples of decomposition of $W$-orbits into $W'$-orbits,
the $W'$-system of simple roots was a subsystem of $W$-system of
simple roots. The present case is simple and the decomposition is
easily fulfilled.

There exist much more complicated cases of the embeddings
$W'\subset W$ (and of the embeddings $L'\subset L$ of the
corresponding Lie algebras). In these cases, the collection of the
unit vectors $\omega_1',\omega_2',\ldots,\omega'_m$ (they are 
such that $2\langle \omega_i',\alpha_j'  \rangle/ \langle
\alpha_j',\alpha_j'  \rangle=\delta_{ij}$ for simple roots
$\alpha_1',\alpha_2',\ldots,\alpha_m'$ corresponding to the Weyl
group $W'$) is not a subsystem of the unit vectors
$\omega_1,\omega_2,\ldots,\omega_n$ (for which $2\langle
\omega_i,\alpha_j  \rangle/ \langle \alpha_j,\alpha_j
\rangle=\delta_{ij}$ for simple roots
$\alpha_1,\alpha_2,\ldots,\alpha_n$ corresponding to the Weyl
group $W$). This fact make the decomposition procedure
complicated.

In these cases a projection matrix, constructed for a fixed pair
$W'\subset W$ and projecting weights
$\lambda=(\lambda_1,\lambda_2, \ldots,\lambda_n)$ in the
$\omega$-coordinates to weights in the $\omega'$-coordinates
(corresponding to the Weyl subgroup $W'$), is used. Such a matrix
is of dimension $m\times n$, where $m$ is the number of
$\omega'$-coordinates. A list of such matrices for a large number
of pairs $W'\subset W$ is given in~\cite{MPS77}.

When performing the decomposition of a $W$-orbit $O(\lambda)$ into
$W'$-orbits we have to project $\omega$-coordinates of all weights
of $O(\lambda)$ to $\omega'$-coordinates and then we have to split
the points obtained under this projection into $W'$-orbits.

Let us give an explicit form of the projection matrices for some
simple cases:
 \begin{gather*}
C_2\to A_1: \ (3\ \ 4);\qquad G_2\to A_1:\ (10\ \ 6);
\\
 C_2\to A_1\times A_1:\
    \left(
 \begin{array}{cc}
 1&1\\ 0&1\
 \end{array} \right);  \qquad
 G_2\to A_2:
\   \left(
 \begin{array}{cc}
 1&1\\ 1&0\
 \end{array} \right);
\\
 C_4\to A_3: \
\left(
 \begin{array}{cccc}
 1&1&0&0\\ 0&0&1&2\\ 0&1&1&0
 \end{array} \right); \qquad
 D_5\to C_2\times C_2: \
\left(
 \begin{array}{ccccc}
 0&0&2&1&1\\ 1&1&0&0&0\\ 0&0&0&1&1\\ 0&1&1&0&0\
 \end{array} \right).
 \end{gather*}

\subsection{Generating functions for multiplicities}\label{Gener-func}

Multiplicities of orbits in products of orbits and in the
restriction of $W$-orbits into $W'$-orbits can be derived by means
of generating functions for these multiplicities. A series of such
generating functions are derived in \cite{GPS1} and \cite{GPS2}.
Let us give some examples in this subsection.

First we consider the restriction $B_3\to B_2\times U(1)$, where
it is supposed that an orbit of $U(1)$ is given by one integer or
half-integer. The generating function for this case is
\begin{gather} \label{gen-1}
\frac{1}{(1-a_1)(1-c_2)(1-e_3)} +
\frac{e_3^*}{(1-a_1)(1-c_2)(1-e^*_3)}+
\frac{d_2}{(1-a_1)(1-d_2)(1-e_3)}
\\
\qquad{}+ \frac{d_2^*}{(1-a_1)(1-d^*_2)(1-e^*_3)} +
\frac{b_1}{(1-b_1)(1-d_2)(1-e_3)} +
\frac{b_1^*}{(1-b^*_1)(1-d^*_2)(1-e^*_3)},\nonumber
 \end{gather}
where
 \begin{gather*}
a_1=A_1B_1,\qquad b_1=A_1Z,\qquad  b_1^*=A_1Z^{-1},\qquad
c_2=A_2B_2^2,
\\
d_2=A_2B_1Z,\qquad d_2^*=A_2B_1Z^{-1},\qquad e_3=A_3B_2Z^{1/2},
\qquad e_3^*=A_3B_2 Z^{-1/2}.
 \end{gather*}
The multiplicities in this case are 0 or 1. In order to obtain a
multiplicity of a $B_2\times U(1)$-orbit $O(\mu_1,\mu_2; z)$ in a
$B_3$-orbit $O(\lambda_1,\lambda_2,\lambda_3)$ (the numbers
$\lambda_1$, $\lambda_2$, $\lambda_3$ and $\mu_1$, $\mu_2$ give
$\omega$-coordinates of the corresponding elements) we have to
expand \eqref{gen-1} into a power series. The multiplicity is
given by the term
$A_1^{\lambda_1}A_2^{\lambda_2}A_3^{\lambda_3}B_1^{\mu_1}B_2^{\mu_2}Z^z$
in the expansion: if this term is absent then the multiplicity is
0; if it exists (i.e. a coefficient of it is equal to 1) then the
multiplicity is 1. Note that multiplicities in the restriction
$B_3\to B_2\times U(1)$ trivially determine multiplicities in the
restriction $B_3\to B_2$.

Multiplicities of $A_2$-orbits in $G_2$-orbits are given in the
same way by the generating function
 \[
\frac{1}{1-a_1} \left[ \frac{1}{1-b_2} + \frac{b_2^*}{1-b^*_2}
\right] ,
 \]
where
 \[
a_1=A_1B_1B_2,\qquad b_2=A_2B_1,\qquad b_2^*=A_2B_2
 \]
and the symbols $A_1$, $A_2$ and $B_1$, $B_2$ are related to
$G_2$-orbits and $A_2$-orbits, respectively.

Multiplicities of $B_3\times A_1$-orbits in $F_4$-orbits are given
by the generating function
 \begin{gather*}
\frac{1}{(1-a_1)(1-e_2)(1-h_3)(1-k_4)}
+\frac{i_3}{(1-a_1)(1-e_2)(1-i_3)(1-k_4)}
\\   \qquad\qquad{}
+ \frac{l_4}{(1-a_1)(1-e_2)(1-i_3)(1-l_4)}
+\frac{b_1}{(1-b_1)(1-e_2)(1-h_3)(1-k_4)}
 \\   \qquad\qquad
{}+\frac{d_2}{(1-b_1)(1-d_2)(1-h_3)(1-k_4)} +
\frac{f_2}{(1-b_1)(1-f_2)(1-i_3)(1-l_4)}
 \\     \qquad\qquad
{}+\frac{j_3}{(1-b_1)(1-f_2)(1-j_3)(1-l_4)}
+\frac{g_2}{(1-b_1)(1-g_2)(1-j_3)(1-l_4)}
 \\      \qquad\qquad
{}+ \frac{b_1i_3}{(1-b_1)(1-e_2)(1-i_3)(1-k_4)}
+\frac{b_1l_4}{(1-b_1)(1-e_2)(1-i_3)(1-l_4)}
 \\      \qquad\qquad
{}+\frac{f_2k_4}{(1-b_1)(1-f_2)(1-i_3)(1-k_4)} +
\frac{c_1}{(1-c_1)(1-g_2)(1-j_3)(1-l_4)} ,
 \end{gather*}
where
\begin{gather*}
a_1=A_1B_1^2,\qquad  b_1=A_1B^2_3C, \qquad c_1=A_1C^2,\qquad
d_2=A_2B_3^4,\qquad
 e_2=A_2B^2_1B_2^2C,
\\
 f_2=A_2B_2^2C^2,\qquad g_2=A_2B_3^2C^3,\qquad h_3=A_3B_1B_3^2,\qquad i_3=A_3B_1B_2C,
\\
 j_3=A_3B_2C^2,\qquad
 k_4=A_4B_2, \qquad l_4=A_4B_1C.
 \end{gather*}
and the symbols $A_1$, $A_2$, $A_3$, $A_4$; $B_1$, $B_2$, $B_3$
and $C$ are related to $F_4$-, $B_3$-  and  $A_1$-orbits,
respectively.

\setcounter{equation}{0}

\section{Affine Weyl group and its fundamental domain}

\subsection{Affine Weyl groups}\label{Aff}

We are interested in orbit functions which are given on the
Euclidean space $E_n$. Orbit functions are invariant with respect
to action of a Weyl group $W$, which is a transformation group of
$E_n$. However, $W$ does not describe all symmetries of orbit
functions. A whole group of invariances of orbit functions is
isomorphic to the so called affine Weyl group $W^{\rm aff}$ which
is an extension of the Weyl group $W$. In this subsection we
define affine Weyl groups.

Let $\alpha_1, \alpha_2,\ldots ,\alpha_n$ be simple roots in the
Euclidean space $E_n$ and let $W$ be the corresponding Weyl group.
The group $W$ is generated by the reflections $r_{\alpha_i}$,
$i=1,2,\ldots ,n$, corresponding to simple roots. We consider also
the reflection $r_{\xi}$ with respect to the $(n-1)$-dimensional
subspace (hyperplane) $X_{n-1}$ containing the origin and
orthogonal to the highest (long) root $\xi$, given
in~\eqref{highestroot}:
\begin{gather} \label{refl-s}
r_{\xi}x=x-\frac{2\langle x,\xi\rangle}{\langle \xi,\xi\rangle}
\xi .
\end{gather}
We shift the hyperplane $X_{n-1}$ by the vector\footnote{Note that
under our conditions for lengths of roots in
Subsection~\ref{roots} we have $\xi^\vee =\xi$.} $\xi^\vee/2$,
where $\xi^\vee =2\xi/\langle\xi,\xi \rangle$. The reflection with
respect to the hyperplane $X_{n-1}+\xi^\vee/2$ will be denoted by
$r_0$. Then in order to fulfill the transformation $r_0$ we have
to fulfill the transformation $r_\xi$ and then to shift the result
by $\xi^\vee$, that is,
\[
r_0x=r_\xi x+\xi^\vee .
 \]
We have $r_00=\xi^\vee$ and it follows from \eqref{refl-s} that
$r_0$ maps $x+\xi^\vee/2$ to
 \[
r_\xi(x+ \xi^\vee/2)+\xi^\vee=x+\xi^\vee/2-\langle x,\xi^\vee
\rangle \xi.
 \]
Therefore,
 \begin{gather*}
r_0(x+\xi^\vee/2)=x+\xi^\vee/2 - \frac{2\langle
x,\xi\rangle}{\langle \xi,\xi\rangle} \xi =x+\xi^\vee/2
-\frac{2\langle x,\xi^\vee\rangle}{\langle
\xi^\vee,\xi^\vee\rangle} \xi^\vee
 \\
 \phantom{r_0(x+\xi^\vee/2)}{}
=x+\xi^\vee/2-\frac{2\langle x+\xi^\vee/2,\xi^\vee\rangle}{\langle
\xi^\vee,\xi^\vee\rangle} \xi^\vee +\frac{2\langle
\xi^\vee/2,\xi^\vee\rangle}{\langle \xi^\vee,\xi^\vee\rangle}
\xi^\vee .
 \end{gather*}
Denoting $x+\xi^\vee/2$ by $y$ we obtain that $r_0$ is given also
by the formula
 \begin{gather} \label{refl-0}
r_0y=y+\left( 1- \frac{2\langle y,\xi^\vee\rangle}{\langle
\xi^\vee,\xi^\vee\rangle}\right) \xi^\vee=\xi^\vee +r_\xi y .
 \end{gather}

The hyperplane $X_{n-1}+\xi^\vee/2$ coincides with the set of
points $y$ such that $r_0y=y$. It follows from \eqref{refl-0} that
this hyperplane is given by the equation
 \begin{gather} \label{hyperpl}
1=\frac{2\langle y,\xi^\vee\rangle}{\langle
\xi^\vee,\xi^\vee\rangle}  =\langle y,\xi\rangle =\sum _{k=1}^n
a_kq_k,
 \end{gather}
where
 \[
y=\sum _{k=1}^n a_k\omega_k,\qquad \xi=\sum _{k=1}^n q_k
\alpha^\vee_k
 \]
(see \eqref{kronecker}). In Subsection \ref{roots}, the numbers
$q_i$ are given in an explicit form.

A group of transformations of the Euclidean space $E_n$ generated
by the reflections $r_0,r_{\alpha_1},\ldots,$ $r_{\alpha_n}$ is
called the {\it affine Weyl group} of the root system $R$ and is
denoted by $W^{\rm aff}$ or by $W^{\rm aff}_R$ (if is necessary to
indicate the initial root system).

\subsection{Properties of affine Weyl groups}\label{Aff-pr}

Adjoining the reflection $r_0$ to the Weyl group $W$ completely
changes the properties of the group~$W^{\rm aff}$.

If $r_{\xi}$ is the reflection with respect to the hyperplane
$X_{n-1}$, then due to \eqref{refl-0} for any $x\in E_n$ we have
 \[
r_0r_{\xi}x=r_0(r_{\xi}x)=\xi^\vee+r_\xi r_\xi x =x+\xi^\vee .
 \]
Clearly, $(r_0r_{\xi})^kx=x+k\xi^\vee$, $k= 0,\pm 1,\pm 2,\ldots
$, that is, the set of elements $(r_0r_{\xi})^k$, $k= 0,\pm 1,\pm
2,\ldots $, is an infinite commutative subgroup of $W^{\rm aff}$.
This means that (unlike to the Weyl group $W$) $W^{\rm aff}$ {\it
is an infinite group}.

Recall that $r_00=\xi^\vee$. For any $w\in W$ we have
 \[
wr_00=w\xi^\vee=\xi^\vee_w,
 \]
where $\xi^\vee_w$ is a coroot of the same length as the coroot
$\xi^\vee$. For this reason, $wr_0$ is the reflection with respect
to the $(n-1)$-hyperplane perpendicular to the root $\xi^\vee_w$
and containing the point $\xi^\vee_w/2$. Moreover,
 \begin{gather} \label{cor-1}
(wr_0)r_{\xi^\vee_w} x=x+\xi^\vee_w .
 \end{gather}
We also have $((wr_0)r_{\xi^\vee_w})^kx=x+k\xi^\vee_w$, $k=0,\pm
1,\pm 2,\ldots$. Since $w$ is any element of $W$, then the set
$w\xi^\vee$, $w\in W$, coincides with the set of coroots of
$R^\vee$, corresponding to all long roots of root system $R$.
Thus, the following proposition is true:

\begin{proposition}\label{proposition5}  The set $W^{\rm aff}\cdot 0$ coincides
with the lattice $Q^\vee_l$ generated by coroots $\alpha^\vee$
taken for all long roots $\alpha$ from $R$.
\end{proposition}

It is easy to see from explicit forms of the root systems $Q$ and
$Q^\vee$ that each coroot $\xi_s^\vee$ for a~short root $\xi_s$ of
$R$ is a linear combination of coroots $w\xi^\vee\equiv \xi_w$,
$w\in W$, with integral coefficients, that is, $Q^\vee =Q_l^\vee$.
Therefore, from Proposition \ref{proposition5} we obtain the
following corollary:

\medskip

\noindent {\bf Corollary.} {\it The set $W^{\rm aff}\cdot 0$
coincides with the coroot lattice $Q^\vee$ of $R$.}

\medskip

Let $\hat Q^\vee$ be the subgroup of $W^{\rm aff}$ generated by
the elements
 \begin{gather} \label{ref-ow}
 (wr_0)r_w, \qquad w\in W,
 \end{gather}
where $r_w\equiv r_{\xi^\vee_w}$ for $w\in W$ (see \eqref{cor-1}).
Since elements \eqref{ref-ow} pairwise commute with each other
(since they are shifts), $\hat Q^\vee$ is a commutative group. The
subgroup $\hat Q^\vee$ can be identified with the coroot lattice
$Q^\vee$. Namely, if for $g\in \hat Q^\vee$ we have $g\cdot
0=\gamma\in Q^\vee$ (that is, $g$ is a shift by $\gamma$), then
$g$ is identified with $\gamma$. This correspondence is
one-to-one.

The subgroups $W$ and $\hat Q^\vee$ generate $W^{\rm aff}$ since a
subgroup of $W^{\rm aff}$, generated by $W$ and $\hat Q^\vee$,
contains the element $r_0$. From the other side, $W\cup \hat
Q^\vee =\{ 1\}$, since $W$ does not contain shifts. Moreover,
$\hat Q^\vee$ is an invariant subgroup of $W^{\rm aff}$ since for
any element $(wr_0)r_w$ from \eqref{ref-ow} and for any element
$w'\in W$ we have
 \[
w'(wr_0r_w){w'}^{-1}x=w'(wr_0r_w)({w'}^{-1}x)=
w'({w'}^{-1}x+\xi^\vee_w)=x+w'\xi^\vee_w,
 \]
where $w'\xi^\vee_w\in Q^\vee$, that is, $w'(wr_0r_w){w'}^{-1}$ is
an element $g'$ of $\hat Q^\vee$ such that $g'x=x+w'\xi^\vee_w$.
Thus, we have proved the following proposition:

\begin{proposition}\label{proposition6}  The group $W^{\rm aff}$ is a semidirect
product of its subgroups $W$ and $\hat Q^\vee$, where $\hat
Q^\vee$ is an invariant subgroup.
\end{proposition}

It follows from this proposition that each element of $W^{\rm
aff}$ can be represented as a product~$wg$, $w\in W$, $g\in \hat
Q^\vee$, or as a product $gw$, $w\in W$, $g\in \hat Q^\vee$.

\subsection{Fundamental domain}\label{Fund}

An open connected simply connected set $D\subset E_n$ is called a
{\it fundamental domain} for the group $W^{\rm aff}$ (for the
group $W$) if it does not contain equivalent points (that is,
points $x$ and $x'$ such that $x'=wx$) and if its closure contains
at least one point from each $W^{\rm aff}$-orbit (from each
$W$-orbit). It is evident that the dominant Weyl chamber (without
the walls of this chamber) is a fundamental domain for the Weyl
group $W$. Recall that this domain consists of all points
$x=a_1\omega_1+\cdots +a_n\omega_n\in E_n$ for which
 \[
a_i=\langle x,\alpha^\vee_i \rangle > 0,\qquad i=1,2,\ldots ,n.
 \]
Let us describe the fundamental domain of the group $W^{\rm aff}$.
Since $W\subset W^{\rm aff}$, it can be chosen as a subset of the
dominant Weyl chamber of $W$.

We have seen that the element $r_0\in W^{\rm aff}$ is a reflection
with respect to the hyperplane $X_{n-1}+\xi^\vee/2$, orthogonal to
the root $\xi$ and containing the point $\xi^\vee/2$. This
hyperplane is given by the equation \eqref{hyperpl}. This equation
shows that the hyperplane $X_{n-1}+\xi^\vee/2$ intersects the
axes, determined by the orbits $\omega_i$, in the points
$\omega_i/q_i$, $i=1,2,\ldots ,n$, where $q_i$ are such as in
\eqref{hyperpl}. We create the simplex with $n+1$ vertices in the
points
 \begin{gather}\label{sympl}
0,\ \frac{\omega_1}{q_1}, \ \ldots , \ \frac{\omega_n}{q_n} .
 \end{gather}
By the definition, this simplex consists of all points $y$ of the
dominant Weyl chamber for which $\langle y,\xi \rangle \le 1$.
Clearly, the interior $F$ of this simplex belongs to the dominant
Weyl chamber.

\begin{theorem}\label{theorem1} The set $F$ is a fundamental domain for the
affine Weyl group $W^{\rm aff}$.
\end{theorem}

\begin{proof} The set $\overline F$ coincides with the set of
points $y$ of the dominant Weyl chamber, for which $\langle y,
\xi\rangle \le 1$. Then
 \[
\langle y, \alpha\rangle \le 1, \qquad \alpha\in R_l,
 \]
where $R_l$ is a subset of the root set $R$ consisting of all long
roots. Indeed, for any $\alpha\in R_l$ we have
$\alpha=\xi-\sum\limits_i a_i\alpha_i$, where $\alpha_i$ are
simple roots and $a_i$ are non-negative integers. Then $\langle
y,\alpha\rangle =\langle y,\xi\rangle -\sum\limits_i a_i \langle
y,\alpha_i\rangle\le 1$, since $\langle y,\alpha_i\rangle\ge 0$
for all $y\in \overline F$. Acting upon $\overline F$ by elements
of the Weyl group $W$ we obtain the domain ${\mathcal D}\subset
E_n$ determined by the equations
 \[
\langle y, \alpha\rangle \le 1, \qquad \alpha\in R_l.
 \]
The hyperplanes $\langle y,\alpha\rangle=1$, $\alpha\in R_l$,
determine walls of the domain ${\mathcal D}$. Since $\langle
\alpha^\vee,\alpha \rangle=2$, for each fixed root $\alpha\in R_l$
a distance between the opposite hyperplanes $\langle
y,\alpha\rangle=1$ and $\langle y,-\alpha\rangle=1$ is
$\alpha^\vee$. Thus, nontrivial elements of $W$ do not map
elements of $F$ inside of $F$ and nontrivial elements of $\hat
Q^\vee$ do not map elements of ${\mathcal D}$ inside of ${\mathcal
D}$. From the other side, successive shifts of~${\mathcal D}$ by
coroots cover the whole space $E_n$.

Thus, we have $\cup _{w\in W^{\rm aff}} w\overline F=E_n$.
Clearly, we cannot remove from $F$ any point $x$ since then $F$ is
not simply connected. Therefore, $F$ is a fundamental domain for
the group $W^{\rm aff}$. The theorem is proved.
\end{proof}

  For the rank 2 cases the fundamental domain is a simplex with
the following vertices:
 \begin{alignat}{2} A_2\
&:&\quad& \{ 0,\ \omega_1,\ \omega_2\} ,
\notag\\
C_2\ &:&\quad& \{ 0,\ \omega_1,\ \omega_2\} , \notag
\\
G_2\ &:&\quad&\{ 0,\ \tfrac{\omega_1}2,\ \omega_2\} .\notag
\end{alignat}

\setcounter{equation}{0}

\section{Orbit functions}\label{Orb}

\subsection{Definition}\label{Orb-1}

The exponential functions $e^{2\pi{\rm i}\langle m,x\rangle}$,
$x\in E_n$, with fixed $m=(m_1,m_2,\ldots ,m_n)$, $m_i\in {\mathbb
R}$, determine the Fourier transform on $E_n$. Orbit functions are
a symmetrized (with respect to a Weyl group) version of
exponential functions. Correspondingly, they determine a
symmetrized version of the Fourier transform.

Orbit functions are defined as follows. Let $W$ be a Weyl group of
transformations of the Euclidean space $E_n$. To each element
$\lambda\in E_n$ from the dominant Weyl chamber (including its
walls) there corresponds an {\it orbit function} $\phi_\lambda$ on
$E_n$, which is given by the formula
 \begin{gather}\label{orb}
\phi_\lambda(x)=\sum_{\mu\in O(\lambda)} e^{2\pi{\rm i}\langle
\mu,x\rangle}, \qquad x\in E_n,
 \end{gather}
where $O(\lambda)$ is the $W$-orbit of the element $\lambda$. It
is also called a C-function (since for the case $A_1$ it coincides
with the cosine). The number of summands in \eqref{orb} is equal
to the size $|O(\lambda)|$ of the orbit $O(\lambda)$ which
coincides with the number $|W|/|W_\lambda|$. Clearly,
$\phi_\lambda (0)=|W|/|W_\lambda|$.

Sometimes (see, for example, \cite{Pat-Z-1} and \cite{Pat-Z-2}),
it is convenient to use a modified definition of orbit functions:
\begin{gather}\label{orb-2}
\hat \phi_\lambda(x)=|W_\lambda| \phi_\lambda (x).
\end{gather}
Then for all orbit functions $\hat\phi_\lambda$ we have $\hat
\phi_\lambda (0)=|W|$.

We are mainly interested in orbit functions $\phi_\lambda$ for
which $\lambda\in P_+$. Namely, such orbit functions determine a
symmetrized Fourier series expansion which will be studied in the
next section.

\medskip

\noindent {\bf Example:} {\it  Orbit functions for $A_1$.} In this
case, there exists only one simple (positive) root~$\alpha$. We
have $\langle \alpha,\alpha \rangle =2$. Therefore, the relation
$2\langle \omega,\alpha \rangle / \langle \alpha,\alpha \rangle
=1$ means that $\langle \omega,\alpha \rangle =1$. This means that
$\omega=\alpha/2$ and $\langle \omega,\omega \rangle =1/2$.
Elements of $P_+$ coincide with $m\omega$, $m\in {\mathbb Z}_+$.
We identify points $x$ of $E_1\equiv {\mathbb R}$ with $\theta
\omega$. Since the Weyl group $W(A_1)$ consists of two elements 1
and $r_\alpha$, and
 \[
r_\alpha x=x-\frac{2\langle \theta\omega,\alpha \rangle}{\langle
\alpha,\alpha \rangle}\alpha=x-\theta\alpha =x-2x=-x ,
 \]
orbit functions $\phi_\lambda(x)$, $\lambda=m\omega$, are given by
the formula
 \[
\phi_\lambda(x)=e^{2\pi{\rm i}\langle m\omega,\theta\omega
\rangle}+e^{2\pi{\rm i}\langle m\omega,-\theta\omega \rangle}
=e^{\pi{\rm i}m\theta}+e^{-\pi{\rm i}m\theta}=2\cos (\pi m\theta),
 \]
where $m\ne 0$. If $m=0$, then $\phi_\lambda(x)\equiv 1$.

\subsection[Orbit functions of $A_2$]{Orbit functions of $\boldsymbol{A_2}$}\label{Orb-2}

Let $\lambda=a\omega_1+b\omega_2\equiv (a\ b),$ with $a>b>0$. Then
taking into account the results of Subsection~\ref{orbits}, for
$\phi_\lambda(x)\equiv \phi_{(a\ b)}(x)$ we have that
\begin{gather*}
 \phi_{(a\ b)}(x)
   =    e^{2\pi i\l(a\ b), x\r}
       + e^{2\pi i\l(-a\ a+b), x\r}
       + e^{2\pi i\l(a+b\ -b), x\r} \\
 \phantom{\phi_{(a\ b)}(x) =}{}  + e^{2\pi i\l(b\ -a-b), x\r}
       + e^{2\pi i\l(-a-b\ a), x\r}
       + e^{2\pi i\l(-b\ -a), x\r} .
\end{gather*}
Using the representation $x=\varphi_1\alpha^\vee_1+
\varphi_2\alpha^\vee_2$, one obtains
\begin{gather}
  \phi_{(a\ b)}(x)=   e^{2\pi i(a\varphi_1+b\varphi_2)}
      + e^{2\pi i(-a\varphi_1+(a+b)\varphi_2)}
      + e^{2\pi i((a+b)\varphi_1-b\varphi_2)} \nonumber\\
\phantom{\phi_{(a\ b)}(x)=}{} + e^{2\pi
i(b\varphi_1-(a+b)\varphi_2)}
      + e^{2\pi i((-a-b)\varphi_1+a\varphi_2)}
      + e^{2\pi i(-b\varphi_1-a\varphi_2)}.
\end{gather}
The actual expression for $ \phi_{(a,\ b)}(x)$ depends on the
choice of coordinate systems for $\lambda$ and $x$. Setting
$x=\theta_1\omega_1+\theta_2\omega_2$ and $\lambda$ as before,
after using formula \eqref{matr} we get
\begin{gather}
  \phi_{(a\ b)}(x)=   e^{\tfrac{2\pi i}3((2a+b)\theta_1+(a+2b)\theta_2)}
      + e^{\tfrac{2\pi i}3((-a+b)\theta_1+(a+2b)\theta_2)}\nonumber\\
\phantom{\phi_{(a\ b)}(x)=}{} + e^{\tfrac{2\pi
i}3((2a+b)\theta_1+(a- b)\theta_2)}
      + e^{-\tfrac{2\pi i}3((a-b)\theta_1+(2a+b)\theta_2)}\nonumber\\
 \phantom{\phi_{(a\ b)}(x)=}{} + e^{-\tfrac{2\pi i}3((a+2b)\theta_1+(-a+b)\theta_2)}
      + e^{-\tfrac{2\pi i}3((a+2b)\theta_1+(2a+b)\theta_2)}.
\end{gather}
Note that the orbit function $\phi_{(a\ a)}(x)$ is real valued for
all $a\in\Z^{\geq0}$ and
\begin{gather}
 \phi_{(a\ a)}(x) = 2\left\{\cos2\pi a( \varphi_1+ \varphi_2)
                +\cos2\pi a(2\varphi_2-\varphi_1)
                +\cos2\pi a(2\varphi_1- \varphi_2) \right\}\nonumber\\
\phantom{\phi_{(a\ a)}(x)}{} = 2\left\{\cos2\pi
a(\theta_1+\theta_2)
                +\cos2\pi a\theta_1+\cos2\pi a\theta_2 \right\}.
\end{gather}

Similarly one finds $ \phi_{(a\ 0)}(x)$ and $ \phi_{(0\ b)}(x)$:
\begin{gather}
 \phi_{(a\ 0)}(x) =e^{\tfrac{2\pi i}3a(2\theta_1+ \theta_2)}
              +e^{\tfrac{2\pi i}3a(-\theta_1+ \theta_2)}
              +e^{\tfrac{2\pi i}3a(-\theta_1-2\theta_2)},\\
 \phi_{(0\ b)}(x) =e^{\tfrac{2\pi i}3b( \theta_1+2\theta_2)}
              +e^{\tfrac{2\pi i}3b( \theta_1- \theta_2)}
              +e^{\tfrac{2\pi i}3b(-2\theta_1-\theta_2)}.
\end{gather}
Note that the pairs $ \phi_{(a\ b)}(x)+ \phi_{(b\ a)}(x)$ are
always real functions.

\subsection[Orbit functions of $C_2$ and $G_2$]{Orbit functions of $\boldsymbol{C_2}$ and $\boldsymbol{G_2}$}

Let $\lambda=a\omega_1+b\omega_2=(a\; b)$ and use the matrices $S$
from \eqref{matr} which are of the form
 \[
S(C_2)=\frac12\begin{pmatrix}1&1\\1&2\end{pmatrix},\qquad
S(G_2)=\frac13\begin{pmatrix}6&3\\3&2\end{pmatrix}.
 \]
Then if $x= (\theta_1\omega_1+\theta_2\omega_2)$ for $C_2$ and
$x=\frac12(\theta_1\omega_1+\theta_2\omega_2)$ for $G_2$ we find
that the orbit functions for $C_2$ and $G_2$ are of the form
\begin{gather}
C_2 :\quad \phi_{(a\ b)}(x)
  =2\cos\pi ((a+b)\theta_1+(a+2b)\theta_2)
   +2\cos\pi (b\theta_1+(a+2b)\theta_2)\nonumber \\
\phantom{C_2 :\quad \phi_{(a\ b)}(x)  =}{}+2\cos\pi
((a+b)\theta_1+a\theta_2)
   +2\cos\pi(b\theta_1-a\theta_2),\nonumber \\
\phantom{C_2 :\quad{}}{}  \phi_{(a\ 0)}(x)
  =2\cos\pi a(\theta_1+\theta_2) +2\cos\pi a\theta_2,\nonumber\\
\phantom{C_2 :\quad{}}{}\phi_{(0\ b)}(x)
 =2\cos\pi b(\theta_1+2\theta_2)
     +2\cos\pi b\theta_1,  \\
G_2 :\quad \phi_{(a\ b)}(x)
  =2\cos\pi (2a+b)\theta_1+(a+\tfrac23b)\theta_2)+2\cos\pi ((a+b)\theta_1+(a+\tfrac23b)\theta_2)\nonumber\\
\phantom{G_2 :\quad \phi_{(a\ b)}(x)
=}{}+2\cos\pi((2a+b)\theta_1+(a+\tfrac13b)\theta_2)
+2\cos\pi((a+b)\theta_1+\tfrac13 b\theta_2)\nonumber\\
\phantom{G_2 :\quad \phi_{(a\ b)}(x)  =}{}
+2\cos\pi(a\theta_1+(a+\tfrac13b)\theta_2)
               +2\cos\pi(a\theta_1-\tfrac13b\theta_2),\nonumber\\
\phantom{G_2 :\quad{}}{} \phi_{(a\ 0)}(x) =2\cos\pi
a(2\theta_1+\theta_2)+2\cos\pi a(\theta_1+\theta_2)
       +2\cos\pi a\theta_1,\nonumber\\
\phantom{G_2 :\quad{}}{} \phi_{(0\ b)}(x)
  =2\cos\pi b(\theta_1+\tfrac23\theta_2)
            +2\cos\pi b(\theta_1+\tfrac13\theta_2)
            +2\cos\pi \tfrac13b\theta_2.
\end{gather}
As we see, orbit functions for $C_2$ and $G_2$ are real.

\subsection[Orbit functions of $A_n$]{Orbit functions of $\boldsymbol{A_n}$}

It is difficult to write down an explicit form for the orbit
functions of $A_n$, $B_n$, $C_n$ and $D_n$ in coordinates with
respect to the $\omega$- and $\alpha$-bases. Thus, in these
particular cases we use the orthogonal coordinate systems
described in Section~\ref{weylABCD}.

Let $\lambda=(m_1,m_2,\ldots ,m_{n+1})$ be an element of
$P_+(A_n)$ represented in the orthogonal coordinates described in
Subsection~\ref{An}. Then $m_1\ge m_2\ge \cdots \ge m_{n+1}$. The
Weyl group in this case coincides with the symmetric group
$S_{n+1}$. We denote by $S_\lambda\equiv W_\lambda$ the subgroup
of $W(A_n)\equiv S_{n+1}$ consisting of elements $w\in S_{n+1}$
such that $w\lambda=\lambda$. Then the orbit $O(\lambda)$ consists
of points $w\lambda$, $w\in W/W_\lambda\equiv S_{n+1}/S_\lambda$.
If we represent the points $x\in E_{n+1}$ in the orthogonal
coordinate system as well, $x=(x_1,x_2,\ldots ,x_{n+1})$, and if
we use the formula \eqref{orb} we find that
\begin{gather}
 \phi_\lambda(x)  =
 \sum _{w\in S_{n+1}/S_\lambda} e^{2\pi{\rm i}\langle w(m_1,
 \ldots ,m_{n+1}),(x_1,\ldots,x_{n+1})\rangle} \nonumber \\
  \phantom{\phi_\lambda(x)}{} =
 \sum _{w\in S_{n+1}/S_\lambda} e^{2\pi{\rm i}((w\lambda)_1x_1+
 \cdots +(w\lambda)_{n+1}x_{n+1})} ,\label{orb-f}
\end{gather}
where $((w\lambda)_{1},(w\lambda)_{2},\ldots ,(w\lambda)_{n+1})$
are orthogonal coordinates of $w\lambda$ if
$\lambda=(m_1,m_2,\ldots ,m_{n+1})$.

Note that $-(m_{n+1},m_n,\ldots ,m_1)\in P_+$ if $(m_1,m_2,\ldots
m_{n+1})\in P_+$. In the Weyl group $W(A_n)$ there exists an
element $w_0$ such that
 \[
w_0(m_{1},m_2,\ldots ,m_{n+1})= (m_{n+1},m_n,\ldots ,m_1).
 \]
It follows from here that in the expressions for
$\phi_{(m_{1},m_2,\ldots ,m_{n+1})}(x)$ and
$\phi_{-(m_{n+1},m_n,\ldots ,m_{1})}(x)$ there are summands
 \[
e^{2\pi{\rm i}\langle w_0\lambda ,x\rangle}= e^{2\pi{\rm
i}(m_{n+1}x_1+
 \cdots +m_{1}x_{n+1})}\qquad
{\rm and}\qquad e^{-2\pi{\rm i}(m_{n+1}x_1+
 \cdots +m_{1}x_{n+1})} ,
 \]
respectively, which are complex conjugate to each other.
Similarly, in the expressions \eqref{orb-f} for
$\phi_{(m_{1},m_2,\ldots ,m_{n+1})}(x)$ and for
$\phi_{-(m_{n+1},m_n,\ldots ,m_{1})}(x)$ all other summands are
pairwise complex conjugate. Therefore,
\begin{gather}\label{compl}
\phi_{(m_{1},m_2,\ldots
,m_{n+1})}(x)=\overline{\phi_{-(m_{n+1},m_n,\ldots ,m_{1})}(x)}.
\end{gather}
If we use for $\lambda$ the coordinates $\lambda_i=\langle
\lambda,\alpha^\vee_i\rangle$ in the $\omega$-basis instead of the
orthogonal coordinates~$m_j$, then this equation can be written as
 \[
\phi_{(\lambda_{1},\ldots
,\lambda_{n})}(x)=\overline{\phi_{(\lambda_{n},\ldots
,\lambda_{1})}(x)}.
 \]
According to \eqref{compl}, if
\begin{gather}
(m_1,m_2,\ldots ,m_{n+1})=-(m_{n+1},m_n,\ldots ,m_{1})
\end{gather}
(that is, the element $\lambda$ has in the $\omega$-basis the
coordinates $(\lambda_1,\lambda_2,\ldots ,\lambda_2,\lambda_1)$,
then {\it the orbit function $\phi_\lambda$ is real.} Here, the
orbit functions can be represented as sums of cosines of the
corresponding angles (as in Subsection \ref{Orb-2}).

\subsection[Orbit functions of $B_n$]{Orbit functions of $\boldsymbol{B_n}$}

Let $\lambda=(m_1,m_2,\ldots ,m_{n})$ be an element of $P_+(B_n)$
in the orthogonal coordinates described in Subsection~\ref{Bn}.
Then $m_1\ge m_2\ge \cdots \ge m_{n}\ge 0$. The Weyl group
$W(B_n)$ consists of permutations of the orthogonal coordinates
with sign alternations of some of them. We denote by $S_\lambda$
the subgroup of the permutation group $S_n$ consisting of elements
$w$ such that $w\lambda=\lambda$. If we represent the points $x\in
E_{n}$ in the orthogonal coordinate system, $x=(x_1,x_2,\ldots
,x_{n})$, and if we use formula \eqref{orb} we find that
\begin{gather}
 \phi_\lambda(x)  =
\sum _{\varepsilon_i=\pm 1}\sum _{w\in S_{n}/S_\lambda}
e^{2\pi{\rm i}\langle w(\varepsilon_1m_1, \ldots
,\varepsilon_nm_{n}),
(x_1,\ldots,x_{n})\rangle} \nonumber \\
\phantom{\phi_\lambda(x)}{} =\sum _{\varepsilon_i=\pm 1}\sum
_{w\in S_{n}/S_\lambda}
 e^{2\pi{\rm i}((w(\varepsilon \lambda))_1x_1+
 \cdots +(w(\varepsilon\lambda))_{n}x_{n})} ,\label{orb-B}
\end{gather}
where $((w(\varepsilon \lambda))_1,
 \ldots ,(w(\varepsilon\lambda))_{n})$ are the coordinates of
$w(\varepsilon \lambda)$ if $\varepsilon
\lambda=(\varepsilon_1m_1, \ldots ,\varepsilon_nm_{n})$. In
\eqref{orb-B} the summation is over those $\varepsilon_i=\pm 1$
for which $m_i\ne -m_i$.

Since in $W(B_n)$ there exists an element that changes signs of
all orthogonal coordinates, for each summand $e^{2\pi{\rm
i}((w(\varepsilon \lambda))_1x_1+  \cdots +(w(\varepsilon
\lambda))_{n}x_{n})}$ in the expressions \eqref{orb-B} for the
orbit function $\phi_{(m_{1},m_2,\ldots ,m_{n})}(x)$ there exists
exactly one summand that is the complex conjugate of it, that is,
the summand $e^{-2\pi{\rm i}(((w(\varepsilon \lambda))_1x_1+
\cdots +(w(\varepsilon \lambda))_{n}x_{n})}$. This means that {\it
all orbit functions of $B_n$ are real.} Each orbit function of
$B_n$ can be represented as a sum of cosines of the corresponding
angles.

\subsection[Orbit functions of $C_n$]{Orbit functions of $\boldsymbol{C_n}$}

Let $\lambda=(m_1,m_2,\ldots ,m_{n})$ be an element of $P_+(C_n)$
in the orthogonal coordinates described in Subsection \ref{Cn}.
Then $m_1\ge m_2\ge \cdots \ge m_{n}\ge 0$. The Weyl group
$W(C_n)$ consists of permutations of the coordinates with sign
alternations of some of them. We denote by $S_\lambda$ the
subgroup of $S_n$ consisting of elements $w$ such that
$w\lambda=\lambda$. If we represent points $x\in E_{n}$ in the
orthogonal coordinate system, $x=(x_1,x_2,\ldots ,x_{n})$, we find
that
\begin{gather}
 \phi_\lambda(x)  =
\sum _{\varepsilon_i=\pm 1}\sum _{w\in S_{n}/S_\lambda}
e^{2\pi{\rm i}\langle w(\varepsilon_1m_1, \ldots
,\varepsilon_nm_{n}),
(x_1,\ldots,x_{n})\rangle} \nonumber\\
 \phantom{\phi_\lambda(x)}{} =\sum _{\varepsilon_i=\pm 1}\sum _{w\in S_{n}/S_\lambda}
 e^{2\pi{\rm i}((w(\varepsilon \lambda))_1x_1+
 \cdots +(w(\varepsilon\lambda))_{n}x_{n})} ,\label{orb-C}
\end{gather}
where, as above, $((w(\varepsilon \lambda))_1,
 \ldots ,(w(\varepsilon\lambda))_{n})$ are the coordinates of the
 points $w(\varepsilon \lambda)$ if
$\varepsilon \lambda=(\varepsilon_1m_1, \ldots
,\varepsilon_nm_{n})$. In \eqref{orb-C} the summation is over
those $\varepsilon_i=\pm 1$ for which $m_i\ne -m_i$.

In the expressions \eqref{orb-C} for $\phi_{(m_{1},m_2,\ldots
,m_{n})}(x)$, for each summand $e^{2\pi{\rm i}((w(\varepsilon
\lambda))_1x_1+  \cdots +(w(\varepsilon \lambda))_{n}x_{n})}$
there exists exactly one summand complex conjugate to it, that is,
$e^{-2\pi{\rm i}((w(\varepsilon \lambda))_1x_1+  \cdots
+(w(\varepsilon \lambda))_{n}x_{n})}$. Therefore, {\it all orbit
functions of $C_n$ are real.} Each orbit function of $C_n$ can be
represented as a~sum of cosines of the corresponding angles.

\subsection[Orbit functions of $D_n$]{Orbit functions of $\boldsymbol{D_n}$}

Let $\lambda=(m_1,m_2,\ldots ,m_{n})$ be an element of $P_+(D_n)$
in the orthogonal coordinates described in Subsection \ref{Dn}.
Then $m_1\ge m_2\ge \cdots \ge m_{n-1}\ge |m_n|$. The Weyl group
$W(D_n)$ consists of permutations of the coordinates with sign
alternations for an even number of them. Let $S_\lambda$ be the
subgroup of the permutation group $S_n$ consisting of elements $w$
such that $w\lambda=\lambda$. Representing points $x\in E_{n}$
also in the orthogonal coordinate system, $x=(x_1,x_2,\ldots
,x_{n})$, and using formula \eqref{orb} we find that
\begin{gather}
 \phi_\lambda(x) =
{\sum_{\varepsilon_i=\pm 1}}'\; \sum _{w\in S_{n}/S_\lambda}
e^{2\pi{\rm i}\langle w(\varepsilon_1m_1, \ldots
,\varepsilon_nm_{n}),
(x_1,\ldots,x_{n})\rangle} \nonumber \\
 \phantom{\phi_\lambda(x)}{} ={\sum_{\varepsilon_i =\pm 1}}'\; \sum _{w\in S_{n}/S_\lambda}
 e^{2\pi{\rm i}((w(\varepsilon\lambda))_1x_1+
 \cdots +(w(\varepsilon\lambda))_nx_{n})} ,\label{orb-D}
\end{gather}
where $((w(\varepsilon \lambda))_1, \ldots
,(w(\varepsilon\lambda))_{n})$ are the coordinates of the points
$w(\varepsilon \lambda)$ and the prime at the sum sign means that
the summation is over values of $\varepsilon_i$ with an even
number of minus signs if $m_n\ne 0$ and, if $m_n=0$, over all the
$\varepsilon_i=\pm 1$ for which $m_i\ne -m_i$.

Note that in the expressions \eqref{orb-D} for the orbit function
$\phi_{(m_{1},m_2,\ldots ,m_{n})}(x)$ of $D_{2k}$ for each summand
$ e^{2\pi{\rm i}((w(\varepsilon\lambda))_1x_1+
 \cdots +(w(\varepsilon\lambda))_nx_{n})}$
there exists exactly one summand complex conjugate to it. This
means that {\it all orbit functions of $D_{2k}$ are real.} Each
orbit function of $D_{2k}$ can be represented as a sum of cosines
of the corresponding angles.

It also follows from \eqref{orb-D} that {\it for $D_{2k+1}$ an
orbit function $\phi_{(m_{1},m_2,\ldots ,m_{n})}(x)$ is real if
and only if the condition $m_{2k+1}=0$ is fulfilled. The orbit
functions $\phi_{(m_{1},\ldots, m_{2k},m_{2k+1})}(x)$ and
$\phi_{(m_{1},\ldots ,m_{2k},-m_{2k+1})}(x)$ of $D_{2k+1}$ are
complex conjugates.} If $m_{2k+1}=0$, then the correspon\-ding
orbit function $\phi_{(m_{1},m_2,\ldots ,m_{n})}(x)$ of $D_{2k+1}$
can be represented as a sum of cosines of the appropriate angles.

\setcounter{equation}{0}

\section{Properties of orbit functions}

\subsection{Invariance with respect to affine Weyl group}

Since the scalar product $\langle\cdot ,\cdot \rangle$ in $E_n$ is
invariant with respect to the Weyl group $W$, that is,
 \[
\langle wx,wy\rangle =\langle x,y\rangle, \qquad
 w\in W,\qquad x,y\in E_n,
 \]
orbit functions $\phi_\lambda$ are invariant with respect to $W$:
 \[
\phi_\lambda (wx)=\phi_\lambda (x), \qquad w\in W.
 \]
Indeed,
 \[
\phi_\lambda (wx)=\sum_{\mu\in O(\lambda)} e^{2\pi{\rm i}\langle
\mu,wx\rangle} = \sum_{\mu\in O(\lambda)} e^{2\pi{\rm i}\langle
w^{-1}\mu,x\rangle} =\sum_{\mu\in O(\lambda)} e^{2\pi{\rm
i}\langle \mu,x\rangle} = \phi_\lambda (x)
 \]
since $wO(\lambda)=O(\lambda)$ for each $w\in W$.

Let us show that $\phi_\lambda$ with $\lambda\in P$ are invariant
with respect to the affine Weyl group $W^{\rm aff}$. Since $W^{\rm
aff}$ is generated by $r_0, r_{\alpha_1},\ldots ,r_{\alpha_n}$
(see Subsection \ref{Aff}), it is enough to show invariance of
$\phi_\lambda$ with respect to $r_0$:
$\phi_\lambda(r_0x)=\phi_\lambda(x)$. Due to \eqref{refl-0}, for
$\mu\in P$ we have
 \[
\langle\mu,r_0x\rangle=\langle \mu ,\xi^\vee + r_{\xi}x\rangle
=\frac{2\langle \mu ,\xi\rangle}{\langle \xi,\xi \rangle} +\langle
\mu, r_{\xi}x\rangle = {\rm integer}+\langle r_{\xi}\mu ,x\rangle
 \]
since $r^2_\xi=1$. Hence,
 \[
\phi_\lambda(r_0x)= \sum_{\mu\in O(\lambda)} e^{2\pi{\rm i}\langle
\mu,r_0x\rangle}=\sum_{\mu\in O(\lambda)} e^{2\pi{\rm i}\langle
r_{\xi}\mu,x\rangle}=\sum_{\mu\in O(\lambda)} e^{2\pi{\rm
i}\langle \mu,x\rangle}=\phi_\lambda(x)
 \]
since $r_{\xi}O(\lambda)=O(\lambda)$.

Note that if $\lambda \not\in P_+$, then $\phi_\lambda$ is not
invariant with respect to $W^{\rm aff}$. It is invariant only
under~$W$.

Due to the invariance of orbit functions $\phi_\lambda$,
$\lambda\in P_+$, with respect to $W^{\rm aff}$, it is enough to
consider them only on the fundamental domain $F\equiv F(W^{\rm
aff})$ of $W^{\rm aff}$. Values of $\phi_\lambda$ on other points
of $E_n$ are determined by using the action of $W^{\rm aff}$ on
$F$ or taking a limit. In particular, {\it orbit functions
$\phi_\lambda$, $\lambda\in P_+$, are invariant under the
reflection of $F$ with respect to any $(n-1)$-dimensional wall of
the fundamental domain $F$.}

\subsection{Continuity}

An orbit function $\phi_\lambda$ is a finite sum of exponential
functions. Therefore it is continuous and has continuous
derivatives of all orders in $E_n$.

The normal derivative of $\phi_\lambda(x)$ to the boundary of $F$
equals zero. It follows from the continuity of $\phi_\lambda(x)$
and of its derivatives, together with its invariance.

\subsection{Realness and complex conjugation}

The results formulated below and concerning orbit functions of the
Coxeter--Dynkin diagrams $A_n$, $B_n$, $C_n$ and $D_n$ were proved
in the previous section. Other cases can be proved, for example,
by means of the representation theory of compact Lie groups; these
proofs are omitted.

Orbit functions of the following Coxeter--Dynkin diagrams are
real:
 \[
A_1,\ B_n,\ C_n,\ D_{2k},\ E_7,\ E_8,\ F_4,\ G_2.
 \]
The orbit functions $\phi_\lambda$ of the remaining
Coxeter--Dynkin diagrams are real, provided the dominant weights
$\lambda$ are invariant with respect to the symmetry
transformation of the diagram. For that the coordinates
$\lambda_1,\lambda_2, \ldots ,\lambda_n$, $\lambda_i=\langle
\lambda,\alpha^\vee_i \rangle$, of $\lambda$ have to be of the
form
\begin{alignat}{3}
&(\lambda_1\ \lambda_2\ \cdots\ \lambda_2\ \lambda_1)
\quad&&\text{for \ }A_n\quad &&(n\geq2),\\
&(\lambda_1\ \lambda_2\ \cdots\ \lambda_{2k-1}\ \lambda_{2k}\
\lambda_{2k})\quad&&\text{for \
}D_{2k+1}\quad &&(k\geq2),\\
&(\lambda_1\ \lambda_2\ \lambda_3\ \lambda_2\ \lambda_1\
\lambda_6)\quad&&\text{for \ }E_6. &&
\end{alignat}

Orbit functions corresponding to the following pairs of dominant
weights are complex conjugate:
\begin{alignat}{3}
&(\lambda_1\ \lambda_2\ \cdots\ \lambda_n)\quad
  &&(\lambda_n\ \lambda_{n-1}\ \cdots\ \lambda_1)
  &&\text{for \ }A_n,  \notag\\
&(\lambda_1\ \lambda_2\ \cdots\ \lambda_{2k-1}\ \lambda_{2k+1}\
\lambda_{2k})\quad
  &&(\lambda_1\ \lambda_2\ \cdots\ \lambda_{2k-1}\ \lambda_{2k}\
  \lambda_{2k+1})\quad
  &&\text{for \ }D_{2k+1},   \notag\\
&(\lambda_1\ \lambda_2\ \lambda_3\ \lambda_4\ \lambda_5\
\lambda_6)\quad
  &&(\lambda_5\ \lambda_4\ \lambda_3\ \lambda_2\ \lambda_1\
  \lambda_6)
  &&\text{for \ }E_6.\notag
\end{alignat}

\subsection{Scaling symmetry}

Let $O(\lambda)$ be an orbit of $\lambda \in E_n$. Since
$w(c\lambda)=cw(\lambda)$ for any $c\in {\mathbb R}$ and for any
$w\in W$, then $O(c\lambda)$ is the orbit consisting of the points
$cw\lambda$, $w\in W$. Let $\phi_\lambda=\sum\limits_{\mu\in
O(\lambda)} e^{2\pi{\rm i}\mu}$ be the orbit function for $\lambda
\in E_n$. Then
 \[
\phi_{c\lambda}(x)=\sum_{c\mu\in O(c\lambda)} e^{2\pi{\rm
i}\langle c\mu ,x\rangle} = \sum_{\mu\in O(\lambda)} e^{2\pi{\rm
i}\langle \mu ,cx\rangle} = \phi_{\lambda}(cx).
 \]
The equality $\phi_{c\lambda}(x)=\phi_{\lambda}(cx)$ expresses the
{\it scaling symmetry of orbit functions.}

If we deal only with orbit functions $\phi_\lambda$, corresponding
to $\lambda\in P_+$, then the scaling symmetry
$\phi_{c\lambda}(x)=\phi_{\lambda}(cx)$ holds for values $c\in
{\mathbb R}\backslash \{ 0\}$ such that $c\lambda\in P_+$.

\subsection{Duality}

The modified orbit functions $\hat \phi_\lambda(x)$ (see formula
\eqref{orb-2}) can be represented in the form
 \[
\hat \phi_\lambda(x)=\sum_{w\in W} e^{2\pi {\rm i}\langle
w\lambda,x \rangle}.
 \]
Due to the invariance of the scalar product $\langle \cdot, \cdot
\rangle$ with respect to the Weyl group $W$, $\langle w\mu ,wy
\rangle= \langle \mu ,y \rangle$, we have
 \[
\hat \phi_\lambda(x)=\sum_{w\in W} e^{2\pi {\rm i}\langle
\lambda,w^{-1}x \rangle} =\sum_{w\in W} e^{2\pi {\rm i}\langle
\lambda,wx \rangle}= \hat \phi_x(\lambda).
 \]
The relation $\hat \phi_\lambda(x)=\hat \phi_x(\lambda)$ expresses
the {\it duality} of orbit functions.

\subsection{Orthogonality}
Orbit functions $\phi_\lambda$, $\lambda\in P_+$, are orthogonal
on $\overline{F}$ with respect to the Euclidean measure:
 \begin{gather}\label{ortog}
|{\overline F}|^{-1}\int_{\overline F} \phi_\lambda(x)
\overline{\phi_{\lambda'}(x)}dx=
|O(\lambda)|\delta_{\lambda\lambda'} ,
 \end{gather}
where the overbar denotes the complex conjugate. This relation
directly follows from the orthogonality of the exponential
functions (entering into the definition of orbit functions) for
different weights $\mu$ and from the fact that a given weight
$\nu\in P$ belongs to precisely one orbit function. In
\eqref{ortog}, $|F|$ means an area of the fundamental domain $F$.

It is evident that for modified orbit functions \eqref{orb-2} the
orthogonality relation is of the form
 \[
|{\overline F}|^{-1}\int_{\overline F} \hat\phi_\lambda(x)
\overline{\hat\phi_{\lambda'}(x)}dx=
|W_\lambda|^2|O(\lambda)|\delta_{\lambda\lambda'}=|W_\lambda| |W|
\delta_{\lambda\lambda'} .
 \]

\subsection[Properties of orbit functions of $A_n$]{Properties of orbit functions of $\boldsymbol{A_n}$}

In this subsection we consider properties of orbit functions of
$A_n$ which follows from properties of symmetric polynomials
(see~\cite{Mac1}).

We represent orbit functions of $A_n$ in the orthogonal coordinate
system as in formula \eqref{orb-f}. Let $\lambda=(m_1,\ldots
,m_{n+1})$ and $\lambda +{\boldsymbol m}=(m_1+m,\ldots
,m_{n+1}+m)$, where $m$ is a fixed real number. If $x=(x_1,\ldots
,x_{n+1})$, $x_1+\cdots +x_{n+1}=0$, and $w\in W$, then we have
 \[
e^{2\pi {\rm i}\langle \lambda+{\boldsymbol m},wx \rangle}=
e^{2\pi {\rm i}\langle \lambda,wx \rangle} e^{2\pi {\rm i}\langle
0+{\boldsymbol m},wx \rangle}=e^{2\pi {\rm i}\langle \lambda,wx
\rangle} .
 \]
It follows from this equality that
\begin{gather}
\phi_\lambda(x)=\phi_{\lambda+{\boldsymbol m}}(x),
 \end{gather}
where $\lambda=(m_1,\ldots ,m_{n+1})$ is given in the orthogonal
coordinate system. This means that instead of $m_i$, $i=1,2,\ldots
,n+1$, determined by formulas of Subsection \ref{An}, we may
assume that $m_1,\ldots ,m_{n+1}$ are integers such that $m_1\ge
m_2\ge \cdots \ge m_{n+1}\ge 0$. We adopt this assumption in this
subsection.

For simplicity we introduce the following notations:
 \[
e^{2\pi {\rm i}x_j}=y_j,\qquad j=1,2,\ldots ,n+1.
 \]
The element
\[
\lambda=(m_1,\ldots ,m_{n+1})\equiv (1,1,\ldots ,1,0,\ldots, 0)
\]
with $r$ units will be denoted by $(1^r)$ (it is the fundamental
weight $\omega_r$). Then from formula \eqref{orb-f} for orbit
functions of $A_n$ one gets
 \[
\phi_{1^r}(x)=\sum_{i_1<i_2<\cdots <i_r} y_{i_1}y_{i_2}\cdots
y_{i_r} .
 \]
It is easy to show using this formula that
 \[
{\sf F}(x;t):=\prod_{i=1}^{n+1} (1+y_it)=\sum _{r=0}^{n+1}
\phi_{1^r}(x)t^r,
 \]
that is, the function ${\sf F}(x;t)\equiv {\sf F}(x_1,\ldots
,x_{n+1};t)$ is a generating function for $n+1$ orbit functions
$\phi_{1^r}(x)$.

Now we denote by $\Phi_r(x)\equiv \Phi_r(x_1,\ldots ,x_{n+1})$,
$r=1,2,\ldots$, the functions
 \[
\Phi_{r}(x)=\sum_{|\lambda |=r} \phi_\lambda (x),
 \]
where $|\lambda|=m_1+m_2+\cdots +m_{n+1}$ if
$\lambda=(m_1,m_2,\ldots ,m_{n+1})$. Since
$(1{-}y_it)^{-1}=\sum\limits _{k=0}^\infty y_i^kt^k$, then
 \[
B(x;t):=\prod_{i=1}^{n+1} (1-y_it)^{-1}=\sum _{r=0}^\infty
\Phi_r(x)t^r,
 \]
where $x_i$ and $y_i$ are related in the same manner as before.
Thus, $B(x;t)$ is a generating function for $\Phi_r(x)$. Since
${\sf F}(x;-t)B(x;t)=1$, after comparing coefficients at fixed
$t^s$ we obtain the equality
 \[
\sum_{r=0}^s (-1)^r\phi_{1^r}(x)\Phi_r(x)=0.
 \]

It follows from formula \eqref{orb-f} that for
$\lambda=(r,0,\ldots ,0)\equiv (r,{\boldsymbol 0})$,
$r=0,1,2,\ldots$, we have
 \[
\phi_{(r,{\boldsymbol 0})}(x)=\sum_{i=1}^{n+1} y_i^r,
 \]
where, as before, $y_i=e^{2\pi {\rm i}x_i}$. The relations
 \[
P(x;t):= \sum_{i=1}^{n+1} \frac{y_i}{1-y_it} =\sum_{i=1}^{n+1}
\sum_{r=1}^\infty y_i^rt^{r-1}=\sum_{r=1}^\infty
\phi_{(r,{\boldsymbol 0})}(x)t^{r-1}
 \]
show that $P(x;t)$ is a generating function for the orbit
functions $\phi_{(r,{\boldsymbol 0})}(x)$.

It is easy to show that
 \[
P(x;t)=B'(x;t)/B(x;t),\qquad P(x;-t)={\sf F}'(x;t)/{\sf F}(x;t),
 \]
where derivatives are taken with respect to variable $t$.
Representing these equalities in the form $P(x;t)B(x;t)=B'(x;t)$
and $P(x;-t){\sf F}(x;t)={\sf F}'(x;t)$, substituting here the
corresponding expressions in terms of orbit functions and then
comparing coefficients at fixed $t^s$, one gets
 \[
s\Phi_s(x)=\sum_{r=1}^s \phi_{(r,{\boldsymbol
0})}(x)\Phi_{s-r}(x), \qquad s\phi_{1^s}(x)=\sum_{r=1}^s
(-1)^{r-1} \phi_{(r,{\boldsymbol 0})}(x) \phi_{1^{s-r}}(x).
 \]

Proofs of the equalities
 \begin{gather*}
\phi_{(r,{\boldsymbol 0})}={\rm det} \begin{pmatrix}
 \phi_{1^1} &   1   &   0 &  \cdots &  0\\
 2\phi_{1^2} & \phi_{1^1} & 1 & \cdots & 0\\
 \cdots   &  \cdots &  \cdots &  \cdots &  \cdots\\
r\phi_{1^r} & \phi_{1^{r-1}} & \phi_{1^{r-2}} &\cdots & 1
\end{pmatrix} ,
\\
\phi_{1^r}=\frac{1}{r!} {\rm det} \begin{pmatrix}
 \phi_{(1,{\boldsymbol 0})} &   1   &   0 &  \cdots &  0\\
 \phi_{(2,{\boldsymbol 0})} & \phi_{(1,{\boldsymbol 0})} & 2 & \cdots & 0\\
 \cdots   &  \cdots &  \cdots &  \cdots &  \cdots\\
\phi_{(r-1,{\boldsymbol 0})} & \phi_{(r-2,{\boldsymbol 0})}&
\phi_{(r-3,{\boldsymbol 0})}
&\cdots & r-1\\
 \phi_{(r,{\boldsymbol 0})} & \phi_{(r-1,{\boldsymbol 0})}& \phi_{(r-2,{\boldsymbol 0})}
&\cdots & \phi_{(1,{\boldsymbol 0})}
\end{pmatrix}
\end{gather*}
are more complicated and we omit them (the corresponding proofs
for symmetric polynomials see in~\cite{Mac1}).

\subsection{Operations with orbit functions}

According to formula \eqref{ortog}, orbit functions are orthogonal
on the fundamental domain. As we shall see in next section, they
form an orthogonal basis in the Hilbert space of squared
integrable functions on $F$. Therefore, we may expand functions on
$F$ as  sums of orbit functions. In parti\-cu\-lar, products of
orbit functions can be uniquely decomposed in orbit functions,
corresponding to the same Weyl group. For such decomposition we
need to use the results on decomposition of products of orbits
(see Section \ref{operat}).

Let elements $\lambda$ and $\lambda'$ be dominant, and let the
product of the orbits $O(\lambda)$ and $O(\lambda')$ decompose as
  \begin{gather}\label{prod-or}
O(\lambda) \otimes O(\lambda')=\bigcup_{\nu} O(\nu).
 \end{gather}
Then for the product of the corresponding orbit functions we have
 \[
\phi_\lambda(x)\phi_{\lambda'}(x)=\sum_\nu \phi_\nu(x),
 \]
where the summation is the same as in \eqref{prod-or}. Indeed,
since according to the definition \eqref{orb} we have
$\phi_\lambda (x)=\sum\limits_{\mu\in O(\lambda)} e^{2\pi{\rm
i}\langle \mu,x \rangle}$, then due to \eqref{prod-or}
 \begin{gather*}
\phi_\lambda(x)\phi_{\lambda'}(x)=\sum_{\mu\in O(\lambda)}
e^{2\pi{\rm i}\langle \mu,x \rangle} \sum_{\mu'\in O(\lambda')}
e^{2\pi{\rm i}\langle \mu',x \rangle}
 \\
\phantom{\phi_\lambda(x)\phi_{\lambda'}(x)}{} =\sum_{\mu\in
O(\lambda)}\sum_{\mu'\in O(\lambda')} e^{2\pi{\rm i}\langle
\mu+\mu',x \rangle}  =\sum_\nu \sum_{\sigma_\nu\in O(\nu)}
e^{2\pi{\rm i}\langle \sigma_\nu,x \rangle} =\sum_\nu \phi_\nu(x).
\end{gather*}

Each of Propositions \ref{proposition1}--\ref{proposition4} of
Subsection \ref{Decomp1} can be formulated in terms of the
corresponding orbit functions. For example, Proposition
\ref{proposition1} says that the product
$\phi_\lambda(x)\phi_{\mu}(x)$ is decomposed into orbits of the
form $\phi_{|w\lambda+\mu|}(x)$, $w\in W/W_\lambda$. Proposition
\ref{proposition2} states that if all elements $w\lambda+\mu$,
$w\in W/W_\lambda$, are strictly dominant, then
 \[
\phi_\lambda(x)\phi_{\mu}(x)=\sum_{w\in W/W_\lambda}
\phi_{w\lambda+\mu}(x).
 \]
According to Proposition \ref{proposition3}, if all elements
$w\lambda+\mu$, $w\in W/W_\lambda$, are dominant, then
 \[
\phi_\lambda(x)\phi_{\mu}(x)=\sum_{w\in W/W_\lambda}
n_{w\lambda+\mu} \phi_{w\lambda+\mu}(x),
 \]
where $n_{w\lambda+\mu}=|W_{w\lambda+\mu}|$.
Proposition~\ref{proposition4} can also be easily formulated in
terms of orbit functions.

In the cases $A_2$, $B_2$ and $G_2$, products
$\phi_\lambda(x)\phi_{\mu}(x)$ are decomposed as sums of orbit
functions by using results of Subsection \ref{Decomp2}. For
example, for orbit functions of $G_2$ we have
 \[
\phi_{(a\ 0)}(x)\phi_{(b\ 0)}(x)=\phi_{(a+b\ 0)}(x)+\phi_{(b-a\
3a)}(x)+\phi_{(2a-b\ 3b-3a)}(x)+\phi_{(b-a\ 0)}(x)
 \]
if $a<b<2a$ and
 \[
\phi_{(a\ 0)}(x)\phi_{(b\ 0)}(x)=\phi_{(a+b\ 0)}(x)+\phi_{(b-a\
3a)}(x)+\phi_{(b-2a\ 3a)}(x)+\phi_{(b-a\ 0)}(x)
 \]
if $b>2a$.

Decomposition of $W$-orbits into $W'$-orbits (see Subsection
\ref{Decomp3}) can also be applied in the decomposition of orbit
functions. Let $R$ be a root system with a Weyl group $W$, and let
$R'$ be another root system which is a subsystem of $R$. Let $R$
span a linear space $E_n$ and $R'$ span its subspace $E_m$ ($E_m$
can coincide with $E_n$). If $\phi_\lambda(x)$ is a $W$-orbit
function and $E_m\ne E_n$, then we may restrict this function onto
the subspace $E_m$. Then the $W$-orbit function $\phi_\lambda(x)$,
considered as a function on $E_m$, can be expanded into a sum of
$W'$-orbit functions. Namely, if $O_W(\lambda)=\cup_{j=1}^s
O_{W'}(\mu_j)$, then
 \[
\phi_\lambda^{(W)}(x)=\sum_{i=1}^s \phi^{(W')}_{\mu_j}(x),\qquad
x\in E_m.
 \]
In particular, all results of Subsections \ref{W-A_n}--\ref{W-D_p}
can be formulated in terms of the corresponding orbit functions.
For example, for $W(A_n)$- and $W(A_{n-1})$-orbit functions we
have
 \[
\phi^{(W(A_{n}))}_{(m_1,\ldots,m_{n+1})}(x)=\sum_{i=1}^{n+1}
\phi^{(W(A_{n-1}))}_{(m_1,\ldots,m_{i-1}\hat{m_i},m_{i+1},
\ldots,m_{n+1})}(x),\qquad x\in E_n,
 \]
where the notations of Subsection \ref{W-A_n} is used.

The results of Subsection \ref{W-A_p} give the following
decomposition of orbit functions of $A_{n-1}$ as a sum of orbit
functions of $A_{p-1}\times A_{q-1}$ with $p+q=n$:
 \[
\phi^{(W(A_{n-1}))}_{(m_1,\ldots,m_{n})}(x)=
\sum_{(m_{i_1},\ldots,m_{i_p})\in \Sigma}
\phi^{(W(A_{p-1}))}_{(m_{i_1},\ldots,m_{i_p})}(y)
\phi^{(W(A_{q-1}))}_{(m_{j_1},\ldots,m_{j_q})}(z),
 \]
where $\Sigma$ is defined in Subsection \ref{W-A_p},
$(m_{j_1},\ldots,m_{j_q})$ is a supplement of the set
$(m_{i_1},\ldots,m_{i_p})$ in $(m_1,m_2,\ldots,m_n)$, and $y\in
E_p$, $z\in E_q$, $x=(y,z)$.

If $E_m=E_n$, then each $W$-orbit function, as a function on
$E_n$, can be represented as a sum of $W'$-orbit functions.
Namely, if $O_W(\lambda)=\cup_{i=1}^s O_{W'}(w_i\lambda)$ (see
Subsection \ref{Decomp3}), then
 \[
\phi^{(W)}_\lambda(x)=\sum_{i=1}^s \phi^{(W')}_{w_i\lambda}(x),
\qquad x\in E_n.
 \]

\subsection{Congruence classes of orbit functions}

The set of all $W$-orbit functions $\phi_{\lambda}(x)$ can be
sorted out into meaningful subsets according to a number of
criteria which may be imposed on either of the two variables
$\lambda$ or $x$. We point out here such example related to the
notion of congruence classes of weights $\lambda$.

Congruence classes of $\lambda\in P$ were introduced in \cite
{LP}. Each $\lambda\in P$ belongs to a single congruence class.
That is, to each $\lambda$ one associates a congruence number
$c(\lambda)$ which is a non-negative integer.

For example, the congruence number $c(\lambda)$ of a point
$\lambda=m\omega$ of the weight lattice of $A_1$, is given by
$c(m\omega)=m \mod 2$. For the rank 2 cases, we have
$\lambda=a\omega_1+b\omega_2$. Then for $\lambda \in P$ the
congruence number $c(\lambda)$ is
 \[
c(a\omega_1+b\omega_2)=\begin{cases} 2a+b \mod3 &\text{ for } A_2,\\
                               a    \mod 2&\text{ for } C_2,\\
                               0          &\text{ for } G_2.
  \end{cases}
 \]

All the roots of a fixed root system $R$ have congruence number 0.
All weights in a $W$-orbit $O(\lambda)$, $\lambda\in P_+$, have
the same congruence number. Under multiplication of orbits
$O(\lambda)$ and $O(\mu)$ of congruence numbers $c(\lambda)$ and
$c(\mu)$, respectively, we obtain a set of weights of the same
congruence number $c(\lambda)+c(\mu)$. Therefore, the product
$O(\lambda)\otimes O(\mu)$, $\lambda\in P_+$, $\mu\in P_+$, is
decomposed into orbits with the same congruence number.

If $\lambda\in P_+$ is of a congruence class $c(\lambda)$, then we
say that the orbit function $\phi_{\lambda}(x)$ is of congruence
class $c(\lambda)$.

We associate with an orbit function $\phi_{\lambda}$, $\lambda\in
P_+$, the congruence number $c(\lambda)$. Then, as in the case of
product of orbits, under multiplication of orbit functions their
congruence numbers are added up. Moreover, all orbit functions in
the decomposition of a product of orbit functions carry the same
congruence number, namely $c(\lambda)+c(\lambda')$.

\setcounter{equation}{0}

\section{Orbit function transform}

The exponential functions $e^{2\pi{\rm i}\langle p,x\rangle}$,
$p=(p_1,p_2,\ldots ,p_n)$, $p_i\in {\mathbb Z}$, given on $E_n$,
determine the Fourier series. As we have seen, symmetrization of
exponential functions leads to orbit functions. The last functions
determine symmetrized Fourier transform, which is also called an
{\it orbit function transform}, and is a generalization of the
decomposition into $\cos$-functions \cite{P-05}. In order to see
this, we first consider a relation of orbit functions to
characters of irreducible representations of compact Lie groups.

\subsection{Orbit functions and representation characters}\label{subsec8.1}

To each Coxeter--Dynkin diagram there corresponds a connected
compact semisimple Lie \linebreak \mbox{group~$G$}. Let us fix a
Coxeter--Dynkin diagram and, therefore, a connected compact Lie
group~$G$. A comp\-lex valued function $f(g)$ on $G$  satisfying
the condition
 \[
f(g)=f\big(hgh^{-1}\big), \qquad h\in G,
 \]
is called a {\it class function}. It is constant on classes of
conjugate elements.

For simplicity, we assume that $G$ is realized by matrices such
that the set of its diagonal matrices constitutes a Cartan
subgroup, which will be denoted by $H$. This subgroup can be
identified with the $n$-dimensional torus ${\sf T}$, where $n$ is
the rank of the group $G$. The subgroup $H$ can be represented as
$H=\exp ({\rm i}{\mathfrak h})$, where ${\mathfrak h}$ is the real
form of an appropriate Cartan subalgebra of the complex semisimple
Lie algebra, determined by the Coxeter--Dynkin diagram.

It is well-known that each element $g$ of $G$ is conjugate to some
element of $H$, that is, class functions are uniquely determined
by their values on $H$.

There exists a one-to-one correspondence between irreducible
unitary representations of the group $G$ and integral highest
weights $\lambda\in P_+$, where $P_+$ is determined by the
Coxeter--Dynkin diagram (see, for example, \cite{Hem-2} and
\cite{Z}). The irreducible representation, corresponding to
a~highest weight $\lambda\in P_+$, will be denoted by $T_\lambda$.
The representation $T_\lambda$ and its properties are determined
by its character $\chi_\lambda(g)$, which is defined as the trace
of $T_\lambda(g)$:
 \[
\chi_\lambda(g)={\rm Tr}\, T_\lambda(g),\qquad g\in G.
 \]
Since ${\rm Tr}\, T_\lambda(g'g{g'}^{-1})={\rm Tr}\,
T_\lambda(g)$, $g'\in G$, the character $\chi_\lambda$ is a class
function, that is, it is uniquely determined by its values on the
subgroup $H$.

All the operators $T_\lambda(h)$, $h\in H$, are diagonal with
respect to an appropriate basis of the representation space (this
basis is called a weight basis) and their diagonal matrix elements
are of the form $e^{2\pi {\rm i}\langle \mu,x \rangle}$, where
$\mu\in P$ is a weight of the representation $T_\lambda$,
$x=(x_1,x_2,\ldots ,x_n)$ are coordinates of an element $t$ of the
Cartan subalgebra ${\mathfrak h}$ in an appropriate coordinate
system (they are coordinates on the torus ${\sf T}$) and $\langle
\cdot ,\cdot \rangle$ is an appropriate bilinear form, which can
be chosen coinciding with the scalar product on $E_n$, considered
above. Then the character $\chi_\lambda(h)$ is a linear
combination of the diagonal matrix elements:
 \begin{gather}\label{trace}
 \chi_\lambda(h)=\sum_{\mu\in D(\lambda)} c_\lambda^\mu
e^{2\pi {\rm i}\langle \mu,x \rangle} ,\qquad h\in H,
 \end{gather}
where $D(\lambda)$ is the set of all weights of the irreducible
representation $T_\lambda$ and $c_\lambda^\mu$ is a multiplicity
of the weight $\mu\in D(\lambda)$ in the representation
$T_\lambda$. It is known from representation theory that the
weight system $D(\lambda)$ of $T_\lambda$ is invariant with
respect to the Weyl group $W$, corresponding to the
Coxeter--Dynkin diagram, and $c_\lambda^{w\mu} =c_\lambda^\mu$,
$w\in W$, for each $\mu\in D(\lambda)$. This means that the
character $\chi_\lambda(h)$ can be represented as
 \begin{gather}\label{char}
 \chi_\lambda(h)=\sum_{\mu\in D_+(\lambda)}
c_\lambda^\mu \phi_\mu(x) ,
 \end{gather}
where $D_+(\lambda)$ is the set of all dominant weights in
$D(\lambda)$ and $\phi_\mu(x)$ is the orbit function,
corresponding to the weight $\mu\in D_+(\lambda)$. Representing
$\chi_\lambda(h)$ as $\chi_\lambda(x)$, where $x=(x_1,x_2,\ldots
,x_n)$ are coordinates, corresponding to the element $t\in
{\mathfrak h}$ such that $h=\exp(2\pi{\rm i}t)$, we can make an
analytic continuation of both sides of \eqref{char} to the
$n$-dimensional Euclidean space $E_n$. Since the right hand side
of \eqref{char} is invariant under transformations from the affine
Weyl group $W^{\rm aff}$, corresponding to the Weyl group $W$, the
function $\chi_\lambda(x)$ is also invariant under the affine Weyl
group $W^{\rm aff}$. That is, it is enough to define
$\chi_\lambda(x)$ only on the fundamental domain $F$ of the group
$W^{\rm aff}$. To this fundamental domain $F$ there corresponds a
fundamental domain (we denote it by $\tilde F$) in the subgroup
$H$ (and in the torus ${\sf T}$).

Many properties of orbit functions follow from properties of
characters $\chi_\lambda$, which we consider known from
representation theory.

As one of the reasons, why characters are rarely used in extensive
applications, one may bring forward the need to know the
multiplicities $c_\lambda^\mu$ in \eqref{trace}. They can be
calculated using a laborious recursive algorithm, starting from
the highest weight $\lambda$. In many situations it is practical
to read off their values from the tables (see \cite{BMP} and
\cite{P}).

\subsection[Orbit function transform on $\overline F$]{Orbit function transform on $\boldsymbol{\overline F}$}

Let $f(g)$ be a continuous class function on $G$. It defines a
continuous function on $H$. We assume that this function on $H$
has continuous partial derivatives of all orders with respect to
analytic parameters on $H$. Such function $f$ can be decomposed in
characters of irreducible unitary representations of $G$:
 \begin{gather}\label{decomp}
f(h)=\sum_{\lambda\in P_+} a_\lambda \chi_\lambda(h).
 \end{gather}
We see from this decomposition that {\it each class function is
uniquely determined by its values on the fundamental domain}
$\tilde F$. Moreover, we can state that each continuous function
with continuous derivatives on $\tilde F$ can be decomposed into
series in characters. In particular, we may decompose an orbit
function (as a continuous function with continuous derivatives)
into series in characters of irreducible representations of $G$:
 \[
\phi_\mu(x)=\sum_{\lambda\in P_+}a^\mu_\lambda \chi_\lambda(x).
 \]

In real, orbit functions are finite linear combinations of
irreducible characters. Indeed, the decomposition \eqref{char} is
the following decomposition
 \begin{gather}\label{char-1}
 \chi_\lambda(x)=\sum_{0\le \mu\le\lambda}
c_\lambda^\mu \phi_\mu(x) ,\qquad c_\lambda^\lambda\ne 0,
 \end{gather}
where $\mu\le\lambda$ means that $\lambda-\mu$ belongs to the
positive root lattice $Q_+$ or $\lambda =\mu$, and $\mu\ge 0$
means that $\mu$ is dominant. Let $\mu_1,\mu_2,\ldots ,\mu_s$ be
the set of all dominant integral weights such that $\mu_i\le
\lambda$. Then the characters $\chi_{\mu_i}(x)$, $i=1,2,\ldots
,s$, are linearly independent. The orbit functions
$\phi_{\mu_i}(x)$, $i=1,2,\ldots ,s$, are also linearly
independent since they are pairwise orthogonal. Due
to~\eqref{char-1} the set~$\chi_{\mu_i}(x)$, $i=1,2,\ldots ,s$, is
a span of $\phi_{\mu_i}(x)$, $i=1,2,\ldots ,s$. Besides, the
set~$\phi_{\mu_i}(x)$, $i=1,2,\ldots ,s$, and the
set~$\chi_{\mu_i}(x)$, $i=1,2,\ldots ,s$, span finite dimensional
linear spaces of functions on $F$ of the same dimension.
Therefore, these spaces coincide and each~$\phi_{\mu_i}(x)$ is
a~linear combination of $\chi_{\mu_j}(x)$, $j=1,2,\ldots ,s$.
Thus, our assertion is proved.

We conclude that each continuous function on ${\overline F}$ with
continuous derivatives can be expanded into orbit functions
$\phi_\lambda$, $\lambda\in P_+$:
 \begin{gather}\label{decom-1}
f(x)=\sum_{\lambda\in P_+} c_\lambda \phi_\lambda(x).
 \end{gather}
Due to the orthogonality relation \eqref{ortog} for orbit
functions, the coefficients $c_\lambda$ in this decomposition are
determined by the formula
 \begin{gather}\label{decom-2}
c_\lambda =| O(\lambda)|^{-1}|{\overline F}|^{-1} \int_{\overline
F} f(x)\overline{\phi_\lambda(x)}dx .
 \end{gather}
Moreover, the Plancherel formula
 \begin{gather}\label{decom-3}
\sum_{\lambda\in P_+} |O(\lambda)| |c_\lambda|^2= |{\overline
F}|^{-1} \int_{\overline F} |f(x)|^2dx
 \end{gather}
holds. Formula \eqref{decom-2} is the symmetrized Fourier
transform of the function $f(x)$. Formula \eqref{decom-1} gives an
inverse transform. Formulas \eqref{decom-1} and \eqref{decom-2}
give the {\it orbit function transforms}.

For the modified orbit functions \eqref{orb-2} the relations
\eqref{decom-1}-\eqref{decom-3} can be written as
 \begin{gather*}
f(x)=\sum_{\lambda \in P_+}c'_\lambda \hat\phi_\lambda (x),
\\
c'_\lambda =|W_\lambda|^{-1} |W|^{-1} |{\overline F}|^{-1}
\int_{\overline F} f(x)\overline{\hat\phi_\lambda(x)} dx,
\\
\sum_{\lambda\in P_+}|W_\lambda| |W||c'_\lambda|^2= |{\overline
F}|^{-1}\int_{\overline F} |f(x)|^2dx,
\end{gather*}
where $c'_\lambda=c_\lambda/|W_\lambda|$.

Let ${\mathcal L}^2({\overline F})$ denote the Hilbert space of
functions on the closure ${\overline F}$ of the fundamental domain
${\overline F}$ with the scalar product
 \[
\langle f_1,f_2\rangle =|{\overline F}|^{-1} \int_{\overline F}
f_1(x)\overline{f_2(x)} dx.
 \]
The set of continuous functions on ${\overline F}$ with continuous
derivatives is dense in ${\mathcal L}^2({\overline F})$.
Therefore, the formulas \eqref{decom-1}--\eqref{decom-3} can be
continued to functions of ${\mathcal L}^2({\overline F})$. These
formulas show that {\it the set of orbit functions $\phi_\lambda$,
$\lambda\in P_+$, form an orthogonal basis of ${\mathcal
L}^2({\overline F})$.}

\subsection{Orbit function transform on the dominant Weyl chamber}

The expansion \eqref{decom-1} of functions on the fundamental
domain $F$ is an expansion in the orbit functions $\phi_\lambda
(x)$, $\lambda\in P_+$. Other orbit functions $\phi_\lambda (x)$,
$\lambda\in E_n$, are not invariant with respect to the
corresponding affine Weyl group $W^{\rm aff}$. They are invariant
only with respect to the Weyl group $W$. A fundamental domain of
$W$ coincides with the dominant Weyl chamber $D_+$. For this
reason, the orbit functions $\phi_\lambda (x)$, $\lambda\in E_n$,
determine another orbit function transform (a~transform on $D_+$).

Let us start with the usual Fourier transforms on ${\mathbb R}^n$:
 \begin{gather}\label{F-1}
\tilde f (\lambda)=\int_{-\infty}^\infty f(x) e^{2\pi {\rm
i}\langle \lambda,x \rangle} dx,
  \\
  \label{F-2}
 f (x)=\int_{-\infty}^\infty \tilde f(\lambda) e^{-2\pi {\rm i}\langle
\lambda,x \rangle} d\lambda.
  \end{gather}
Let the function $f(x)$ be invariant with respect to a Weyl group
$W$. It is easy to check that the function $\tilde f (\lambda)$ is
also $W$-invariant. Replace in \eqref{F-1} $\lambda$ by
$w\lambda$, $w\in W$, and sum both side of \eqref{F-1} over $w\in
W$. Then instead of \eqref{F-1} we obtain
 \begin{gather}\label{F-3}
\tilde f (\lambda)= \int_{D_+} f(x) \hat\phi_\lambda(x) dx,
  \end{gather}
where we have taken into account that both $f(x)$ and
$\hat\phi_\lambda(x)$ are $W$-invariant. Note that
$\hat\phi_\lambda(x)$ are modified orbit functions defined by
\eqref{orb-2}.

Similarly, starting from \eqref{F-2}, we obtain the inverse
formula:
 \begin{gather}\label{F-4}
 f (x)= \int_{D_+} \tilde f(\lambda)
 \overline{\hat\phi_\lambda(x)} d\lambda .
  \end{gather}
For the transforms \eqref{F-3} and \eqref{F-4} the Plancherel
formula
 \[
 \int_{D_+} |f(x)|^2 dx=
\int_{D_+} |\tilde f(\lambda) |^2  d\lambda
 \]
holds.

\setcounter{equation}{0}
\section{Finite orbit function transform}\label{sec9}

Along with the usual Fourier transform there exists a finite
Fourier transform. Similarly, it is possible to introduce a finite
orbit function transform. A finite orbit function transform in the
one-dimensional case is well known and widely used, it is called
the {\it discrete cosine transform} (see~\cite{strang} and
references therein). The basis for construction of the finite
orbit function transform was given in~\cite{MP87}.

This transform is used (see \cite{AP, Pat-Z-1} and \cite{Pat-Z-2})
to find an approximate values of a function $f(x)$ on the whole
space $E_n$, if its values on some finite set is known.

In order to describe the finite orbit function transform we first
consider the finite Fourier transform.

\subsection{Finite Fourier transform}\label{subsec9.1}
Let us fix a positive integer $N$ and consider the numbers
 \begin{gather}\label{f-F-1}
e_{mn}:=N^{-1/2}\exp (2\pi {\rm i}mn/N),\qquad m,n=1,2,\ldots,N.
 \end{gather}
The matrix $(e_{mn})_{m,n=1}^N$ is unitary, that is,
 \begin{gather}\label{f-F-2}
\sum_k e_{mk}\overline{e_{nk}} =\delta_{mn},\qquad \sum_k
e_{km}\overline{e_{kn}} =\delta_{mn}.
 \end{gather}
Indeed, according to the formula for a sum of a geometric
progression we have
 \begin{gather*}
t^a+t^{a+1}+\cdots +t^{a+r}=(1-t)^{-1}t^a(1-t^{r+1}),\qquad t\ne
1,
\\
t^a+t^{a+1}+\cdots +t^{a+r}=r+1,\qquad t=1.
\end{gather*}
Setting $t=\exp (2\pi{\rm i}(m-n)/N)$, $a=1$ and $r=N-1$, we
prove~\eqref{f-F-2}.

Let $f(n)$ be a function of $n\in \{ 1,2,\ldots ,N\}$. We may
consider the transform
 \begin{gather}\label{f-F-3}
\sum_{n=1}^N f(n)e_{mn}\equiv N^{-1/2} \sum_{n=1}^N f(n) \exp
(2\pi{\rm i}mn/N) =\tilde f (m).
 \end{gather}
Then, since the matrix $(e_{mn})_{m,n=1}^N$ is unitary, we express
$f(n)$ as a linear combination of functions~\eqref{f-F-1}:
 \begin{gather}\label{f-F-4}
f(n)= N^{-1/2} \sum_{m=1}^N {\tilde f}(m) \exp (-2\pi{\rm i}mn/N)
.
 \end{gather}
The function ${\tilde f}(m)$ is a {\it finite Fourier transform}
of $f(n)$. The finite Fourier transform is a linear map. The
formula \eqref{f-F-4} gives an inverse transform. The Plancherel
formula
 \[
\sum_{m=1}^N |\tilde f(m)|^2=\sum_{n=1}^N | f(n)|^2
 \]
holds for transforms \eqref{f-F-3} and \eqref{f-F-4}.

The finite Fourier transform on the $r$-dimensional linear space
$E_r$ is defined in a similar way. We again fix a positive integer
$N$. Let ${\boldsymbol m}=(m_1,m_2,\ldots,m_r)$ be an $r$-tuple of
integers such that each $m_i$ runs over the integers
$1,2,\ldots,N$. Then the finite Fourier transform on $E_r$ is
given by the kernel
 \[
e_{\boldsymbol{mn}}:=e_{m_1n_1}e_{m_2n_2}\cdots
e_{m_rn_r}=N^{-r/2} \exp (2\pi{\rm i}{\boldsymbol m}\cdot
{\boldsymbol n}/N),
 \]
where ${\boldsymbol m}\cdot {\boldsymbol n}=m_1n_1+m_2n_2+\cdots
+m_rn_r$. If $F({\boldsymbol m})$ is a function of $r$-tuples
${\boldsymbol m}$, $m_i\in\{ 1,2,\ldots,N\}$, then the finite
Fourier transform of $F$ is given by
 \[
{\tilde F}({\boldsymbol n})=N^{-r/2}\sum_{\boldsymbol
m}F({\boldsymbol m}) \exp (2\pi{\rm i}{\boldsymbol m}\cdot
{\boldsymbol n}/N).
 \]
The inverse transform is
 \[
F({\boldsymbol m})=N^{-r/2}\sum_{\boldsymbol n}{\tilde
F}({\boldsymbol n}) \exp (-2\pi{\rm i}{\boldsymbol m}\cdot
{\boldsymbol n}/N).
 \]
The corresponding Plancherel formula is of the form
$\sum\limits_{\boldsymbol m} |F({\boldsymbol
m})|^2=\sum\limits_{\boldsymbol n} |\tilde F ({\boldsymbol
n})|^2$.

\subsection[$W$-invariant lattices]{$\boldsymbol{W}$-invariant lattices}

In order to determine an analogue of the finite Fourier transform,
based on orbit functions, we need a symmetrized analogue of the
set
 \[
\{ {\boldsymbol m}=\{ m_1,\ldots,m_r\}\ |\  m_i\in
\{1,2,\ldots,N\}\},
 \]
used for multidimensional finite Fourier transform. Such a set has
to be invariant with respect to the Weyl group $W$. It was
constructed in~\cite{MP87}. Let us briefly describe it.

The lattice $Q^\vee$ is a discrete $W$-invariant subset of $E_n$.
Clearly, the set $\frac 1m Q^\vee$ is also $W$-in\-variant, where
$m$ is a fixed positive integer. Then the set
 \[
T_m={\textstyle \frac 1m} Q^\vee / Q^\vee
 \]
is finite and $W$-invariant. If $\alpha_1,\alpha_2,\ldots
,\alpha_n$ is the set of simple roots for the Weyl group $W$, then
$T_m$ can be identified with the set of elements
 \begin{gather}\label{lat-1}
\frac1m \sum_{i=1}^n d_i\alpha_i^\vee,\qquad d_i=0,1,2,\ldots,
m-1.
 \end{gather}

We need to select from $T_m$ a set of elements which belongs to
the closure $\overline{F}$ of the fundamental domain $F$. These
elements lie in the collection $\frac 1m Q^\vee \cap
\overline{F}$.

Let $\mu\in \frac 1m Q^\vee \cap \overline{F}$ be an element
determining an element of $T_m$ and let $M$ be the least positive
integer such that $M\mu \in P^\vee$. (Then there exists a least
positive integer $N$ such that $N\mu\in Q^\vee$. One has $M | N$
and $N | m$; see~\cite{MP87}.)

The collection of points of $T_m$ which belong to $\overline{F}$
(we denote the set of these points by $F_M$) can be derived from
the results of V.~Kac in~\cite{K}. It coincides with the set of
elements
\begin{gather}\label{fin-orb-1}
   s=\frac{s_1}M \omega^\vee_1+\cdots +\frac{s_n}M \omega^\vee_n,\qquad
   \omega^\vee_i :=\frac{2\omega_i}{\langle \alpha_i,\alpha_i  \rangle}  ,
 \end{gather}
where $s_1,s_2,\ldots,s_n$ run over the values from ${\mathbb
Z}^{\ge}$ which satisfy the following condition: there exists a
non-negative integer $s_0$ such that
\begin{gather}\label{fin-orb-2}
s_0+\sum_{i=1}^n s_im_i=M,
 \end{gather}
where $m_1,\ldots, m_n$ are positive integers taken from the
formula \eqref{highestroot}. (In Subsection \ref{roots} one can
find values of $m_i$ for all simple Lie algebras.)

Indeed, the fundamental domain consists of all points $y$ from the
dominant Weyl chamber for which $\langle y,\xi \rangle \le 1$,
where $\xi$ is the highest (long) root, $\xi=\sum\limits_{i=1}^n
m_i\alpha_i$. Since $\langle \alpha_i,\hat\omega_j
\rangle=\delta_{ij}$ and for elements $s$ of \eqref{fin-orb-1} one
has $s_i/M \ge 0$ and
 \[
\langle s,\xi \rangle= \frac 1M \sum_{i=1}^n s_im_i=  \frac 1M
(M-s_0)\le 1,
 \]
then $s\in \overline{F}$. The converse reasoning shows that points
of $\frac 1m Q^\vee \cap \overline{F}$ must be of the form
\eqref{fin-orb-1}.

The numbers $s_0,s_1,s_2,\ldots,s_n$ can be viewed as attached to
the corresponding nodes of the extended Coxeter--Dynkin diagram.

To every positive integer $M$ there corresponds a grid $F_M$ of
points \eqref{fin-orb-1} in $\overline{F}$. This grid is related
to some set $T_m$ such that $M | m$. The precise relation between
$M$ and $m$ can be defined by the grid $F_M$ (see~\cite{MP87}) .
Acting upon the grid $F_M$ by elements of the Weyl group $W$ we
obtain the whole set $T_m$.

Remark that for fulfilling decompositions in orbit functions on a
finite set we need a grid $F_M$ and do not need the number $m$.
This number is needed for proving the corresponding results.

\subsection{Expending in orbit functions through finite sets}

The aim of this subsection is to give an analogue of the finite
Fourier transform when, instead of exponential functions, we use
orbit functions. This analogue is not so simple as the finite
Fourier transform. For this reason, we consider some weak form of
the transform. In fact, we consider this weak form in order to be
able to recover the decomposition $f(x)=\sum\limits_\lambda
a_\lambda \phi_\lambda(x)$ for all values $x\in E_n$ by values of
$f(x)$ on a finite set of point.

When considering the finite Fourier transform of Subsection
\ref{subsec9.1}, we restrict the exponential function to a finite
set. Similarly, in order to determine finite orbit function
transform we have to restrict orbit functions $\phi_\lambda(x)$ to
an appropriate finite set of values of $x$. Candidates for such
finite sets are sets $T_m$. However, orbit functions
$\phi_\lambda(x)$, $\lambda\in P_+$, are invariant with respect to
the affine Weyl group $W^{\rm aff}$. For this reason, we consider
these orbit functions $\phi_\lambda(x)$ on grids $F_M$.

On the other hand, we also have to choose a finite number of orbit
functions, that is, a finite number of dominant elements
$\lambda\in P_+$. The best choice is when the number of orbit
functions coincides with $|F_M|$. These orbit functions must be
selected in such a way that the matrix
 \begin{gather}\label{*}
\left( \phi_{\lambda_i}(x_j)\right)_{\lambda_i\in \Omega,x_j\in
F_M}
 \end{gather}
(where $\Omega$ is our finite set of dominant elements $\lambda\in
P_+$) is not singular. In order to have non-singularity of this
matrix some conditions must be satisfied. In general, they are not
known. For this reason, we consider some weaker form of the
transform (when $|\Omega|\ge |F_M|$) and then explain how the set
$|\Omega|$ of $\lambda\in P_+$ can be chosen in such a way that
$|\Omega|= |F_M|$.

Let $O(\lambda)$ and $O(\mu)$ be two different $W$-orbits. We say
that the group $T_m$ {\it separates} $O(\lambda)$ and $O(\mu)$ if
for any two different elements $\lambda_1\in O(\lambda)$ and
$\mu_1\in O(\mu)$ there exists an element $x\in T_m$ such that
$\exp (2\pi{\rm i}\langle \lambda_1,x \rangle) \ne \exp (2\pi{\rm
i}\langle \mu_1,x \rangle)$. Note that $\lambda$ may coincides
with $\mu$.

Let $f_1$ and $f_2$ be two functions on $E_n$ which are finite
linear combinations of orbit functions. We introduce a
$T_m$-scalar product by the formula
 \[
\langle f_1,f_2 \rangle_{T_m}=\sum_{x\in T_m}
f_1(x)\overline{f_2(x)} .
 \]

\begin{proposition}\label{proposition7} If $T_m$ separates $O(\lambda)$ and
$O(\mu)$, then
\begin{gather}\label{Fuir}
\langle \phi_\lambda,\phi_\mu \rangle_{T_m}=m^n
|O(\lambda)|\delta_{\lambda\mu}.
 \end{gather}
 \end{proposition}

 \begin{proof} We have
\begin{gather}
\langle \phi_\lambda,\phi_\mu \rangle_{T_m}=\sum_{x\in T_m}
\sum_{\sigma\in O(\lambda)} \sum_{\tau\in O(\mu)} \exp (2\pi{\rm
i}\langle \sigma-\tau,x\rangle)\nonumber
\\
\phantom{\langle \phi_\lambda,\phi_\mu \rangle_{T_m}}{}
=\sum_{\sigma\in O(\lambda)} \sum_{\tau\in O(\mu)}
\left(\sum_{x\in T_m} \exp (2\pi{\rm i}\langle
\sigma-\tau,x\rangle)\right).\label{**}
\end{gather}
Since $T_m$ separates $O(\lambda)$ and $O(\mu)$, then none of the
non-zero differences $\sigma-\tau$ in the last sum vanishes on
$T_m$. Since $T_m$ is a group and $|T_m|=m^n$, one has
 \[
\sum_{x\in T_m} \exp (2\pi{\rm i}\langle
\sigma-\tau,x\rangle)=m^n\delta_{\sigma \tau}.
 \]
Then it follows from \eqref{**} that $\langle
\phi_\lambda,\phi_\mu \rangle_{T_m}=m^n
|O(\lambda)|\delta_{\lambda\mu}$. The proposition is proved.
\end{proof}

Let $f$ be a $W^{\rm aff}$-invariant function on $E_n$ which is a
finite linear combination of orbit functions:
\begin{gather}\label{Fur-2}
 f(x)=\sum_{\lambda_j\in
P_+} a_{\lambda_j}\phi_{\lambda_j}(x).
 \end{gather}
Our aim is to determine $f(x)$, $x\in E_n$, by its values on a
finite subset of $E_n$, namely, on $T_m$.

We suppose that $T_m$ separates orbits on the right hand side of
\eqref{Fur-2}. Then taking the $T_m$-scalar product of both sides
of \eqref{Fur-2} with $\phi_{\lambda_i}$ and taking into account
the relation \eqref{Fuir} we obtain
 \[
a_{\lambda_i}=\left( m^n |O(\lambda_i)|\right)^{-1}\langle f,
\phi_{\lambda_i} \rangle_{T_m}.
 \]
Let $s^{(1)},s^{(2)},\ldots,s^{(h)}$ be all elements of
$\overline{F}\cap \frac 1m Q^\vee$. As before, by $W_{s^{(i)}}$ we
denote the subgroup of $W$ whose elements leave $s^{(i)}$
invariant. Then
\begin{gather}
a_{\lambda_j}=m^{-n}|O(\lambda_j)|^{-1}\sum_{x\in T_m} f(x)
\overline{\phi_{\lambda_j}(x)}= m^{-n}\frac{|W_{\lambda_j}|}{|W|}
\sum_{i=1}^h \frac{|W|}{|W_{s^{(i)}}|} f(s^{(i)})
\overline{\phi_{\lambda_j}(s^{(i)})}\nonumber
\\
\phantom{a_{\lambda_j}}{}
 =m^{-n} |W_{\lambda_j}|\sum_{i=1}^h |W_{s^{(i)}}|^{-1}
f(s^{(i)}) \overline{\phi_{\lambda_j}(s^{(i)})},\label{Fuir-3}
 \end{gather}
where $W_{\lambda_j}$ is a stabilizer subgroup of $\lambda_j$
in~$W$.

Thus, {\it a finite number of values $f(s^{(i)})$,
$i=1,2,\ldots,h$, of the function $f(x)$ determines the
coefficients $a_{\lambda_j}$ and, therefore, the function $f(x)$
on the whole space $E_n$.}

This means that we can reconstruct a $W^{\rm aff}$-invariant
function $f(x)$ on the whole space $E_n$ by its values on the
finite set $F_M$ under an appropriate value of $M$. Namely, we
have to expand this function, taken on $F_M$, into the series
\eqref{Fur-2} by means of the coefficients $a_{\lambda_j}$,
determined by formula \eqref{Fuir-3}, and then to continue
analytically the expansion \eqref{Fur-2} to the whole fundamental
domain $\overline{F}$ (and, therefore, to the whole space $E_n$),
that is, to consider the decomposition \eqref{Fur-2} for all $x\in
E_n$.

We have assumed that the function $f(x)$ is a finite linear
combination of orbit functions. If $f(x)$ expands into infinite
sum of orbit functions, then for applying the above procedure we
have to approximate the function $f(x)$ by taking a finite number
of terms in this infinite sum and then apply the procedure. That
is, in this case we obtain an approximate expression of the
function $f(x)$ by using a finite number of its values.

At last, we explain how to choose a set $\Omega$ in formula
\eqref{*}. The set $F_M$ consists of the points~\eqref{fin-orb-1}.
This set determines the set of points
 \[
\lambda=s_1\omega_1+\cdots +s_n\omega_n,
 \]
where $s_1,\ldots,s_n$ run over the same values as for the set
$F_M$. The set of these weights can be taken as the set $\Omega$.
The corresponding considerations for rank~2 cases can be seen
in~\cite{Pat-Z-1} and~\cite{Pat-Z-2}.

\subsection{Orbit functions at rational points}

In the fundamental domain $F$ there exist a finite number of
points (we denote them $x_j$) such that all orbit functions take
integer values at these points:
 \[
\phi_\lambda(x_j)\in{\mathbb Z}\qquad\text{for all}\quad\lambda\in
P.
 \]
Elements of a Lie group, corresponding to these points (see
Subsection~\ref{subsec8.1}), are called rational~\cite{MP82}.

More generally, in each compact simple Lie group there exist few
conjugacy classes of ele\-ments, which have all integer-valued
characters (hence also integer-valued orbit functions). Points of
these conjugacy classes are rational elements. Their orders $M$
are relatively low in the Lie group. The points of the fundamental
domain, representing such elements, are listed in Tables~1--6 
for compact simple Lie groups of rank 2 and
3. Generally such points are not found in the literature except
for the group $E_8$, see~\cite{MP-Con}.

The simplest example of a rational element in a compact simple Lie
group is the identity element of the group. The character at the
identity element is equal to the dimension of the corresponding
representation; the orbit function at the identity element is
equal to the size of the corresponding Weyl group orbit.

Lines of Tables 1--6 identify (conjugacy classes of) rational
elements of the Lie group.  The first entry $M$  at each line is
the adjoint order of the element, i.e.\ its order in the adjoint
representation of the group or, more generally, in any
representation where the center of the Lie group coalesces to
identity.

The second entry $N$ on a line is the full order of the element,
that is its order in representations where the center of the Lie
group is faithfully represented. $M$ always divides $N$ and
$1\leq\tfrac{N}M\leq|Z|$, where $|Z|$ is the order of the center
of the compact simple Lie group.

There is an interesting general one-to-one correspondence between
rational elements of $A_{2k-1}$ and of $A_{2k}$, which we
illustrate in Tables~1 and~4.

\begin{center}
\small

{\bf Table 1.} Rational elements in $A_2$ and in $A_1$, their
adjoint orders $M$ and full orders $N$.
 \medskip

\begin{tabular}
{|c|c||c|c|||c|c||c|c|} \hline
$M$&$N$&$[s_0,s_1,s_2]$&$(\tfrac{s_1}M,\tfrac{s_2}M)$&
$M$&$N$&$[s_0,s_1]$&$(\tfrac{s_1}M)$\\
\hline\hline 1&1&[1,0,0]&(0,0)
       &1&1&[1,0]&(0)\\
\hline 2&2&[0,1,1]&$(\tfrac12,\tfrac12)$
       &1&2&[0,1]&$(\tfrac12)$\\
\hline 3&3&[1,1,1]&$(\tfrac13,\tfrac13)$
       &3&3&$[1,2]$&$(\tfrac23)$\\
\hline 4&4&[2,1,1]&$(\tfrac14,\tfrac14)$
       &2&4&[1,1]&$(\tfrac14)$\\
\hline 6&6&[4,1,1]&$(\tfrac16,\tfrac16)$
       &3&6&[2,1]&$(\tfrac16)$\\
\hline
\end{tabular}
\end{center}

\vspace{-3mm}

\begin{center}
\small

{\bf Table 2.} Rational elements in $C_2$, their adjoint order
$M=s_0+2s_1+s_2$ and full order $N$.

\medskip

\begin{tabular}{|c|c||c|c|}
\hline
$M$&$N$&$[s_0,s_1,s_2]$&$(\tfrac{s_1}M,\tfrac{s_2}M)$\\
\hline\hline 1&1&[1,\ 0,\ 0]&(0,\ 0)\\\hline 1&2&[0,\ 0,\ 1]&(0,\
1)\\\hline\hline 2&2&[0,\ 1,\ 0]&$(\tfrac12,\ 0)$\\\hline
2&4&[1,\ 0,\ 1]&$(0,\ \tfrac12)$\\
\hline\hline 3&3&[1,\ 0,\ 2]&$(0,\ \tfrac23)$\\\hline 3&6&[2,\ 0,\
1]&$(0,\ \tfrac13)$\\\hline
3&6&[0,\ 1,\ 1]&$(\tfrac13,\ \tfrac13)$\\
\hline\hline 4&4&[2,\ 1,\ 0]&$(\tfrac14,\ 0)$\\\hline 4&4&[0,\ 1,\
2]&$(\tfrac14,\tfrac12)$\\\hline
4&8&[1,\ 1,\ 1]&$(\tfrac14,\ \tfrac14)$\\
\hline\hline 5&5&[1,\ 1,\ 2]&$(\tfrac15,\ \tfrac25)$\\\hline
5&10&[2,\ 1,\ 1]&$(\tfrac15,\ \tfrac15)$ \\
\hline\hline 6&6&[4,\ 1,\ 0]&$(\tfrac16,\ 0)$\\\hline 6&6&[2,\ 1,\
2]&$(\tfrac16,\ \tfrac13)$\\\hline 6&6&[0,\ 1,\ 4]&$(\tfrac16,\
\tfrac23)$\\\hline
6&12&[1,\ 2,\ 1]&$(\tfrac13,\ \tfrac16)$\\
\hline\hline 12&12&[6,\ 1,\ 4]&$(\tfrac1{12},\ \tfrac13)$\\\hline
12&12&[4,\ 1,\ 6]&$(\tfrac1{12},\ \tfrac12)$\\\hline
\end{tabular}
\end{center}

\newpage

\begin{center}
\small

{\bf Table 3.} Rational elements in $G_2$, their adjoint order
$M=s_0+2s_1+3s_2$ and the full order $N$.
 \medskip

\begin{tabular}{|c|c||c|c|}
\hline
$M$&$N$&$[s_0,s_1,s_2]$&$(\tfrac{s_1}M,\tfrac{s_2}M)$\\
\hline\hline 1&1&[1,\ 0,\ 0]&(0,\ 0)\\\hline\hline 2&2&[0,\ 1,\
0]&$(\tfrac12,\ 0)$\\\hline\hline 3&3&[1,\ 1,\ 0]&$(\tfrac13,\
0)$\\\hline 3&3&[0,\ 0,\ 1]&$(0,\ \tfrac13)$\\\hline\hline
4&4&[2,\ 1,\ 0]&$(\tfrac14,\ 0)$\\\hline 4&4&[1,\ 0,\ 1]&$(0,\
\tfrac14)$\\\hline\hline 6&6&[4,\ 1,\ 0]&$(\tfrac16,\ 0)$\\\hline
6&6&[3,\ 0,\ 1]&$(0,\ \tfrac16)$\\\hline
6&6&[1,\ 1,\ 1]&$(\tfrac16,\ \tfrac16)$\\
\hline\hline
7&7&[2,\ 1,\ 1]&$(\tfrac17,\ \tfrac17)$\\
\hline\hline 8&8&[3,\ 1,\ 1]&$(\tfrac18,\ \tfrac18)$\\\hline
8&8&[1,\ 2,\ 1]&$(\tfrac14,\ \tfrac18)$\\
\hline\hline 12&12&[3,\ 3,\ 1]&$(\tfrac14,\ \tfrac1{12})$\\\hline
12&12&[1,\ 4,\ 1]&$(\tfrac13,\ \tfrac1{12})$\\\hline
\end{tabular}
\end{center}

\begin{center}
\small

{\bf Table 4.} Rational elements in $A_3$ and in $A_4$, their
adjoint orders $M$ and the full orders $N$.

\medskip

\begin{tabular}
{|c|c||c|c|||c|c||c|c|} \hline $M$&$N$&$[s_0,s_1,s_2,s_3]$
        &$(\tfrac{s_1}M,\tfrac{s_2}M,\tfrac{s_3}M)$&
$M$&$N$&$[s_0,s_1,s_2,s_3,s_4]$
        &$(\tfrac{s_1}M,\tfrac{s_2}M,\tfrac{s_3}M
                         ,\tfrac{s_2}M)$\\
\hline\hline 1&1&[1,\ 0,\ 0,\ 0]&(0,\ 0,\ 0)
         &1&1&[1,\ 0,\ 0,\ 0,\ 0]&(0,\ 0,\ 0,\ 0)\\
\hline 1&2& [0,\ 0,\ 1,\ 0]&$(0,\ 1,\ 0)$
&2&2&[0,\ 0,\ 1,\ 1,\ 0]&$(\ 0,\ \tfrac12,\ \tfrac12,\ 0)$\\
\hline\hline 2&2& [0,\ 1,\ 0,\ 1]&$(\tfrac12,\ 0,\ \tfrac12)$
&2&2& [0,\ 1,\ 0,\ 0,\ 1]&$(\tfrac12,\ 0,\ 0,\ \tfrac12)$\\
\hline 2&4&[1,\ 0,\ 1,\ 0]&$(0,\ \tfrac12,\ 0)$
&4&4& [2,\ 0,\ 1,\ 1,\ 0]&$(0,\ \tfrac14,\ \tfrac14,\ 0)$\\
\hline\hline 3&3&[1,\ 0,\ 2,\ 0]&$(0,\ \tfrac23,\ 0)$
&3&3& [1,\ 0,\ 1,\ 1,\ 0]&$(0,\ \tfrac13,\ \tfrac13,\ 0)$\\
\hline 3&3&[1,\ 1,\ 0,\ 1]&$(\tfrac13,\ 0,\ \tfrac13)$
&3&3& [1,\ 1,\ 0,\ 0,\ 1]&$(\tfrac13,\ 0,\ 0,\ \tfrac13)$\\
\hline 3&6&[2,\ 0,\ 1,\ 0]&$(0,\ \tfrac13,\ 0)$
&6&6& [4,\ 0,\ 1,\ 1,\ 0]&$(0,\ \tfrac16,\ \tfrac16,\ 0)$\\
\hline 3&6&[0,\ 1,\ 1,\ 1]&$(\tfrac13,\ \tfrac13,\ \tfrac13)$
&6&6& [0,\ 2,\ 1,\ 1,\ 2]&$(\tfrac13,\ \tfrac16,\ \tfrac16,\ \tfrac13)$  \\
\hline\hline 4&4&[2,\ 1,\ 0,\ 1]&$(\tfrac14,\ 0,\ \tfrac14)$
&4&4& [2,\ 1,\ 0,\ 0,\ 1]&$(\tfrac14,\ 0,\ 0,\ \tfrac14)$\\
\hline 4&4&[0,\ 1,\ 2,\ 1]&$(\tfrac14,\ \tfrac12,\ \tfrac14)$
&4&4& [0,\ 1,\ 1,\ 1,\ 1]&$(\tfrac14,\ \tfrac14,\ \tfrac14,\ \tfrac14)$\\
\hline 4&8&[1,\ 1,\ 1,\ 1]&$(\tfrac14,\ \tfrac14,\ \tfrac14)$
&8&8& [2,\ 2,\ 1,\ 1,\ 2]&$(\tfrac14,\ \tfrac18,\ \tfrac18,\ \tfrac14)$\\
\hline\hline 5&5&[1,\ 1,\ 2,\ 1]&$(\tfrac15,\ \tfrac25,\
\tfrac15)$
&5&5& [1,\ 1,\ 1,\ 1,\ 1]&$(\tfrac15,\ \tfrac15,\ \tfrac15,\ \tfrac15)$\\
\hline 5&10&[2,\ 1,\ 1,\ 1]&$(\tfrac15,\ \tfrac15,\ \tfrac15)$
&10&10& [4,\ 2,\ 1,\ 1,\ 2]&$(\tfrac15,\ \tfrac1{10},\ \tfrac1{10},\ \tfrac15)$\\
\hline\hline 6&6&[4,\ 1,\ 0,\ 1]&$(\tfrac16,\ 0,\ \tfrac16)$
&6&6& [4,\ 1,\ 0,\ 0,\ 1]&$(\tfrac16,\ 0,\ 0,\ \tfrac16)$\\
\hline 6&6&[2,\ 1,\ 2,\ 1]&$(\tfrac16,\ \tfrac13,\ \tfrac16)$
&6&6& [2,\ 1,\ 1,\ 1,\ 1]&$(\tfrac16,\ \tfrac16,\ \tfrac16,\ \tfrac16)$\\
\hline 6&6&[0,\ 1,\ 4,\ 1]&$(\tfrac16,\ \tfrac23,\ \tfrac16)$
&6&6& [0,\ 1,\ 2,\ 2,\ 1]&$(\tfrac16,\ \tfrac13,\ \tfrac13,\ \tfrac16)$\\
\hline 6&12&[1,\ 2,\ 1,\ 2]&$(\tfrac13,\ \tfrac16,\ \tfrac13)$
&12&12& [2,\ 4,\ 1,\ 1,\ 4]&$(\tfrac13,\ \tfrac1{12},\ \tfrac1{12},\ \tfrac13)$\\
\hline\hline 12&12&[6,\ 1,\ 4,\ 1]&$(\tfrac1{12},\ \tfrac13,\
\tfrac1{12})$
&12&12& [6,\ 1,\ 2,\ 2,\ 1]&$(\tfrac1{12},\ \tfrac16,\ \tfrac16,\ \tfrac1{12})$\\
\hline 12&12&[4,\ 1,\ 6,\ 1]&$(\tfrac1{12},\ \tfrac12,\
\tfrac1{12})$
&12&12& [4,\ 1,\ 3,\ 3,\ 1]&$(\tfrac1{12},\ \tfrac14,\ \tfrac14,\ \tfrac1{12})$ \\
\hline\hline
\end{tabular}
\end{center}

\newpage

\begin{center}
\small

{\bf Table 5.} Rational elements in $B_3$, their adjoint order
$M=s_0{+}s_1{+}2s_2{+}2s_3$ and the full order $N$.
\medskip

\begin{tabular}
{|c|c||c|c|} \hline $M$&$N$&$[s_0,s_1,s_2,s_3]$
        &$(\tfrac{s_1}M,\tfrac{s_2}M,\tfrac{s_3}M)$\\
\hline\hline
1&1&[1,\ 0,\ 0,\ 0]&(0,\ 0,\ 0)\\
\hline
1&2&[0,\ 1,\ 0,\ 0]&$(1,\ 0,\ 0)$\\
\hline\hline
2&2&[0,\ 0,\ 1,\ 0]&$(0,\ \tfrac12,\ 0)$\\
\hline
2&4&[1,\ 1,\ 0,\ 0]&$(\tfrac12,\ 0,\ 0)$\\
\hline
2&4&[0,\ 0,\ 0,\ 1]&$(0,\ 0,\ \tfrac12,)$\\
\hline\hline
3&3&[1,\ 2,\ 0,\ 0]&$(\tfrac23,\ 0,\ 0)$\\
\hline
3&3&[1,\ 0,\ 1,\ 0]&$(0,\ \tfrac13,\ 0,$\\
\hline
3&3&[0,\ 1,\ 0,\ 1]&$(\tfrac13,\ 0,\ \tfrac13)$  \\
\hline
3&6&[2,\ 1,\ 0,\ 0]&$( \tfrac13,\ 0,\ 0)$\\
\hline
3&6&[0,\ 1,\ 1,\ 0]&$( \tfrac13,\ \tfrac13,\ 0)$  \\
\hline
3&6&[1,\ 0,\ 0,\ 1]&$(0,\ 0,\ \tfrac13)$  \\
\hline\hline
4&4&[2,\ 0,\ 1,\ 0]&$(0,\ \tfrac14,\  0)$ \\
\hline
4&4&[0,\ 2,\ 1,\ 0]&$(\tfrac12,\ \tfrac14,\ 0)$ \\
\hline
4&4&[1,\ 1,\ 0,\ 1]&$(\tfrac14,\ 0,\ \tfrac14)$ \\
\hline
4&8&[1,\ 1,\ 1,\ 0]&$(\tfrac14,\ \tfrac14,\ 0)$ \\
\hline
4&8&[0,\ 0,\ 1,\ 1]&$(0,\ \tfrac14,\ \tfrac14)$ \\
\hline\hline
5&5&[1,\ 2,\ 1,\ 0]&$(\tfrac25,\ \tfrac15,\ 0)$ \\
\hline
5&10&[2,\ 1,\ 1,\ 0]&$(\tfrac15,\ \tfrac15,\ 0)$ \\
\hline\hline
6&6&[4,\ 0,\ 1,\ 0]&$(0,\ \tfrac16,\ 0)$ \\
\hline
6&6&[2,\ 2,\ 1,\ 0]&$(\tfrac13,\ \tfrac16,\ 0)$ \\
\hline
6&6&[0,\ 4,\ 1,\ 0]&$(\tfrac23,\ \tfrac16,\ 0)$ \\
\hline
6&6&[3,\ 1,\ 0,\ 1]&$(\tfrac16,\ 0,\ \tfrac16)$ \\
\hline
6&6&[1,\ 3,\ 0,\ 1]&$(\tfrac12,\ 0,\ \tfrac16)$ \\
\hline
6&6&[1,\ 1,\ 1,\ 1]&$(\tfrac16,\ \tfrac16,\ \tfrac16)$ \\
\hline
6&6&[0,\ 0,\ 1,\ 2]&$(0,\ \tfrac16,\ \tfrac13)$ \\
\hline
6&12&[1,\ 1,\ 2,\ 0]&$(\tfrac16,\ \tfrac13,\ 0)$ \\
\hline
6&12&[2,\ 2,\ 0,\ 1]&$(\tfrac13,\ 0,\ \tfrac16)$ \\
\hline
6&12&[0,\ 0,\ 2,\ 1]&$(0,\ \tfrac13,\ \tfrac16)$ \\
\hline
6&12&[1,\ 1,\ 0,\ 2]&$(\tfrac16,\ 0,\ \tfrac13)$ \\
\hline \hline
7&7&[2,\ 1,\ 1,\ 1]&$(\tfrac17,\ \tfrac17,\ \tfrac17)$ \\
\hline
7&14&[1,\ 2,\ 1,\ 1]&$(\tfrac27,\ \tfrac17,\ \tfrac17)$ \\
\hline \hline
8&8&[3,\ 1,\ 1,\ 1]&$(\tfrac18,\ \tfrac18,\ \tfrac18)$ \\
\hline
8&8&[1,\ 3,\ 1,\ 1]&$(\tfrac38,\ \tfrac18,\ \tfrac18)$ \\
\hline
8&8&[1,\ 1,\ 2,\ 1]&$(\tfrac18,\ \tfrac14,\ \tfrac18)$ \\
\hline \hline
9&9&[2,\ 3,\ 1,\ 1]&$(\tfrac13,\ \tfrac19,\ \tfrac19)$ \\
\hline
9&18&[3,\ 2,\ 1,\ 1]&$(\tfrac29,\ \tfrac19,\ \tfrac19)$ \\
\hline \hline
10&20&[2,\ 2,\ 2,\ 1]&$(\tfrac15,\ \tfrac15,\ \tfrac1{10})$ \\
\hline
10&20&[1,\ 1,\ 2,\ 2]&$(\tfrac1{10},\ \tfrac1{5},\ \tfrac15)$ \\
\hline \hline
12&12&[6,\ 4,\ 1,\ 0]&$(\tfrac1{3},\ \tfrac1{12},\ 0)$ \\
\hline
12&12&[4,\ 6,\ 1,\ 0]&$(\tfrac1{2},\ \tfrac1{12},\ 0)$ \\
\hline
12&12&[3,\ 1,\ 3,\ 1]&$(\tfrac1{12},\ \tfrac14,\ \tfrac1{12})$ \\
\hline
12&12&[1,\ 3,\ 3,\ 1]&$(\tfrac1{4},\ \tfrac14,\ \tfrac1{12})$ \\
\hline
12&12&[1,\ 1,\ 4,\ 1]&$(\tfrac1{12},\ \tfrac13,\ \tfrac1{12})$ \\
\hline
12&12&[5,\ 1,\ 0,\ 3]&$(\tfrac1{12},\ 0,\ \tfrac14)$ \\
\hline
12&12&[1,\ 5,\ 0,\ 3]&$(\tfrac5{12},\ 0,\ \tfrac14)$ \\
\hline
12&24&[3,\ 3,\ 1,\ 2]&$(\tfrac1{4},\ \tfrac1{12},\ \tfrac16)$ \\
\hline
12&24&[2,\ 2,\ 1,\ 3]&$(\tfrac1{6},\ \tfrac1{12},\ \tfrac14)$ \\
\hline \hline
15&15&[4,\ 1,\ 2,\ 3]&$(\tfrac1{15},\ \tfrac2{15},\ \tfrac15)$ \\
\hline
15&30&[1,\ 4,\ 2,\ 3]&$(\tfrac4{15},\ \tfrac2{15},\ \tfrac15)$ \\
\hline \hline
\end{tabular}
\end{center}

\newpage

\begin{center}
\small

{\bf Table 6.} Rational elements in $C_3$, their adjoint order
$M=s_0{+}2s_1{+}2s_2{+}s_3$ and the full order $N$.
\medskip

\begin{tabular}
{|c|c||c|c|} \hline $M$&$N$&$[s_0,s_1,s_2,s_3]$
        &$(\tfrac{s_1}M,\tfrac{s_2}M,\tfrac{s_3}M)$\\
\hline\hline
1&1&[1,\ 0,\ 0,\ 0]&(0,\ 0,\ 0)\\
\hline
1&2&[0,\ 0,\ 0,\ 1]&$(0,\ 0,\ 1)$\\
\hline\hline
2&2&[0,\ 1,\ 0,\ 0]&$(\tfrac12,\ 0,\ 0)$\\
\hline
2&2&[0,\ 0,\ 1,\ 0]&$(0,\ \tfrac12,\ 0)$\\
\hline
2&4&[1,\ 0,\ 0,\ 1]&$(0,\ 0,\ \tfrac12,)$\\
\hline\hline
3&3&[1,\ 1,\ 0,\ 0]&$(\tfrac13,\ 0,\ 0)$\\
\hline
3&3&[1,\ 0,\ 1,\ 0]&$(0,\ \tfrac13,\ 0,$\\
\hline
3&3&[1,\ 0,\ 0,\ 2]&$(0,\ 0,\ \tfrac23)$  \\
\hline
3&6&[2,\ 0,\ 0,\ 1]&$( 0,\ 0,\ \tfrac13)$\\
\hline
3&6&[0,\ 1,\ 0,\ 1]&$( \tfrac13,\ 0,\ \tfrac13)$  \\
\hline
3&6&[0,\ 0,\ 1,\ 1]&$(0,\ \tfrac13,\ \tfrac13)$  \\
\hline\hline
4&4&[2,\ 1,\ 0,\ 0]&$( \tfrac14,\ 0,\ 0)$ \\
\hline
4&4&[2,\ 0,\ 1,\ 0]&$(0,\ \tfrac14,\ 0)$ \\
\hline
4&4&[0,\ 1,\ 1,\ 0]&$(\tfrac14,\  \tfrac14,\ 0)$ \\
\hline
4&4&[0,\ 1,\ 0,\ 2]&$(\tfrac14,\ 0,\ \tfrac12)$ \\
\hline
4&4&[0,\ 0,\ 1,\ 2]&$(0,\ \tfrac14,\ \tfrac12)$ \\
\hline\hline
5&5&[1,\ 1,\ 1,\ 0]&$(\tfrac15,\ \tfrac15,\ 0)$ \\
\hline
5&10&[0,\ 1,\ 1,\ 1]&$(\tfrac15,\ \tfrac15,\ \tfrac15)$ \\
\hline\hline
6&6&[4,\ 1,\ 0,\ 0]&$( \tfrac16,\ 0,\ 0)$ \\
\hline
6&6&[4,\ 0,\ 1,\ 0]&$(0,\ \tfrac16,\ 0)$ \\
\hline
6&6&[2,\ 1,\ 1,\ 0]&$(\tfrac16,\ \tfrac16,\ 0)$ \\
\hline
6&6&[0,\ 2,\ 1,\ 0]&$(\tfrac13,\  \tfrac16, 0,\ )$ \\
\hline
6&6&[0,\ 1,\ 2,\ 0]&$(\tfrac16,\  \tfrac13, 0,\ )$ \\
\hline
6&6&[2,\ 1,\ 0,\ 2]&$(\tfrac16,\ 0,\ \tfrac13)$ \\
\hline
6&6&[2,\ 0,\ 1,\ 2]&$(0,\ \tfrac16,\ \tfrac13)$ \\
\hline
6&6&[0,\ 1,\ 1,\ 2]&$(\tfrac16,\ \tfrac16,\ \tfrac13)$ \\
\hline
6&6&[0,\ 1,\ 0,\ 4]&$(\tfrac16,\ 0,\ \tfrac23)$ \\
\hline
6&6&[0,\ 0,\ 1,\ 4]&$(0,\ \tfrac16,\ \tfrac23)$ \\
\hline
6&12&[1,\ 1,\ 1,\ 1]&$(\tfrac16,\ \tfrac16,\ \tfrac16)$ \\
\hline \hline
7&7&[1,\ 1,\ 1,\ 2]&$(\tfrac17,\ \tfrac17,\ \tfrac27)$ \\
\hline
7&14&[2,\ 1,\ 1,\ 1]&$(\tfrac17,\ \tfrac17,\ \tfrac17)$ \\
\hline \hline
8&8&[2,\ 2,\ 1,\ 0]&$(\tfrac14,\ \tfrac18,\ 0)$ \\
\hline
8&8&[2,\ 1,\ 1,\ 2]&$(\tfrac18,\ \tfrac18,\ \tfrac14)$ \\
\hline
8&8&[0,\ 1,\ 2,\ 2]&$(\tfrac18,\ \tfrac14,\ \tfrac14)$ \\
\hline \hline
9&9&[1,\ 2,\ 1,\ 2]&$(\tfrac29,\ \tfrac19,\ \tfrac29)$ \\
\hline
9&18&[2,\ 1,\ 2,\ 1]&$(\tfrac19,\ \tfrac29,\ \tfrac19)$ \\
\hline \hline
10&10&[4,\ 2,\ 1,\ 0]&$(\tfrac15,\ \tfrac1{10},\ 0)$ \\
\hline
10&10&[0,\ 1,\ 2,\ 4]&$(\tfrac1{10},\ \tfrac1{5},\ \tfrac25)$ \\
\hline \hline
12&12&[2,\ 4,\ 1,\ 0]&$(\tfrac1{3},\ \tfrac1{12},\ 0)$ \\
\hline
12&12&[6,\ 1,\ 2,\ 0]&$(\tfrac1{12},\ \tfrac16,\ 0)$ \\
\hline
12&12&[4,\ 1,\ 3,\ 0]&$(\tfrac1{12},\ \tfrac14,\ 0)$ \\
\hline
12&12&[2,\ 3,\ 1,\ 2]&$(\tfrac1{4},\ \tfrac1{12},\ \tfrac16)$ \\
\hline
12&12&[2,\ 1,\ 3,\ 2]&$(\tfrac1{12},\ \tfrac14,\ \tfrac16)$ \\
\hline
12&12&[0,\ 1,\ 4,\ 2]&$(\tfrac1{12},\ \tfrac13,\ \tfrac16)$ \\
\hline
12&12&[6,\ 1,\ 0,\ 4]&$(\tfrac1{12},\ 0,\ \tfrac13)$ \\
\hline
12&12&[6,\ 0,\ 1,\ 4]&$(0,\ \tfrac1{12},\ \tfrac13)$ \\
\hline
12&12&[4,\ 1,\ 1,\ 4]&$(\tfrac1{12},\ \tfrac1{12},\ \tfrac13)$ \\
\hline
12&12&[0,\ 3,\ 1,\ 4]&$(\tfrac1{4},\ \tfrac1{12},\ \tfrac13)$ \\
\hline
12&12&[4,\ 1,\ 0,\ 6]&$(\tfrac1{12},\ 0,\ \tfrac12)$ \\
\hline
\end{tabular}
\end{center}

\newpage

\begin{center}
\small {\bf Table 6.} Continuation.
 \medskip

\begin{tabular}
{|c|c||c|c|} \hline $M$&$N$&$[s_0,s_1,s_2,s_3]$
        &$(\tfrac{s_1}M,\tfrac{s_2}M,\tfrac{s_3}M)$\\
\hline\hline
12&12&[4,\ 0,\ 1,\ 6]&$(0,\ \tfrac1{12},\ \tfrac12)$ \\
\hline
12&12&[0,\ 2,\ 1,\ 6]&$(\tfrac1{6},\ \tfrac1{12},\ \tfrac12)$ \\
\hline \hline
15&15&[3,\ 1,\ 2,\ 6]&$(\tfrac1{15},\ \tfrac2{15},\ \tfrac6{15})$ \\
\hline
15&30&[6,\ 2,\ 1,\ 3]&$(\tfrac2{15},\ \tfrac1{15},\ \tfrac15)$ \\
\hline \hline
20&20&[8,\ 1,\ 3,\ 4]&$(\tfrac1{20},\ \tfrac3{20},\ \tfrac1{5})$ \\
\hline
20&20&[4,\ 3,\ 1,\ 8]&$(\tfrac3{20},\ \tfrac1{20},\ \tfrac25)$ \\
\hline \hline
24&24&[6,\ 5,\ 1,\ 6]&$(\tfrac5{24},\ \tfrac1{24},\ \tfrac1{4})$ \\
\hline
24&24&[6,\ 1,\ 5,\ 6]&$(\tfrac1{24},\ \tfrac5{24},\ \tfrac14)$ \\
\hline \hline
30&30&[10,\ 1,\ 6,\ 6]&$(\tfrac1{30},\ \tfrac15,\ \tfrac1{5})$ \\
\hline
30&30&[6,\ 6,\ 1,\ 10]&$(\tfrac1{5},\ \tfrac1{30},\ \tfrac13)$ \\
\hline \hline
\end{tabular}
\end{center}

\setcounter{equation}{0}

\section[Solutions of the Neumann boundary value problem\\
on $n$-dimensional simplexes]{Solutions of the Neumann boundary value problem\\
on $\boldsymbol{n}$-dimensional simplexes}

\subsection[The case of $n$-dimensional simplexes related to
$A_n$, $B_n$, $C_n$ and $D_n$]{The case of
$\boldsymbol{n}$-dimensional simplexes related to
$\boldsymbol{A_n}$, $\boldsymbol{B_n}$, $\boldsymbol{C_n}$ and
$\boldsymbol{D_n}$}

Let $F$ be the fundamental domain of one of the affine Weyl groups
$W^{\rm aff}(A_n)$, $W^{\rm aff}(B_n)$, $W^{\rm aff}(C_n)$,
$W^{\rm aff}(D_n)$. We use the orthogonal coordinates
$x_1,x_2,\ldots ,x_{n+1}$ in $F$ in the case of $W^{\rm aff}(A_n)$
and the orthogonal coordinates $x_1,x_2,\ldots ,x_n$ in other
cases (see Section \ref{weylABCD}). Thus the fundamental domain
$F$ for $W^{\rm aff}(A_n)$ is placed in the hyperplane
$x_1+x_2+\cdots +x_{n+1}=0$.

We consider the Laplace operator
 \[
\Delta=\frac{\partial^2}{\partial
x^2_1}+\frac{\partial^2}{\partial x^2_2}+\cdots
+\frac{\partial^2}{\partial x^2_r}
 \]
on $F$, where $r=n+1$ for $A_n$ and $r=n$ for $B_n$, $C_n$ and
$D_n$. Let us take a summand from the expression \eqref{orb-B} for
the orbit function $\phi_\lambda(x)$ of $B_n$ and act upon it by
the operator $\Delta$. We get
 \begin{gather*}
\Delta  e^{2\pi{\rm i}((w(\varepsilon \lambda))_1x_1+
 \cdots + (w(\varepsilon \lambda))_nx_{n})}
=-4\pi^2[(\varepsilon_1m_1)^2+
 \cdots +(\varepsilon_nm_{n})^2]e^{2\pi{\rm i}((w(\varepsilon \lambda))_1x_1+
 \cdots + (w(\varepsilon \lambda))_nx_{n})}\!\!
 \\
 \qquad\qquad{}
=-4\pi^2(m_1^2+\cdots +m_n^2)e^{2\pi{\rm i}((w(\varepsilon
\lambda))_1x_1+
 \cdots + (w(\varepsilon \lambda))_nx_{n})}
 \\   \qquad\qquad{}
=-4\pi^2 \langle \lambda,\lambda \rangle\, e^{2\pi{\rm
i}((w(\varepsilon \lambda))_1x_1+
 \cdots + (w(\varepsilon \lambda))_nx_{n})},
 \end{gather*}
where $\lambda=(m_1,m_2,\ldots ,m_n)$ is the weight, determining
$\phi_\lambda(x)$, in the orthogonal coordinates and $w\in
S_n/S_\lambda$. Since this action does not depend on a summand
from \eqref{orb-B}, we have
\begin{gather}\label{Lap}
\Delta \phi_\lambda(x)= -4\pi^2\langle \lambda,\lambda \rangle
\phi_\lambda(x).
 \end{gather}
For $A_n$, $C_n$ and $D_n$ this formula also holds and the
corresponding proofs are the same. Remark that in the case $A_n$
the scalar product $\langle \lambda,\lambda \rangle$ is equal to
 \[
\langle \lambda,\lambda \rangle =m_1^2+m_2^2+\cdots +m_{n+1}^2.
 \]
Thus, orbit functions are eigenfunctions of the Laplace operator
on the fundamental domain $F$ satisfying the Neumann boundary
condition
 \begin{gather}\label{Neum}
 \left.\frac{\partial \phi_\lambda(x)}{\partial
m}\right|_{\partial F}=0\,, \qquad \lambda\in P_+\,,
\end{gather}
where $\partial F$ is the $(n-1)$-dimensional boundary of $F$ and
$m$ is the normal to the boundary.

\subsection[The Laplace operator in the $\omega$-basis]{The Laplace operator in the $\boldsymbol{\omega}$-basis}

Now we parametrize elements of $F$ by the coordinates in the
$\omega$-basis: $x=\theta_1\omega_1+\cdots+ \theta_2\omega_2$.
Denoting by $\p_k$ the partial derivative with respect to
$\theta_k$, we have the Laplace operator $\Delta$ in the form
\begin{gather}\label{operator}
\Delta = \sum_{i,j=1}^n\l\alpha_i\mid\alpha_i\r^{-1}
       M_{ij}\p_i\p_j,
\end{gather}
where $(M_{ij})$ is the corresponding Cartan matrix. One can see
that it is indeed the Laplace operator as follows. The matrix
$(S_{ij})=(\l\alpha_j\mid\alpha_j\r M_{ij})$ is symmetric with
respect to transposition and its determinant is positive. Hence it
can be diagonalized, so that $\Delta$ becomes a sum of second
derivatives (with respect to new variables) with no mixed
derivative terms.

\subsection{Rank two and three special cases}

Problems in solving the Neumann boundary value problem on $F$ is
most often encountered in dimensions 2 and 3. We write down the
explicit form of the Laplace operators in coordinates relative to
the $\omega$-basis for ranks 2 and 3 derived from formula
\eqref{operator}. For rank two the operator $\Delta$ is of the
form
\begin{alignat}{3}
A_2\,&:&\ (\p_1^2-\p_1\p_2+\p_2^2)\phi
     &=-\tfrac{4\pi^2}3(a^2+ab+b^2)\phi,\quad
     &&F=\{0,\omega_1,\omega_2\},
\\
C_2\,&:&\ (2\p_1^2-2\p_1\p_2+\p_2^2)\phi
     &=-2\pi^2(a^2+4ab+4b^2)\phi,\quad
     &&F=\{0,\omega_1,\omega_2\},
\\
G_2\,&:&\ (\p_1^2-3\p_1\p_2+3\p_2^2)\phi
     &=-\tfrac{4\pi^2}3(3a^2+3ab+b^2)\phi,\quad
     &&F=\{0,\tfrac{\omega_1}2,\omega_2 \}.
\end{alignat}
Here, in order to simplify the notation, $\phi$ stands for
$\phi_\lambda(x)$, $\lambda=(a\,b)$ and $x=(\theta_1\,\theta_2)$.
Although the same symbols are used for analogous objects in the
three cases, their geometric meaning is very different. It is
given by the appropriate matrix $M$ in \eqref{Mmatrix}. In
particular, the vertices of~$F$ form an equilateral triangle in
case of $A_2$, for $C_2$ the triangle is half of a square, and it
is a half of an equilateral triangle for $G_2$. In the semisimple
case $A_1\times A_1$ one has
$M=2\left(\begin{smallmatrix}1&0\\0&1\end{smallmatrix}\right)$,
therefore $\Delta=2\p_1^2+2\p_2^2$, and $\phi_\lambda(x)$ is the
product of two orbit functions, one from each $A_1$. The
fundamental domain is the square.

There are three 3-dimensional cases to consider, namely $A_3$,
$B_3$, and $C_3$. In addition there are four cases involving
non-simple groups of the same rank. For $A_3$, $B_3$, and $C_3$
the result can be represented by the formulas
 \begin{alignat}{2} A_3\
&:&\quad\Delta&=\p_1^2+\p_2^2+\p_3^2-\p_1\p_2-\p_2\p_3,
\notag\\
B_3\ &:&\quad\Delta&=\p_1^2+\p_2^2+2\p_3^2-\p_1\p_2-2\p_2\p_3,
\notag
\\
C_3\ &:&\quad\Delta&= 2\p_1^2+ 2\p_2^2+\p_3^2-
2\p_1\p_2-2\p_2\p_3. \notag
\end{alignat}

\subsection{Orbit functions as eigenfunctions of other operators}
Orbit functions are eigenfunctions of many other operators. We
consider examples of such operators.

With each $y\in E_n$ we associate the shift operator $T_y$ which
acts on the exponential functions $e^{2\pi{\rm i}\langle \lambda,x
\rangle}$ as
 \[
T_y e^{2\pi{\rm i}\langle \lambda,x \rangle}=e^{2\pi{\rm i}\langle
\lambda,x+y \rangle}=e^{2\pi{\rm i}\langle \lambda,y
\rangle}e^{2\pi{\rm i}\langle \lambda,x \rangle}.
 \]
We define an action of elements of the Weyl group $W$ on
functions, given on $E_n$, as $wf(x)=f(wx)$. Now for each $y\in
E_n$ we define an operator acting on orbit functions by the
formula
 \[
D_y=\sum_{w\in W} wT_y.
 \]
Then
\begin{gather*}
D_y\phi_\lambda(x)=D_y\sum _{w\in W/W_\lambda} e^{2\pi{\rm
i}\langle w\lambda,x \rangle} =\sum_{w'\in W} \sum _{w\in
W/W_\lambda} e^{2\pi{\rm i}\langle w\lambda,y \rangle} e^{2\pi{\rm
i}\langle w\lambda,w'x \rangle}
\\
\phantom{D_y\phi_\lambda(x)}{} =\sum _{w\in W/W_\lambda}
e^{2\pi{\rm i}\langle w\lambda,y \rangle} \sum _{w'\in W}
e^{2\pi{\rm i}\langle w\lambda,w'x \rangle}
\\
\phantom{D_y\phi_\lambda(x)}{} =\sum _{w\in W/W_\lambda}
e^{2\pi{\rm i}\langle w\lambda,y \rangle} \sum _{w'\in W}
e^{2\pi{\rm i}\langle {w'}^{-1}w\lambda,x \rangle}
\\
\phantom{D_y\phi_\lambda(x)}{} =|W_\lambda| \sum _{w\in
W/W_\lambda} e^{2\pi{\rm i}\langle w\lambda,y
\rangle}\phi_\lambda(x)
=|W_\lambda|\phi_\lambda(y)\phi_\lambda(x),
\end{gather*}
that is, $\phi_\lambda(x)$ is an eigenfunction of the operator
$D_y$ with eigenvalue $|W_\lambda|\phi_\lambda(y)$.

It is shown similarly that in the cases of $A_n$, $B_n$, $C_n$,
$D_n$ orbit functions $\phi_\lambda(x)$ are eigenfunctions of the
operators
 \[
\sum_{w\in W} w\frac{\partial^2}{\partial^2 x_i}, \qquad
 i=1,2,\ldots ,r,
 \]
where $x_1,x_2,\ldots,x_r$ are orthogonal coordinates of the point
$x$, $r=n+1$ for $A_n$ and $r=n$ for other cases. In fact, these
operators are multiple to the Laplace operator $\Delta$.

It is easy to show that in the cases of $A_n$, $B_n$, $C_n$, and
also $D_n$ with even $n$, orbit functions~$\phi_\lambda(x)$ are
solutions of the equations
 \[
\sum_{w\in W} w\frac{\partial}{\partial x_i} f=0, \qquad
 i=1,2,\ldots ,r.
 \]

 \setcounter{equation}{0}

\section{Orbit functions and symmetric polynomials}

\subsection{Orbit functions and monomial symmetric polynomials}

As is mentioned in Introduction, orbit functions are a certain
modification of monomial symmetric polynomials $m_\lambda(y)$,
$\lambda\in P_+$. For simplicity, we restrict ourselves to the
case of root systems and Weyl groups of $A_{n-1}$, $B_n$, $C_n$
and $D_n$. We use for $x\in E_n$ and for elements $\lambda\in P$
the orthogonal coordinate systems described in Section
\ref{weylABCD} (moreover we assume that orthogonal coordinates
$m_1,m_2,\ldots, m_n$ of $\lambda\in P$ take only integral values
in the cases of $A_{n-1}$ and $B_n$). Then elements of $W$ have a
natural description in term of permutations and changes of signs.

In the expression for orbit functions
$\phi_\lambda(x)=\sum\limits_{\mu \in O(\lambda)} e^{2\pi{\rm
i}\sum\limits_i x_i\mu_i}$, $\lambda\in P_+$, we replace each
$e^{2\pi{\rm i}x_j}$  by $y_j$. Then orbit functions
$\phi_\lambda(x)$ turn into the Laurent polynomials (that is,
polynomials in $y_1,y_2,\ldots,y_n,y_1^{-1},y_2^{-1},\ldots,
y_n^{-1}$)
 \begin{gather}\label{mon}
m_\lambda(y)=\sum_{\mu\in O(\lambda)} y^\mu \equiv \sum_{\mu\in
O(\lambda)} y_1^{\mu_1}y_2^{\mu_2}\cdots y_n^{\mu_n},
 \end{gather}
where $\mu_1,\mu_2,\ldots ,\mu_n$ are orthogonal coordinates of
$\mu\in P$. They are called {\it monomial symmetric polynomials}.
They are very useful for studying symmetric (with respect to $W$)
Laurent polynomials, which constitute orthogonal bases of the
space ${\mathbb C}[y_1,y_2,\ldots ,y_n]^W$ of all symmetric (under
the Weyl group $W$) Laurent polynomials in $y_1,y_2,\ldots ,y_n$
with respect to some scalar products.

For studying symmetric orthogonal polynomials in ${\mathbb
C}[y_1,y_2,\ldots ,y_n]^W$ one usually replaces $y^\mu$ by
$e^\mu$, where $e^\mu$ is considered as a function on $E_n$:
 \[
e^\mu(x)=e^{\langle \mu,x \rangle} =e^{\mu_1x_1+\cdots +\mu_nx_n}.
 \]
Then instead of the orbit functions $\phi_\lambda(x)$ considered
above we obtain the modified orbit functions
 \begin{gather}\label{mod}
\tilde\phi_\lambda(x)=\sum_{\mu\in O(\lambda)} e^{\langle \mu,x
\rangle} = \sum_{\mu\in O(\lambda)}e^{\mu_1x_1+\cdots +\mu_nx_n}.
 \end{gather}

Usually in the theory of symmetric polynomials the functions
\eqref{mod} are denoted by $m_\lambda(x)$ (see, for example,
\cite{Mac3}). We used the symbol $m_\lambda$ for polynomials
\eqref{mon}. In order to be closer to the notations of the theory
of symmetric polynomials, we denote functions~\eqref{mod}, which
are Laurent polynomials in $e^{x_j}$, $j=1,2,\ldots ,n$, by $\hat
m_\lambda(x)$. It is shown in the same way as in
Section~\ref{sec9} that the functions $\hat m_\lambda(x)$ are
eigenfunctions of the Laplace operator
$\Delta=\frac{\partial^2}{\partial
x^2_1}+\frac{\partial^2}{\partial x^2_2}+\cdots
+\frac{\partial^2}{\partial x^2_n}$, namely,
 \[
\Delta \hat m_\lambda(x)=\langle \lambda,\lambda \rangle \hat
m_\lambda(x)=(\lambda_1^2+\lambda_2^2+\cdots +\lambda_n^2)\, \hat
m_\lambda(x).
 \]

As a rule, different types of orthogonal symmetric Laurent
polynomials in $e^{x_j}$, $j=1,2,\ldots ,n$, are eigenfunctions of
operators, which are obtained from $\Delta$ by adding some terms.
We shall see this below. We shall also see how monomial symmetric
polynomials $\hat m_\lambda$ are used for construction of such
eigenfunctions.

Note that if we take integral orthogonal coordinates
$m_1,m_2,\ldots, m_n$ in the $A_n$ case in such a~way that $m_1\ge
m_2\ge \cdots\ge m_n\ge 0$, then Laurent polynomials
$m_\lambda(y)$ and $\hat m_\lambda (x)$ turn into usual (not
Laurent) polynomials.

\subsection{Jacobi symmetric polynomials}

Jacobi polynomials in one variable are well-known orthogonal
polynomials of the theory of special functions of mathematical
physics. Jacobi polynomials of many variables are symmetric
(Laurent) polynomials which are defined by means of polynomials
$\hat m_\lambda$, $\lambda\in P_+$. We fix for every root
$\alpha\in R$ a positive integer $k_\alpha$ such that
$k_{w\alpha}=k_\alpha$ for each $w\in W$. Since there exist only
one or two $W$-orbits of roots in $R$, we have one or two numbers
$k$, respectively. We introduce the notation
 \[
\rho_k=\frac 12 \sum_{\alpha\in R} k_\alpha \alpha.
 \]
Next we construct the operator
 \[
M_2=\Delta-\sum_{\alpha\in R} k_\alpha
\frac{1+e^\alpha}{1-e^\alpha} \partial_\alpha ,
 \]
where $\partial_\alpha$ is the derivative in direction of the root
$\alpha$ and $e^\alpha$ is the function on $E_n$, defined in the
previous subsection. The following properties of the operator
$M_2$ are proved in \cite{Heck, HeckO}:
\medskip

(i) $M_2$ preserves the space ${\mathbb C}[e^{x_1},e^{x_2},\ldots
,e^{x_n}]^W$.

(ii) The action of $M_2$ on functions $\hat m_\lambda(x)$ is
triangular:
 \[
M_2 \hat m_\lambda(x)=\langle \lambda,\lambda+2\rho_k \rangle \hat
m_\lambda(x) +\sum_{\mu<\lambda} \hat m_\mu(x) ,
 \]
where, as before, $\mu<\lambda$ means that $\lambda-\mu\in Q_+$
and $\mu\ne\lambda$.

\medskip

The following theorem is crucial in the definition of Jacobi
symmetric polynomials, proof of which can be found in  \cite{Heck}
and \cite{HeckO}.

\begin{theorem}\label{theorem2}  To every $\lambda\in P_+$ there corresponds a
unique polynomial $J_\lambda\in {\mathbb C}[e^{x_1},\ldots
,e^{x_n}]^W$ such that
\begin{gather}\label{Jac}
J_\lambda(x)=\hat m_\lambda(x)+{\rm lower\ order\ terms},
 \\
 \label{Jaco}
M_2J_\lambda(x)=\langle \lambda,\lambda+2\rho_k \rangle
J_\lambda(x) ,
 \end{gather}
where under lower order terms linear combinations of functions
$\hat m_\mu(x)$, $\mu\in P_+$, with $\mu<\lambda$ are understood.
\end{theorem}

Replacing $e^{x_j}$ by $y_j$, $j=1,2,\ldots ,n$, in $J_\lambda$ we
obtain symmetric (Laurent) polynomials which are called {\it
Jacobi polynomials of many variables}. They are orthogonal with
respect to a certain positive measure which will be given in the
next subsection.

Replacing $y_j$ by $e^{2\pi{\rm i}x_j}$, $j=1,2,\ldots ,n$, in
Jacobi polynomials we obtain functions of $x_1,x_2,\ldots,$ $x_n$
which are linear combinations of orbit functions and, therefore,
are invariant with respect to the affine Weyl group $W^{\rm aff}$.
This means that, as in the case of orbit functions, they are
uniquely determined by their values on the fundamental domain of
the group $W^{\rm aff}$.

\subsection{Macdonald symmetric polynomials}

Macdonald symmetric (Laurent) polynomials are also constructed by
means of monomial symmetric polynomials. They are a quantum
analogue of Jacobi symmetric polynomials, considered in the
previous subsection.

We introduce a variable $q$ and with every root $\alpha\in R$
associate a variable $t_\alpha$ such that $t_\alpha=t_{w\alpha}$,
$w\in W$ (therefore, there exist one or two variables $t$). Let
${\mathbb C}_{q,t_\alpha}\equiv {\mathbb C}(q,t_\alpha)$ be the
field of rational functions in $q$ and $t_\alpha$. If
$e^{x_1},e^{x_2},\ldots ,e^{x_n}$ are such as in the previous
subsection, then
 \[
{\mathbb C}_{q,t_\alpha}[e^{x_1},e^{x_2},\ldots ,e^{x_n}]
 \]
will denote the set of (Laurent) polynomials in
$e^{x_1},e^{x_2},\ldots ,e^{x_n}$ with coefficients from ${\mathbb
C}(q,t_\alpha)$.

For each function $\sum\limits_\lambda a_\lambda e^\lambda$ on
$E_n$ we define the constant term $[\sum\limits_\lambda a_\lambda
e^\lambda]_0$ coinciding with
 \[
[\sum_\lambda a_\lambda e^\lambda]_0 =a_0.
 \]
Now one can determine an inner product $\langle \cdot,\cdot
\rangle_{q,t_\alpha}$ on ${\mathbb
C}_{q,t_\alpha}[e^{x_1},e^{x_2},\ldots ,e^{x_n}]$ by the formula
 \begin{gather}\label{scal}
 \langle p_1,p_2 \rangle_{q,t_\alpha}=|W|^{-1}
[p_1{\overline{p_2}} \Delta_{q,t_\alpha}]_0 ,
 \end{gather}
 where the bar over $p_2$ denotes the linear involution which is
 uniquely determined by $\overline{e^\lambda}=e^{-\lambda}$, and
 \[
\Delta_{q,t_\alpha}=\prod_{\alpha\in R} \prod_{i=1}^\infty
\frac{1-q^{2i}e^\alpha}{1-t_\alpha^2q^{2i}e^\alpha} .
 \]
Here $\Delta_{q,t_\alpha}$ must be considered as a Laurent series
in $q$ and $t_\alpha$ with coefficients from the space ${\mathbb
C}[e^{x_1},e^{x_2},\ldots ,e^{x_n}]$. The inner product
\eqref{scal} is non-degenerate and invariant with respect to~$W$.
I.~Macdonald \cite{Mac3} proved the following theorem:

\begin{theorem}\label{theorem3} There exists a unique family $P_\lambda \in
{\mathbb C}_{q,t_\alpha}[e^{x_1},e^{x_2},\ldots ,e^{x_n}]^W$,
$\lambda\in P_+$, satisfying the conditions
 \begin{gather*}
P_\lambda=\hat m_\lambda +\sum_{\mu<\lambda} a_\lambda^\mu \hat
m_\mu ,\qquad  a_\lambda^\mu\in {\mathbb C}_{q,t_\alpha},
\\
 \langle P_\lambda,P_\mu \rangle_{q,t_\alpha}=0,\qquad
 {\rm if}\qquad \lambda\ne\mu .
\end{gather*}
\end{theorem}

Replacing $e^{x_j}$ by $y_j$, $j=1,2,\ldots ,n$, in $P_\lambda$ we
obtain (for each fixed values of $q$ and $t_\alpha$) orthogonal
symmetric polynomials which are called {\it Macdonald symmetric
polynomials}.

Replacing $y_j$ by $e^{2\pi{\rm i}x_j}$, $j=1,2,\ldots ,n$, in
Macdonald polynomials we obtain orthogonal functions which are
finite linear combinations of orbit functions and are invariant
with respect to the affine Weyl group $W^{\rm aff}$. This means
that, as in the case of orbit functions, they are uniquely
determined by their values on the fundamental domain of $W^{\rm
aff}$.

For some special values of $q$ and $t_\alpha$ (see \cite{Mac3})
Macdonald polynomials reduce to more simple sets of polynomials:

(a) If $t_\alpha=1$, then $P_\lambda=\hat m_\lambda$ independently
of $q$.

(b) If $t_\alpha=q$ for all $\alpha\in R$, then
$P_\lambda=\chi_\lambda$, where $\chi_\lambda$ are characters of
finite dimensional irreducible representations of the
corresponding simple Lie groups.

(c) If $q,t_\alpha\to 1$ in such a way that
$t_\alpha=q^{k_\alpha}$, $k_\alpha\in {\mathbb Z}_+$ are fixed,
then $P_\lambda\to J_\lambda$, where $J_\lambda$ are Jacobi
symmetric polynomials. In this case the inner product for
Macdonald polynomials turns into the scalar product
 \[
\langle p_1,p_2\rangle =|W|^{-1}
[p_1\overline{p_2}\delta^k\overline{\delta^k}]_0
 \]
with respect to which Jacobi polynomials are orthogonal. Here
$\delta^k=\prod\limits_{\alpha\in R_+}
(e^{\alpha/2}-e^{-\alpha/2})^{k_\alpha}$.

(d) If $q=0$ and $t_\alpha=1/p$, where $p$ is a prime integer,
then $P_\lambda$ are zonal spherical polynomials for the
corresponding $p$-adic group.

(e) For the root system $A_1$, Macdonald polynomials reduce to
continuous $q$-ultraspherical orthogonal polynomials of one
variable (for definition and properties of these polynomials
see~\cite{GR}).

\subsection*{Acknowledgements}

The first author (AK) acknowledges CRM of University of Montreal
for hospitality when this paper was under preparation. We are
grateful for partial support for this work to the National
Research Council of Canada and to MITACS.

\LastPageEnding

\end{document}